\documentclass[useAMS,usenatbib,structabstract]{aa}

\usepackage{graphicx,latexsym}
\usepackage{natbib}
\usepackage{rotating,ams}
\usepackage[ps2pdf,linktocpage]{hyperref}
\hypersetup{%
  colorlinks=false,
  bookmarksopen=true,
  bookmarksnumbered=true,
  pdfstartpage={1},  
  pdftitle = {The photometric evolution of
dissolving star clusters. II. Realistic models. Colours and M/L ratios},
  pdfauthor = {P. Anders, H.J.G.L.M Lamers, H. Baumgardt} ,
  pdfkeywords = {star clusters, N-body simulations, evolutionary 
  synthesis models, cluster age determination},
  pdfproducer = {LaTeX with hyperref}
}

\def\spose#1{\hbox to 0pt{#1\hss}}
\def\lta{\mathrel{\spose{\lower 3pt\hbox{$\mathchar"218$}}
     \raise 2.0pt\hbox{$\mathchar"13C$}}}
\def\gta{\mathrel{\spose{\lower 3pt\hbox{$\mathchar"218$}}
     \raise 2.0pt\hbox{$\mathchar"13E$}}}

\begin{document} 

\vspace{-1.5cm}
\title{The photometric evolution of
dissolving star clusters}

\subtitle{II. Realistic models. Colours and M/L ratios}

\author{P. Anders \inst{1}\fnmsep\thanks{E-mail:
P.Anders@uu.nl} \and H.J.G.L.M Lamers \inst{1}  \and H. Baumgardt
\inst{2}}

\institute{Sterrenkundig Instituut, Universiteit Utrecht, P.O. Box
80000,  NL-3508 TA Utrecht, The Netherlands
\and
Argelander Institut f\"ur Astronomie, Universit\"at Bonn, Auf dem
H\"ugel 71, 53121 Bonn, Germany
}

\date{Accepted ---. Received ---; in original form ---.}


\abstract
{Evolutionary synthesis models are the prime method to construct
spectrophotometric models of stellar populations, and to derive physical
parameters from observations through comparison with such models. One of
the basic assumptions for evolutionary synthesis models so far has been
the time-independence of the stellar mass function (except for the
successive removal of high-mass stars due to stellar evolution).
However, dynamical simulations of star clusters in tidal fields have
shown the mass function to change due to the preferential removal of
low-mass stars from clusters.}
{Here we combine the results from dynamical simulations of star clusters
in tidal fields (and especially the derived parametrisation of the
changing mass function) with our evolutionary synthesis code GALEV to
extend the models by a new dimension: the total cluster disruption
time.}
{Following up on our earlier work, which was based on simplifying
assumptions of the mass function time dependence, we reanalyse the mass
function evolution found in N-body simulations of star clusters in tidal
fields, parametrise it as a function of age and total disruption time of
the cluster and use this parametrisation to compute  GALEV models as a
function of age, metallicity and the total cluster disruption time.}  
{We study the impact of cluster dissolution on the colour (generally,
they become redder) and magnitude (they become fainter) evolution of
star clusters, their mass-to-light ratios (which can deviate by a factor
of $\sim$2 -- 4 from predictions of standard models without cluster
dissolution), and quantify the effect on the cluster age determination
from integrated photometry (in most cases, clusters appear to be older
than they are. Depending on the filter set available and the
evolutionary stage of the cluster, the age difference can range from 20
to 200\%). By comparing our model results with observed M/L ratios for
old compact objects in the mass range 10$^{4.5}$ -- 10$^8$ M$_\odot$, we
find a strong discrepancy for objects more massive than 10$^7$
M$_\odot$, in a sense that observed M/L ratios are higher than predicted
by our models. This could be either caused by differences in the
underlying stellar mass function or be an indication for the presence of
dark matter in these objects. Less massive objects are well represented
by the models.\\The models for a range of total cluster disruption times
and metallicities are available online, at
http://www.phys.uu.nl/$\sim$anders/data/SSP\_varMF/ and 
http://data.galev.org, and will be made available via CDS.}
{}


\maketitle

\section{Introduction}

Since the pioneering work by Tinsley
(\citealt{1968ApJ...151..547T,1976ApJ...203...52T,1980FCPh....5..287T}),
evolutionary synthesis modelling has become the method-of-choice to
predict spectrophotometric properties of stellar populations. Currently
popular models include {\sc starburst99}
(\citealt{1999ApJS..123....3L}), {\sc galaxev}
(\citealt{2003MNRAS.344.1000B}), {\sc galev}
(\citealt{2003A&A...401.1063A,2004A&A...413...37B}), {\sc pegase}
(\citealt{1997A&A...326..950F}), and the Maraston models
(\citealt{2005MNRAS.362..799M}), all giving predictions for single-age
populations , so-called ``Simple Stellar Populations'' (SSPs). In
addition, {\sc galaxev}, {\sc pegase} and {\sc galev} also provide
models for populations with arbitrary extended star formation histories
(SFH, like galaxies), while {\sc starburst99} only allows for an
extended constant SFH. Comparing predictions from these models with
observations allows to derive basic physical parameters of the studied
system (e.g., among many others, \citealt{2002A&A...387..412B,
2003AJ....126.1276K, 2004MNRAS.347...17A, 2004MNRAS.352..263D,
2005ApJ...634L..41K, 2006MNRAS.366..295D, 2007ApJ...667L.145S}).

While the specific input physics (e.g. the choice of stellar isochrones
and spectral libraries, the inclusion of gaseous emission) and
implementation varies among the models, some basic techniques and
limitations are inherent to all of them: assigning a spectrum to each
star along the isochrone, weighting them according to a chosen stellar
initial mass function (IMF) and integrating along the isochrone (and
over the SFH, if applicable) results in the integrated properties of the
stellar population at a given age. For all currently available models,
the stellar mass function (MF) is time-independently fixed at its
initial value, the IMF.

Cluster disruption\footnote{With {\sl disruption} we encompass all kinds
of different cluster mass loss and destruction {\sl events} (e.g. single
disruptive encounters with giant molecular clouds, infant mortality,
final cluster ``death''). {\sl Dissolution} stands for any gradual
destruction {\sl process}, e.g.  mass lost due to stellar evolution,
tidal dissolution or multiple weak encounters with giant molecular
clouds.} has become a well-studied phenomenon. It can be observed both
in the earliest phases in a cluster's life (the so-called ``infant
mortality'' caused by the removal of gas left over from the cluster
formation process by stellar winds and/or the first supernovae, see e.g.
\citealt{2003ARA&A..41...57L, 2006MNRAS.369L...9B}) and for old clusters
(e.g. the prominent tidal tails of the Milky Way globular cluster
Palomar 5, \citealt{2003AJ....126.2385O}). Age and mass distributions of
a whole star cluster {\sl system} can be used to determine the typical
disruption time of clusters of a given mass in this cluster system
(\citealt{2003MNRAS.338..717B, 2005A&A...429..173L,
2005A&A...441..949G}). This cluster disruption time is predominantly
determined  by the external tidal field the cluster is experiencing (see
\citealt{2005A&A...429..173L}), the local density of giant molecular
clouds (\citealt{2006MNRAS.371..793G}) and the occurrence of spiral arm
passages (\citealt{2007MNRAS.376..809G}). In addition, the cluster loses
mass due to stellar evolution. While in the case of ``infant mortality''
the cluster is likely (almost) completely disrupted (although a bound
core might remain, see e.g. \citealt{2006MNRAS.369L...9B} for ``infant
weight loss''), {\sl cluster dissolution} in a smooth external tidal
field is a more gradual process and accompanied by perpetual dynamical
readjustment within the cluster. The latter is characterised by a
mass-dependent probability to remove a star from a cluster: due to
energy equipartition massive stars tend to sink towards the cluster
center, while low-mass stars are driven outwards where they are more
easily removed by the surrounding tidal field
(\citealt{1969A&A.....2..151H, 1975ApJ...201..773S,
1997MNRAS.286..709G}). The resulting radial dependence of the mean
stellar mass inside a cluster is called ``mass segregation''. Mass
segregation established by the very star formation process itself is
referred to as ``primordial mass segregation'' (for observational
evidence of ``primordial mass segregation'' see e.g.
\citealt{2004A&A...416..137G, 2007AJ....134.1368C}).

\citet{2003MNRAS.340..227B} (from now on: BM03) performed the first
(and, so far, most extensive) quantitative large-scale study of how the
stellar MF inside a star cluster changes due to dynamical cluster
evolution in a tidal field. They confirmed earlier findings for a
preferential loss of low-mass stars and derived a formula describing the
change in MF slope for low-mass stars. However, their derived formula
(formula (13) in BM03) only applies to stars with masses $\le$ 0.5
M$_\odot$, while the effect is pronounced also for higher-mass stars
(see BM03 Fig. 7), which dominate the flux emerging from the cluster
(for ages shorter than a Hubble time). BM03 performed their simulations
for clusters not  primordially mass-segregated. Recently, in
\citet{2008ApJ...685..247B}, they studied also the dissolution of
initially mass-segregated clusters (with a simplified initial setup
different from the BM03 simulations, hence we can not combine these sets
of simulations), finding an even stronger MF evolution than BM03.
\citet{2008MNRAS.386.2047M} studied the evolution of the stellar MF
inside star clusters  during the gas removal/``infant weight loss''
phase, and found it to also preferentially remove low-mass stars,
leading to a flattening or even turning-over of the MF. This effect is
most pronounced for initially mass-segregated clusters, and would be
amplified by the later dynamical cluster evolution, as presented in
BM03. Although their results can not be straightforwardly combined with
the BM03 results (due to differences in model setups), both studies
suggest even further enhancement of the effects studied in this paper.

In \citet{2006A&A...452..131L} we constructed simplified evolutionary
synthesis models for solar metallicity, based on the {\sc galev} models
and the results from BM03. The main simplification made concerned the
description of the changing (logarithmic) MF, which we modelled with
fixed slopes, but a time-dependent lower mass limit (i.e. assuming that
only the lowest-mass stars are removed from the cluster, while
higher-mass stars might only be removed by stellar evolution). We scaled
our models to match the total mass in stars with M $<$ 2 M$_\odot$ with
the BM03 simulations.

Recently, this approach has been improved by \citet{2008A&A...490..151K}
by incorporating the effects of stellar remnants for clusters of
different initial masses and different total disruption times for a
range of metallicities. They showed that the presence of stellar
remnants plays a dominant role in the mass evolution of the clusters and
therefore also in the evolution of the mass-to-light ratio. They also
found that metallicity affects the colour evolution of the clusters, not
only by the difference in the colours of the stars, but also by
influencing the cluster dynamics due to the sensitivity of stellar mass
and remnant formation on metallicity. They determined colours and
mass-to-light ratios for a range of metallicities.
\citet{2008A&A...486L..21K} compared these predicted mass-to-light
ratios with the observed ones for cluster samples in different galaxies
(Milky Way, Cen A, M31 and LMC) and found that the effects of mass
segregation (and the associated preferential loss of low-mass stars) can
explain the observed range much better than the range predicted by
standard SSP models. As the models of \citet{2008A&A...490..151K} are
based on the simplified assumption of only the {\sl lowest-mass stars}
being removed from the cluster, they can be improved by models in which
the mass function changes in a physically more realistic manner, i.e. by
changing the {\sl slope} of the (logarithmic) mass function as derived
from dynamical N-body simulations. This is the purpose of this paper.


We describe our input physics in Sect. \ref{sec:input}, in particular we
reanalyse the data presented in BM03 to derive formulae parametrising
the changing mass function (Sect. \ref{sec:MF}). In Sect.
\ref{sec:results} we present our new evolutionary synthesis models, and
discuss the implications they have for mass-to-light (M/L) ratios and
cluster age determinations from observations. In Sect.
\ref{sec:comparison} we present a comparison with previous models
(\citealt{2006A&A...452..131L, 2008A&A...490..151K}) and investigate the
impact of model uncertainties (fit uncertainties, initial-final mass
relations, isochrones). We finish with our conclusions in Sect.
\ref{sec:conclusions}.

\section{Input physics}
\label{sec:input}
\subsection{N-Body simulations by BM03}

BM03 carried out a parameter study for dynamical evolution of clusters
dissolving in a tidal field. They studied clusters with a range of
particle numbers (8k -- 128k, i.e. a range in cluster mass) on circular
and elliptical orbits at different Galactocentric distances (i.e.
strengths of the surrounding gravitational field). They accounted for
mass lost due to stellar evolution (using fit formulae by
\citealt{2000MNRAS.315..543H}), two-body relaxation and the external
tidal field.

They initialised their clusters with a universal
\citet{2001MNRAS.322..231K} IMF, which is of the form: 

\begin{equation}
\xi(m){\rm d}m \sim \left\{
\begin{array}{l l}
  m^{-1.3}{\rm d}m & \quad \mbox{$m < 0.5 {\rm M}_\odot$}\\
  m^{-2.3}{\rm d}m & \quad \mbox{$m \ge 0.5 {\rm M}_\odot$}\\ 
 \end{array} 
 \right.
\end{equation}

\noindent
with masses in the range $0.1 {\rm M}_\odot \le m \le 15 {\rm M}_\odot$.
This rather low upper mass limit was chosen to account for the uncertain
kick velocities of neutron stars (or equivalently, their ejection
probability from the cluster, see BM03 for details). 


The MFs provided by BM03 are for single stars, where dynamically created
binaries are resolved in their components. They give the MF for the
whole cluster (i.e. for all stars within the tidal radius). This MF
compares well with the MF around the half-mass radius of the cluster, as
shown by BM03.

BM03 do {\sl not} take into account primordial mass segregation and
primordial binaries. However, primordial mass segregation is found to
even further increase the changes in the MFs found by BM03 (see
\citealt{2008ApJ...685..247B}). Primordial binaries seem to have little
impact on the stars {\sl evaporating} slowly from a cluster (see
\citealt{2008MNRAS.389..889K}), but enhance the number of stars
violently {\sl ejected} during strong binary interactions. However, the
latter are still only a small fraction of the stars leaving the cluster,
hence we expect little changes in our conclusions if simulations with
primordial binaries would have been included in our studies.

BM03 do not include a possibly present intermediate-mass black hole
(IMBH) in the cluster. Recently, \citet{2008ApJ...686..303G} found that
the presence of an IMBH reduces mass segregation in the centre, which
might also influence the mass loss from star clusters, although this has
still not been shown.  In addition, the existence of IMBHs in star
clusters is still under debate (see e.g. \citealt{2008MNRAS.389..379M}).

\subsection{Total cluster disruption time and the total cluster mass}
\label{sec:timemass}

Similar to BM03 we will identify the ``total cluster disruption time''
with the time when only 5\% of the initial cluster mass remains bound.
In order to avoid confusion, we will specifically label this time
t$_{\rm 95\%}$, i.e. the time when the cluster has lost 95\% of its
initial mass. However, we will provide our models for ages up to the
point where a cluster with initially 10$^6$ M$_\odot$ has lost all but
10$^2$ M$_\odot$ of its luminous mass (or to a maximum age of 16 Gyrs,
whichever occurs first). This termination age of the cluster models is
$\sim$ 20 -- 26\% longer than the cluster disruption time t$_{\rm
95\%}$ (for models with termination ages $<$ 16 Gyr, see Fig.
\ref{fig:MLratios}, bottom panel).

As pioneered by \citet{2003MNRAS.338..717B} and
\citet{2005A&A...429..173L}, we will use t$_4$ = t$_{\rm dis}^{\rm
total}(M = 10^4 M_\odot$), the total disruption time of a 10$^4$
M$_\odot$ star cluster, as a rough proxy to characterise the strength of
the gravitational field surrounding a cluster: the stronger the field
the faster the cluster will dissolve, and the shorter t$_4$. Using the
implicit equation

\begin{equation}
t_{\rm dis}^{\rm total}(M_{\rm i}) = t_4 \cdot \left (\frac{M_{\rm
i}}{10^4 M_\odot}\right )^\gamma \cdot \left (\frac{\mu_{\rm ev}(t_{\rm
dis}^{\rm total})}{\mu_{\rm ev}(t_4)}\right )^\gamma
\end{equation}

\noindent 
for the total disruption time $t_{\rm dis}^{\rm total}$ (see
\citealt{2005A&A...441..117L}), t$_4$ can be translated into the total
disruption time of clusters with arbitrary initial mass M$_{\rm i}$. For
example, a gravitational field characterised by t$_4$ = 1.3 Gyr (the
value found by \citealt{2005A&A...429..173L} for the Solar
Neighbourhood) leads to complete disruption of a cluster with 10$^3$
M$_\odot$ within approx. 300 Myr, while a 10$^6$ M$_\odot$ cluster would
survive for 22.6 Gyr. $\mu_{\rm ev}(t)$ describes the fraction of the
mass that the cluster would have at time $t$ if stellar evolution was
the only mass loss mechanism, and $\gamma$=0.62 (as determined from
observations by \citealt{2003MNRAS.338..717B,2005A&A...429..173L}, and
in agreement with N-body simulations, see BM03 and 
\citealt{2008MNRAS.389L..28G}).

The total cluster mass as a function of the fractional age t/t$_{\rm
95\%}$ is derived from formula (6) in \citet{2005A&A...441..117L} (who
also show the good agreement with the data from BM03): 

\begin{equation}
M_{\rm tot}(t) = M_{\rm i} \cdot \left\{\mu_{\rm ev}(t)^\gamma -
\frac{t}{t_{\rm 95\%}} \cdot \left[\mu_{\rm ev}(t_{\rm 95\%})^\gamma -
0.05^\gamma
\right]\right\}^{1/\gamma}
\label{eq:massloss}
\end{equation}

\noindent

with $M_{\rm i}$ = 10$^6$ M$_\odot$ the initial cluster mass. The
stellar evolution part of this equation was taken directly from the {\sc
galev} models used in the remainder of this work (for details see
below).

Since the total disruption time of clusters in a given environment
(e.g. tidal field) depends on the initial cluster mass, Eq.
\ref{eq:massloss} can also be used to calculate the initial cluster mass
for an observed present-day total mass and adopted t$_{95\%}$.

The mass fraction in stellar remnants is taken from BM03 (their formula
(16)):

\begin{equation}
f_{\rm rem}(t) = f^{\rm se}_{\rm rem}(t) + 0.18 \cdot
\left(\frac{t}{t_{\rm 95\%}}\right)^2 + 0.16 \cdot
\left(\frac{t}{t_{\rm 95\%}}\right)^3
\end{equation}

\noindent
with $f^{\rm se}_{\rm rem}$ the mass fraction in stellar remnants from
stellar evolution only (i.e. without dynamical cluster evolution
effects) taken from our {\sc galev} models, and the other two terms
describe the increase in the mass fraction of the remnants due to the
preferential loss of low-mass non-remnant stars.

The luminous mass is then: 

\begin{equation}
M_{\rm lum}(t) = M_{\rm tot}(t) \cdot (1 - f_{\rm rem}(t)).
\end{equation}

\subsection{Parametrising the changing mass function}
\label{sec:MF}

Throughout the paper we look at the logarithm of the
logarithmically binned mass function (MF). Hence, for a
\citet{1955ApJ...121..161S} MF, the {\sl power-law index} -2.35 becomes
a {\sl linear slope} of -1.35.

In order to parametrise the changes in the (logarithmic) mass function,
we

\begin{itemize}

\item took the MF data from BM03

\item divided them by the IMF (by doing so we remove the power law break
at 0.5 M$_\odot$ of the \citealt{2001MNRAS.322..231K} IMF)

\item skipped the 2 highest not-empty mass bins (as those are affected
and partially emptied by stellar evolution), and

\item fitted the remainder with a piecewise power law, independently for
every simulation and age. 

\end{itemize}

\begin{figure}
\begin{center}
 \vspace{-0.5cm}
 \hspace{1.2cm}
 \includegraphics[angle=270,width=0.9\linewidth]{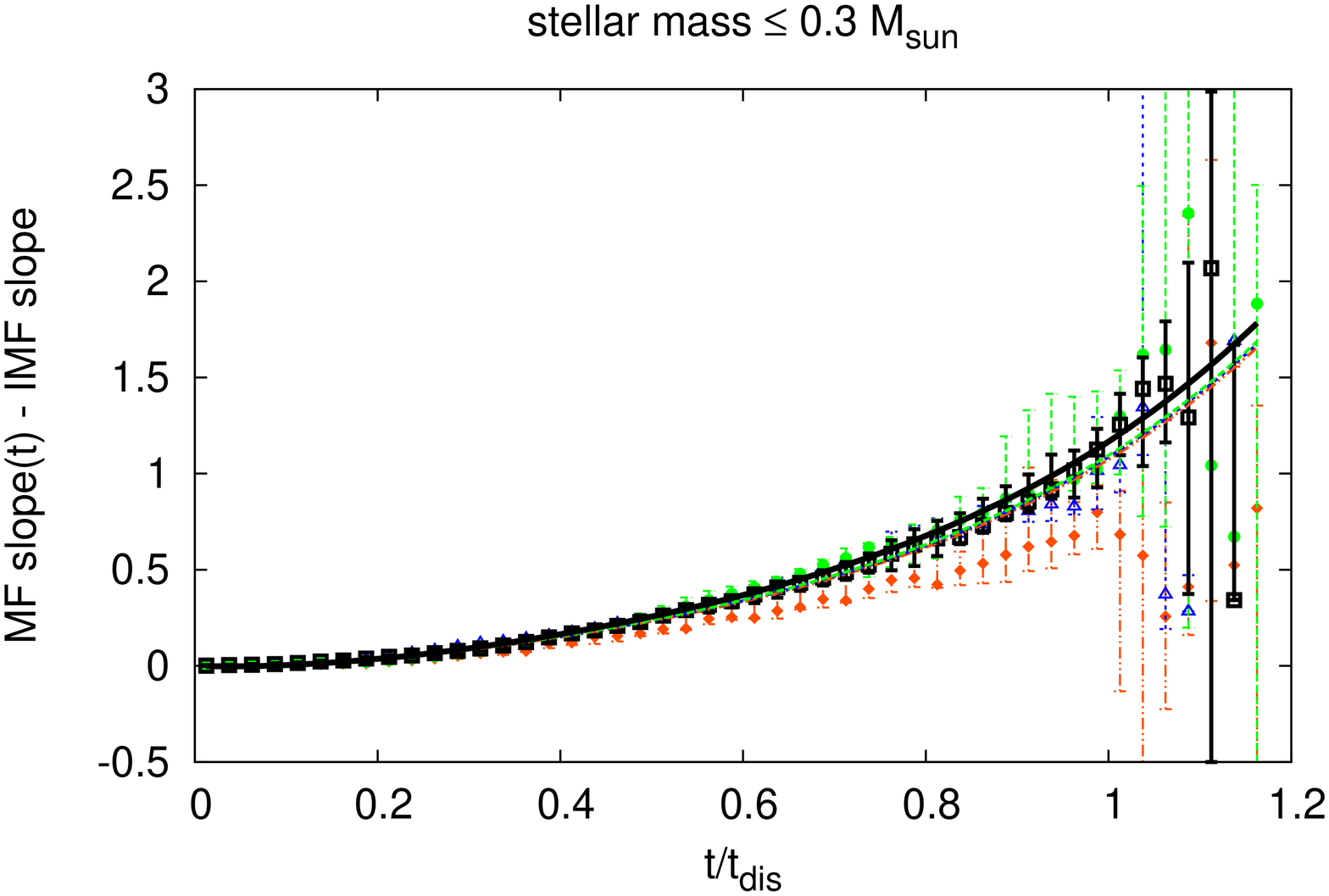}
 \includegraphics[angle=270,width=0.9\linewidth]{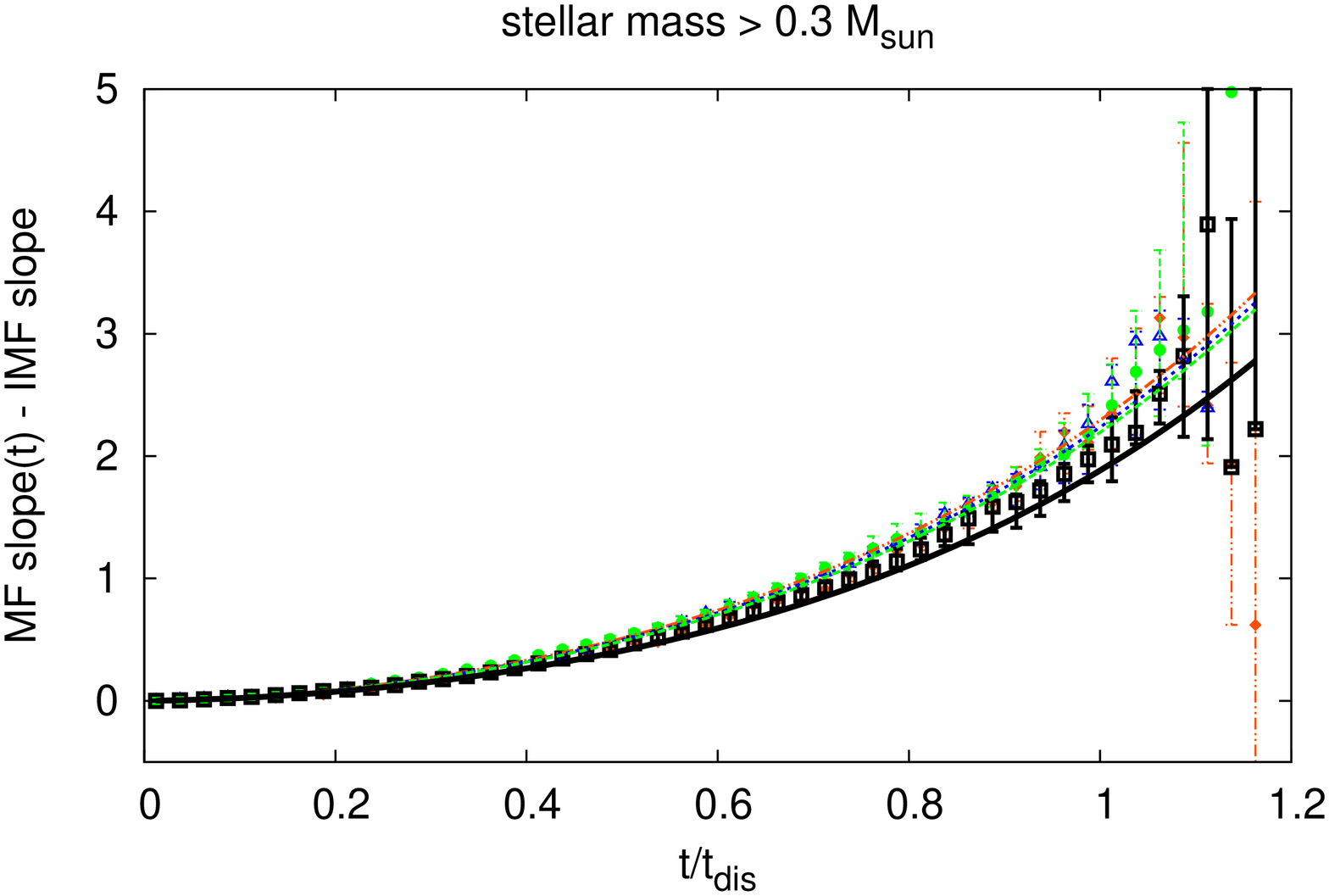}
\end{center}

\caption{Dependence of MF slopes (top panel: for stellar masses
$\le$ 0.3 M$_\odot$, bottom panel: for stellar masses $>$ 0.3 M$_\odot$)
on the fractional age t/t$_{\rm 95\%}$. Symbols represent median slopes
(and 16\% and 84\% percentiles uncertainty ranges) derived from
individual data  from BM03 runs, and binned in intervals of
$\Delta(t/t_{\rm 95\%})$ = 0.025. The data are grouped according to
their disruption times: t$_{\rm 95\%}$ $\le$ 3.5 Gyr (filled orange
diamonds), 3.5 Gyr $<$ t$_{\rm 95\%}$ $\le$ 6 Gyr (open blue triangles),
6 Gyr $<$ t$_{\rm 95\%}$ $\le$ 10 Gyr (filled green circles), 10 Gyr $<$
t$_{\rm 95\%}$ (open black squares). Smooth lines represent the fit formula
(\ref{form:fit}) for the respective age ranges: t$_{\rm 95\%}$ = 1 Gyr
(orange dot-dot-dashed line), 5 Gyr (blue, dotted), 8 Gyr (green,
dashed) and 30 Gyr (black, solid). Shown is the difference between the
time-dependent slope of the MF and the slope of the IMF
(i.e. $\alpha(t)-\alpha(0)$).}

\label{fig:slopefrac}
\end{figure}

We tried fitting slopes and break points simultaneously, finding
the results for the break points scattering in the range 0.25 -- 0.35
M$_\odot$. This scatter was found to be uncorrelated with any other
quantity, confirming it to be of random nature. Hence, we chose a double
power-law with a break point fixed to 0.3 M$_\odot$, and only fitted the
slopes below and above this break point independently for every
simulation and age. In Fig. \ref{fig:slopefrac} we show the slopes of
the changing MF, relative to the slopes of the IMF (i.e.
$\alpha(t)-\alpha(0)$). The points are derived by fitting the data from
BM03, using the abovementioned scheme, and then the median in bins of
$\Delta(t/t_{\rm 95\%})$ = 0.025 was taken. The error bars represent the
16\% and 84\% percentiles of individual data points in each bin,
equivalent to 1$\sigma$ ranges for a Gaussian distribution. These data
are grouped according to their disruption time t$_{\rm 95\%}$: t$_{\rm
95\%}$ $\le$ 3.5 Gyr  = filled orange diamonds, 3.5 Gyr $<$ t$_{\rm
95\%}$ $\le$ 6 Gyr = open blue triangles, 6 Gyr $<$ t$_{\rm 95\%}$ $\le$
10 Gyr = filled green circles, and 10 Gyr $<$ t$_{\rm 95\%}$ = open
black squares.

This fit formula, in conjunction with the \citet{2001MNRAS.322..231K}
IMF, leads to a 3-component power-law MF with break points at 0.3
M$_\odot$ and 0.5 M$_\odot$ and time-dependent slopes.

%
%

\begin{figure*}
\begin{center}
  \vspace{-0.5cm}
  \hspace{1.2cm}
  \begin{tabular}{cc}
	 \includegraphics[angle=270,width=0.4\linewidth]{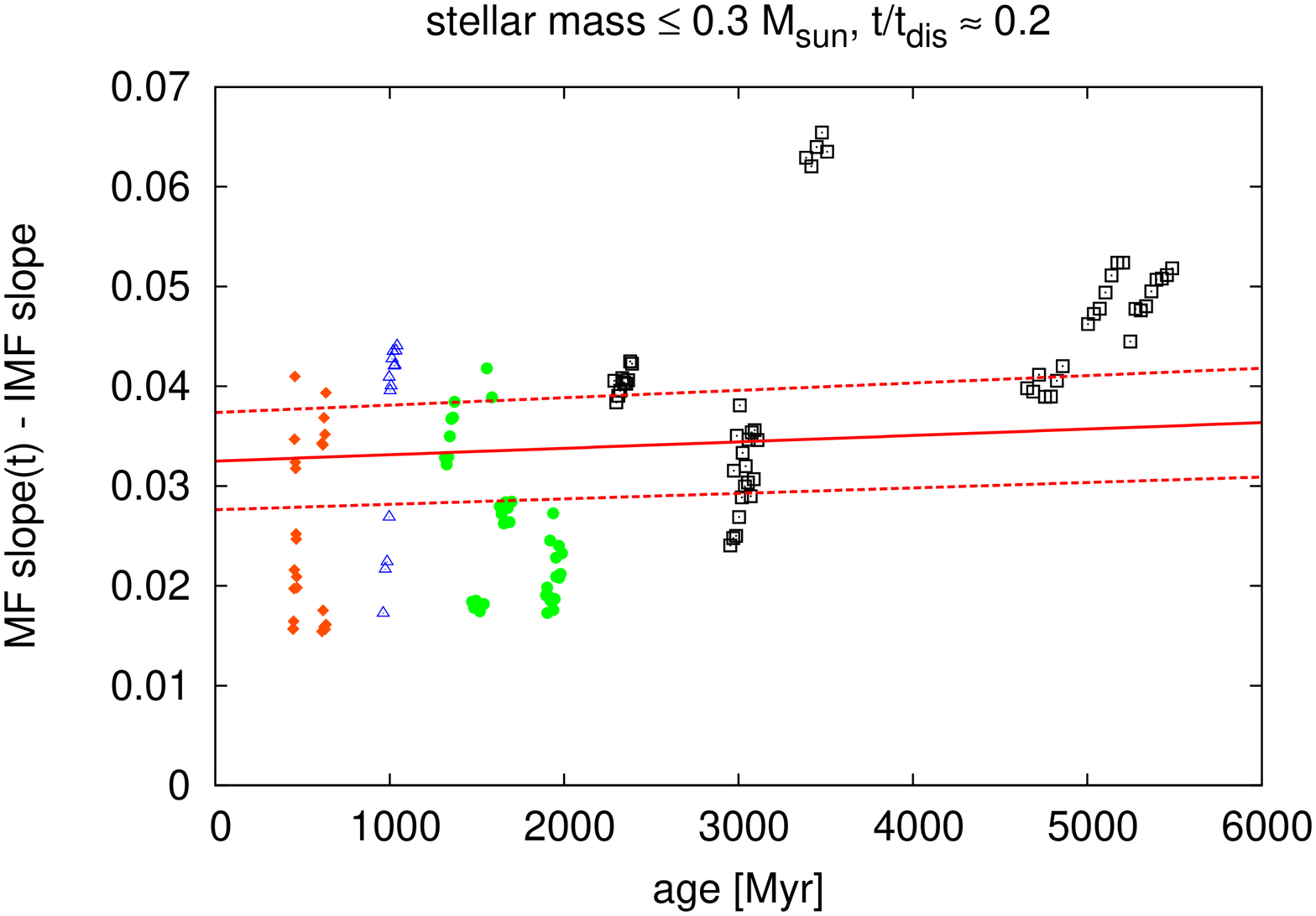} &
	 \includegraphics[angle=270,width=0.4\linewidth]{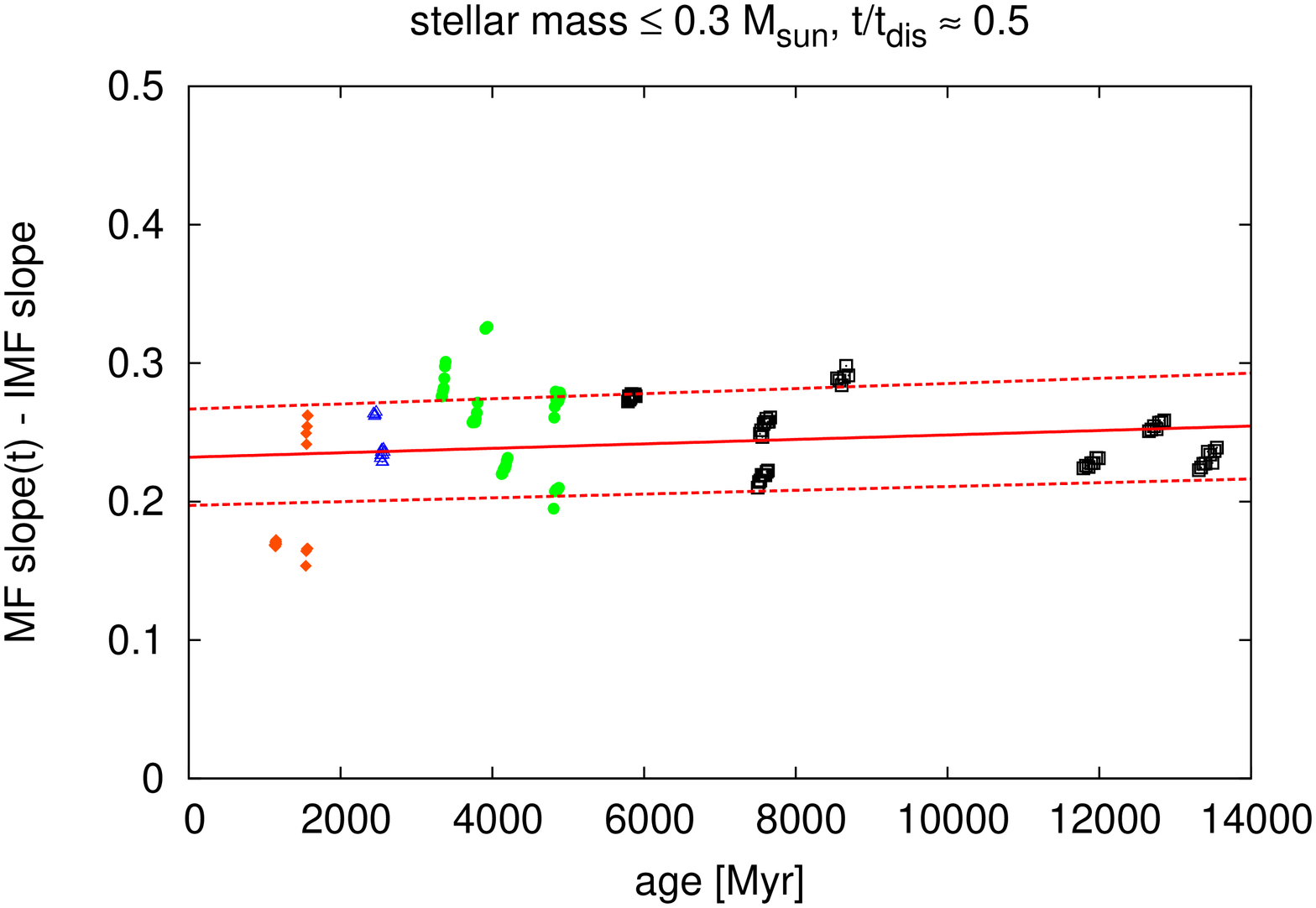} \\  
	 \includegraphics[angle=270,width=0.4\linewidth]{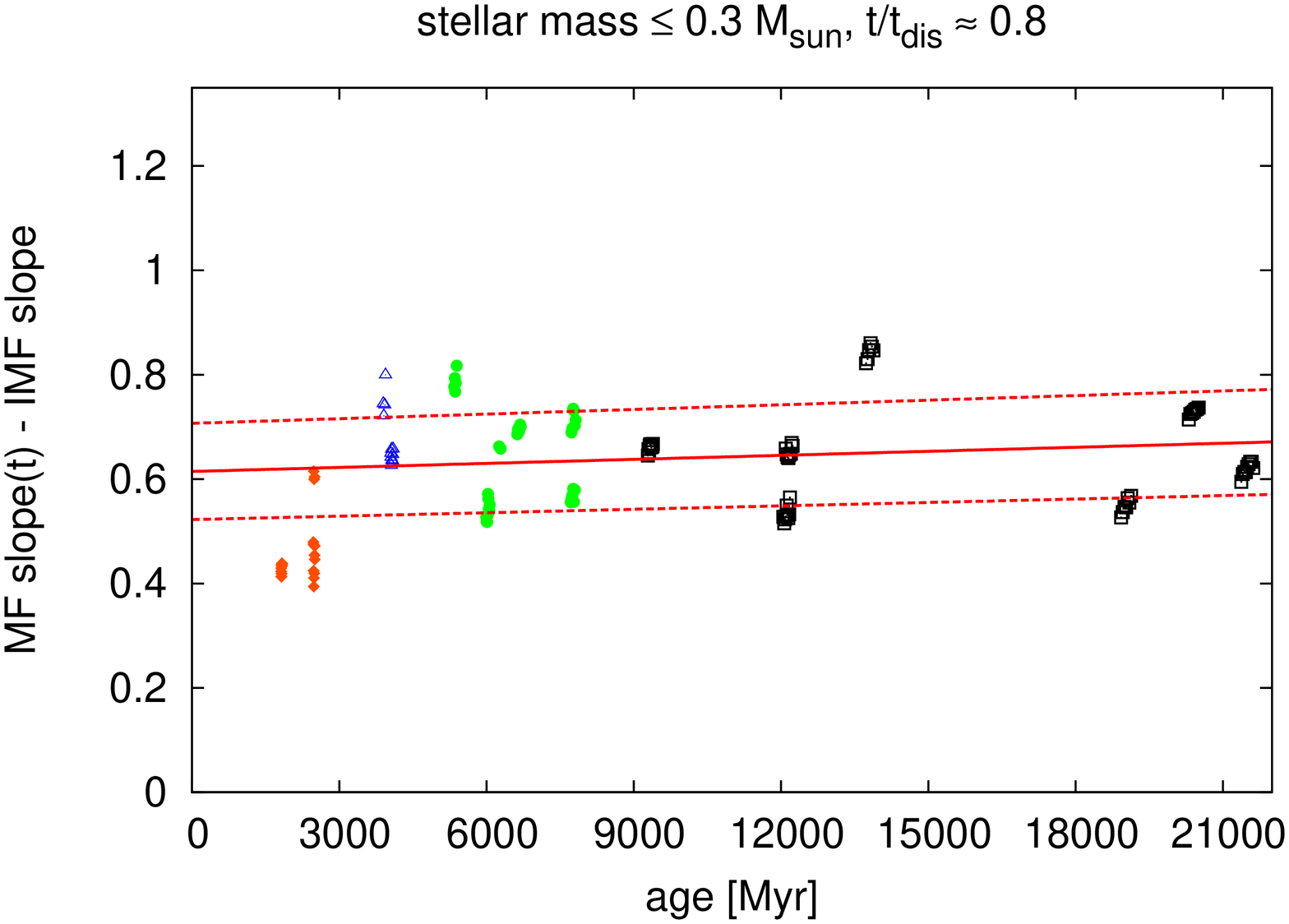} &
	 \includegraphics[angle=270,width=0.4\linewidth]{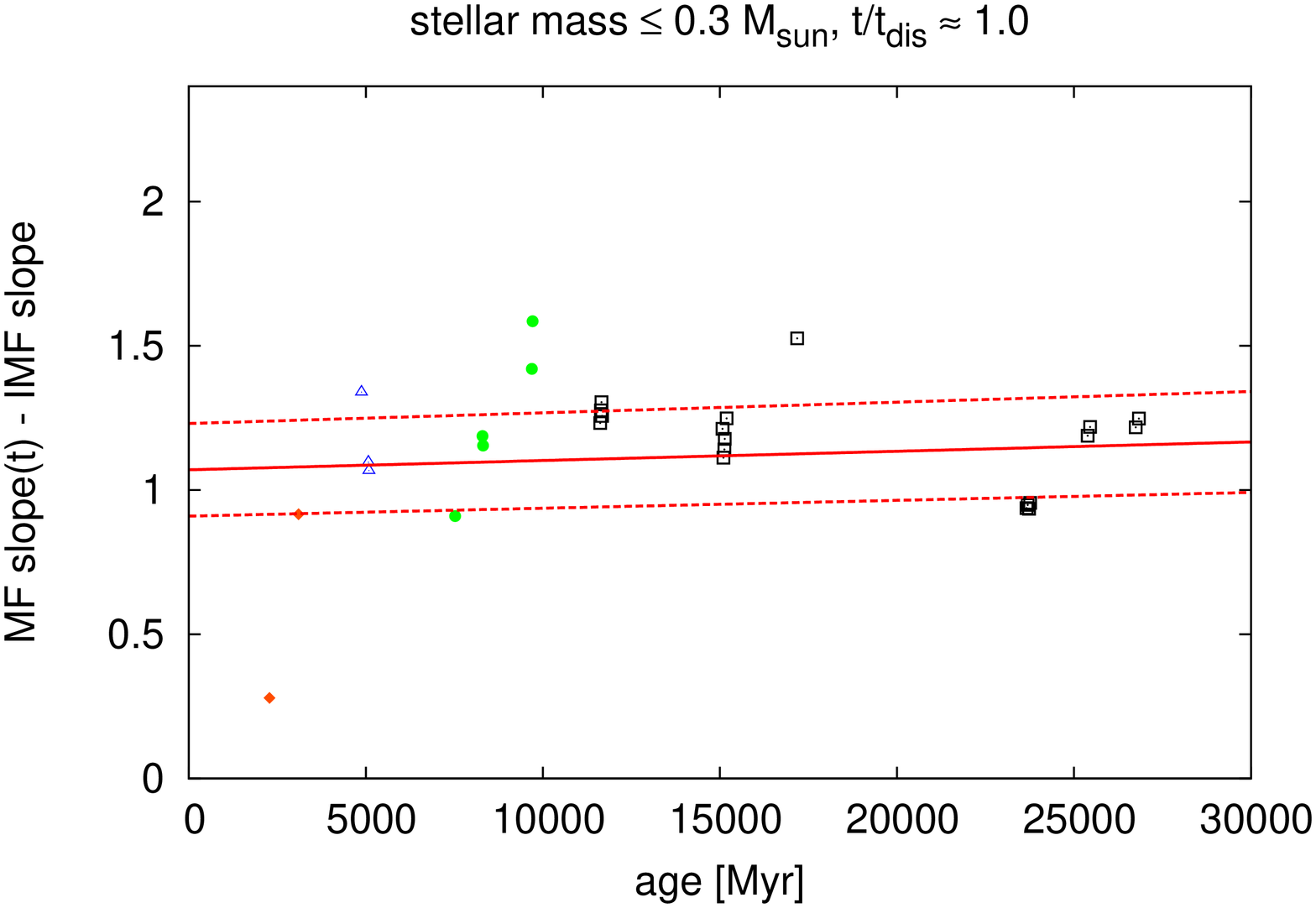} \\
  \end{tabular}
\end{center}

\caption{Dependence of MF slope on the age (in Myr) of the cluster for
stellar masses $\le$ 0.3 M$_\odot$ and for 4 different fractional ages
t/t$_{\rm 95\%}$: t/t$_{\rm 95\%}$ = 0.15-0.25 (top left panel),
t/t$_{\rm 95\%}$ = 0.45-0.55 (top right), t/t$_{\rm 95\%}$ = 0.75-0.85
(bottom left), and t/t$_{\rm 95\%}$ = 0.95-1.05 (bottom right). In each
panel, the solid line represents the fit formula (\ref{form:fit}) for
the appropriate t/t$_{\rm 95\%}$, the dashed lines represent the median spread ranges
(85\% and 115\% of fit line).  Points are the same as Fig.
\ref{fig:slopefrac}. {\bf Please beware of the very different vertical
axes.} The y-axes are chosen to go from 0 to 2$\times$maximum fit value.}

\label{fig:slope1change}

\begin{center}
  \vspace{-0.5cm}
  \hspace{1.2cm}
  \begin{tabular}{cc}
	 \includegraphics[angle=270,width=0.4\linewidth]{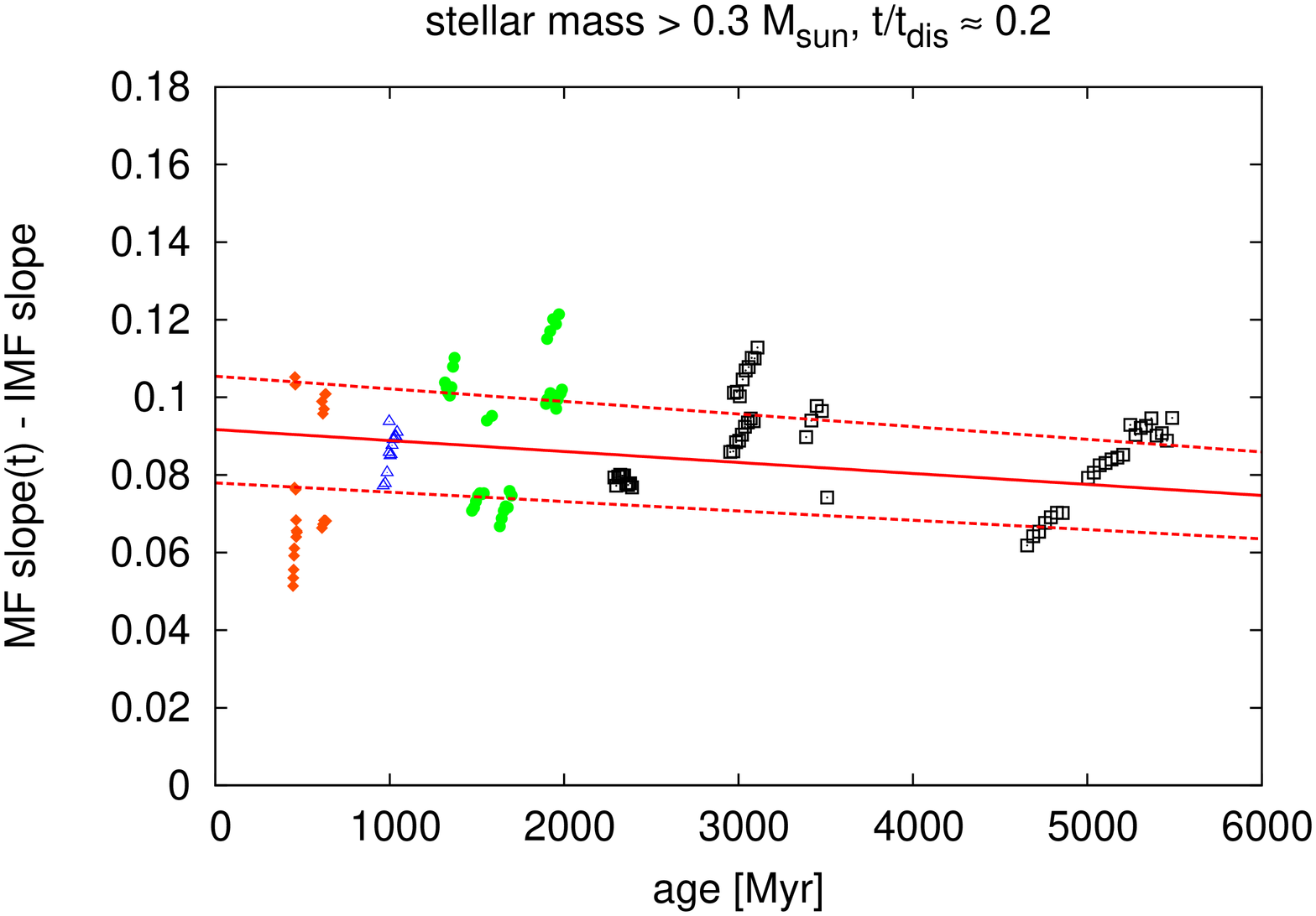} &
	 \includegraphics[angle=270,width=0.4\linewidth]{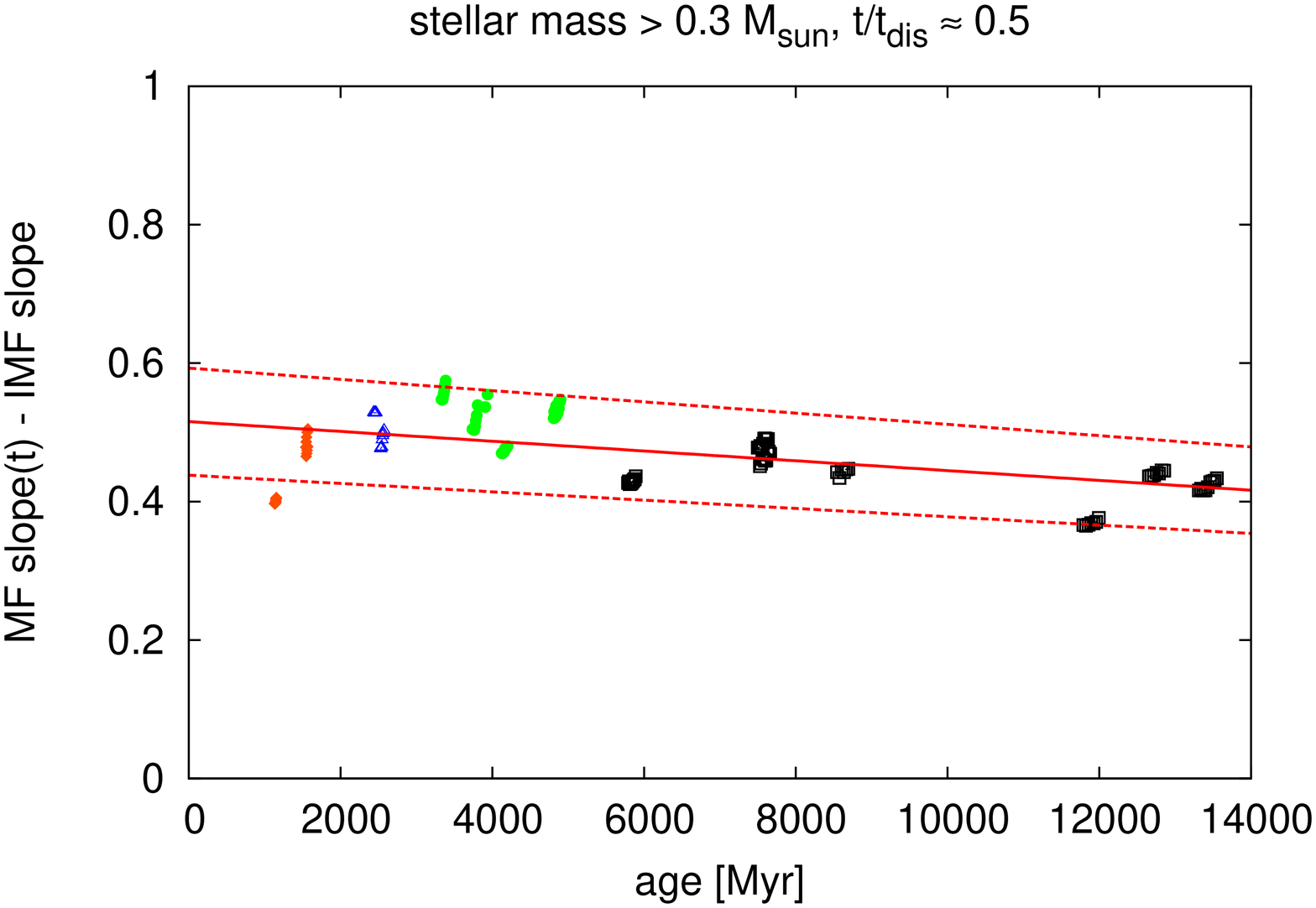} \\  
	 \includegraphics[angle=270,width=0.4\linewidth]{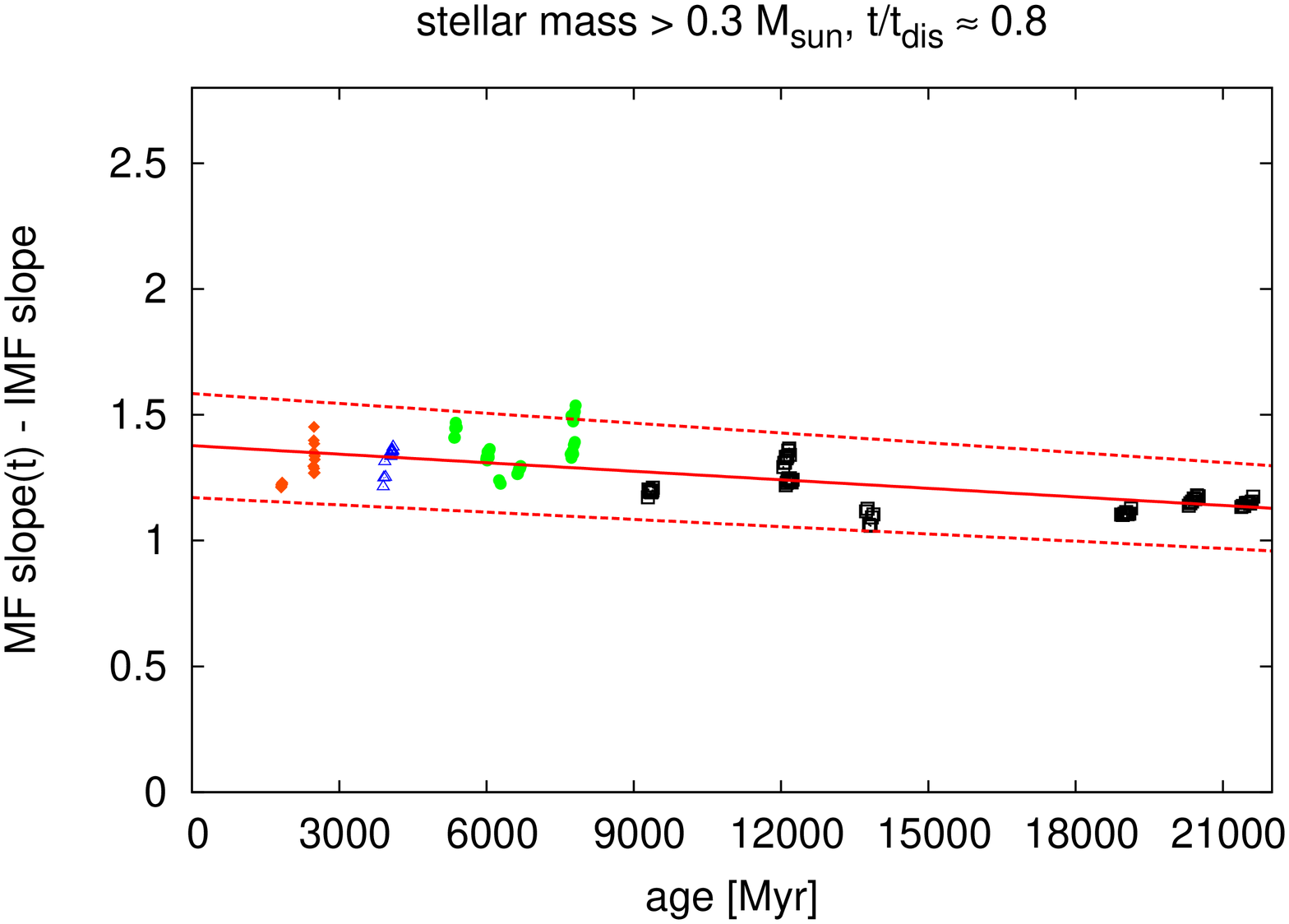} &
	 \includegraphics[angle=270,width=0.4\linewidth]{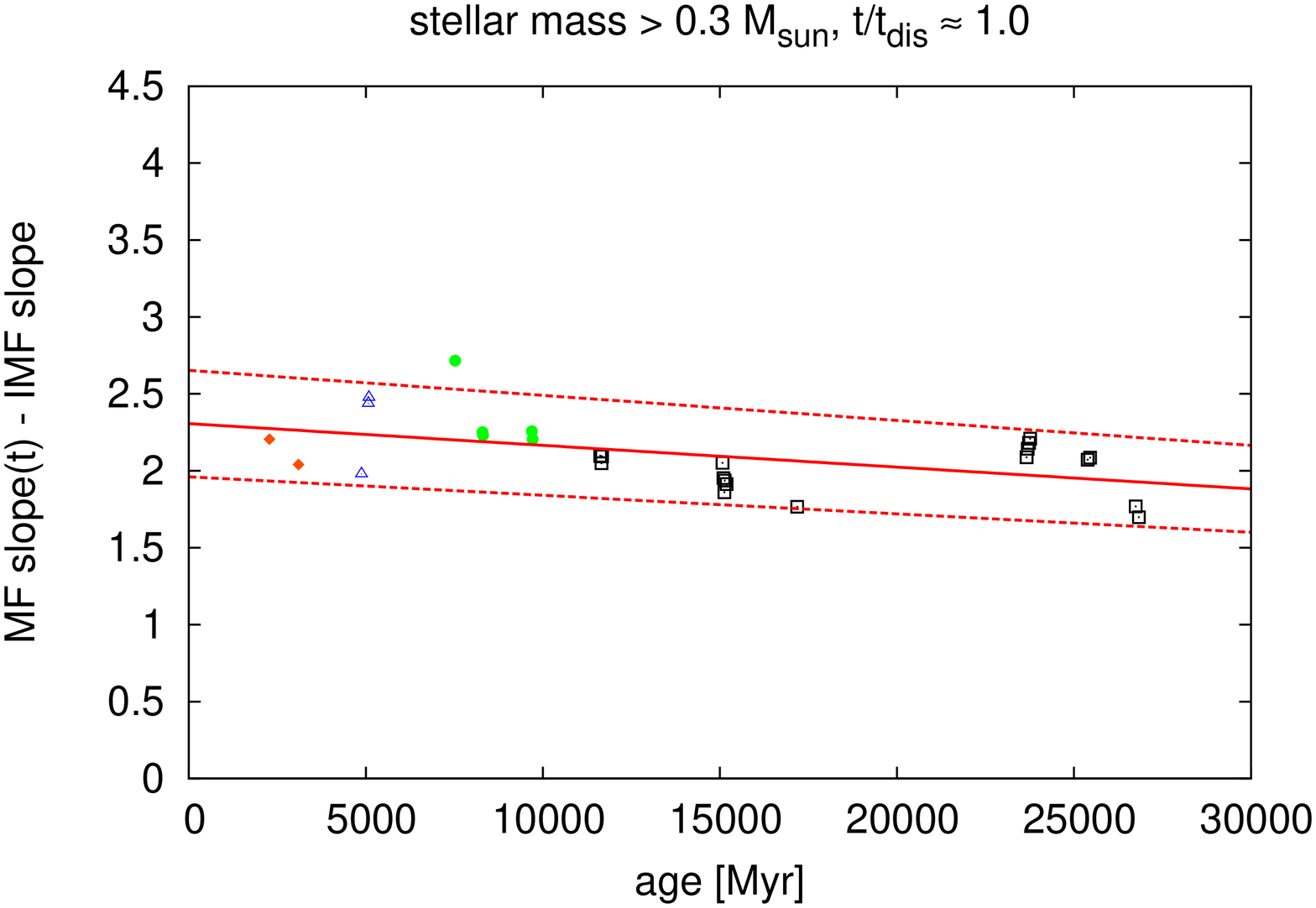} \\
  \end{tabular}
\end{center}
\caption{Same as Fig. \ref{fig:slope1change}, but for stellar masses $>$
0.3 M$_\odot$.}
\label{fig:slope2change}
\end{figure*}

The time evolution of the changing MF slopes can be expressed as:

\begin{equation}
\begin{array}{l l}
\alpha(t) = & \alpha_{\rm IMF}(0)\\
            & + a_1 \cdot x + a_2 \cdot x^2 + a_3 \cdot x^3 + a_4 \cdot	x^4\\
				& + b \cdot x \cdot t
 \end{array} 
\label{form:fit}
\end{equation}

\noindent  

where ``t'' is the cluster age in Myrs, while ``x'' = t/t$_{\rm
95\%}$ is the fractional cluster age in units of its total disruption
time. The best-fit coefficients are provided in Table \ref{tab:fit_par}.
These fit parameters have a very high formal accuracy, due to the large
number of data points used for the fit. However, the spread of N-body
models around our best fit is the  dominant source of uncertainty (see
below and Sect. \ref{sec:spread}). We therefore omit the formal fit
uncertainties in Table \ref{tab:fit_par}.

In Fig. \ref{fig:slopefrac} we overplot our fit formulae
for 4 disruption times t$_{\rm 95\%}$ (1 Gyr, 5 Gyr, 8 Gyr and
30 Gyr), representative for the chosen grouping in disruption time.

We restrict the fitting to ages $\le$ t$_{\rm 95\%}$, as in many cases
clusters with larger ages do not contain enough stars to determine the
MF slopes with reasonable accuracy. However, the general trends continue
beyond t$_{\rm 95\%}$, following the fitted relation further on,
allowing an extrapolation for ages $>$ t$_{\rm 95\%}$ (see Fig.
\ref{fig:slopefrac}).

In addition, we take only simulations into account which started with
32k or more particles (these simulations have total disruption times in the
range 2.3 -- 25.5 Gyr), as many simulations with lower particle numbers
show substantial uncertainties in the determined MF slopes.

We would like to emphasize that considering all simulations with 16k or
more particles or with 64k or more particles yield fitted slopes which
deviate from the 32k results only by less than $\pm$ 0.1 for ages up to
at least 1.3 $\times$ t$_{\rm 95\%}$. Considering also 8k simulations or only
the 128k simulations yields larger deviations, due to large run-to-run
scatter and small number of data/coverage of parameter space,
respectively.

On average, the spread of the BM03 simulation results around the
fitted relation Eq. \ref{form:fit} is of the order of 15\%, as
shown in Fig. \ref{fig:slope1change} and \ref{fig:slope2change}. For
ages $\la$ 1/3 $\times$ t$_{\rm 95\%}$ these {\sl relative} deviations
are larger, however, at those times the {\sl absolute} deviations of the
data from the fit formulae are small (of the order of $\Delta$slope =
0.02-0.03).  The impact of this uncertainty will be further discussed in
Sect. \ref{sec:spread}.

Although the BM03 simulations are performed for a metallicity Z=0.001
(using the fitting formulae from \citealt{2000MNRAS.315..543H}), we will
use Eq. \ref{form:fit} for all metallicities, assuming
the metallicity to -- at most -- introduce second-order effects on
cluster dynamics. This is supported by \citet{2004MNRAS.355.1207H}, who
find metallicity effects largely cancelling each other, resulting in a
weak overall metallicity dependence of cluster dynamics (although
details are metallicity-sensitive).

\begin{table}
\begin{center}

\caption{The best-fit coefficients for Eq. (\ref{form:fit}). The middle
column gives the coefficients for the low-mass end of the MF, for masses
$\le$ 0.3 M$_\odot$. The right column gives the coefficients for masses
$>$ 0.3 M$_\odot$.}

\begin{tabular}{@{}*{3}{|l}{|}@{}}
\hline
coefficient & m $\le$ 0.3 M$_\odot$ & m $>$ 0.3 M$_\odot$\\
\hline
$a_1$ & -0.1345 & 0.08389\\
$a_2$ & 1.7986 & 1.9324\\
$a_3$ & -1.8121 & -0.4435\\
$a_4$ & 1.2181 & 0.734\\
$b$ & 3.215e-6 & -1.4143e-5\\
\hline
\end{tabular}
\label{tab:fit_par}
\end{center}
\end{table}

\subsection{The {\sc galev} models}

The {\sc galev} models are extensively described in
\citet{2002A&A...392....1S}, \citet{2003A&A...401.1063A} and
\citet{2004A&A...413...37B}. Here we give only a brief summary of the
relevant input physics used.

The {\sc galev} models used in this work are based on isochrones
from the Padova group, first presented in \citet{1994A&AS..106..275B},
and subsequently updated to include the TP-AGB phase\footnote{The
models of the Padova group are available at their webpage:
http://pleiadi.pd.astro.it/}. This update, although not documented in a
refereed publication, was made publicly available approx. 1999, and
treats the TP-AGB phase as later described in
\citet{2000A&AS..141..371G}. As we will mainly concentrate on the
evolution of old stellar clusters, the Padova isochrones are preferred
over the Geneva isochrones (\citealt{1992A&AS...96..269S}). We want to
emphasise (and stimulate the various groups of stellar evolution
modellers) that the isochrone sets by \citet{1994A&AS..106..275B} and
\citet{1992A&AS...96..269S}  (and associated papers) are the {\sl only}
available isochrones which cover stellar evolution (in a {\sl
consistent} way) up to its final stages as well as a mass range up to
$\sim$ 120 M$_\odot$ required to correctly model ongoing star formation
in galaxies. For further discussion of this point see Sect.
\ref{sec:isochrones}.

For consistency with the BM03 simulations we use a
\citet{2001MNRAS.322..231K} IMF. 

At each age, the time-dependent MF is evaluated from equation
(\ref{form:fit}) for the requested total disruption time t$_{\rm 95\%}$.
To each star from the isochrones an appropriate spectrum from the BaSeL
library (\citealt{1997A&AS..125..229L,1998A&AS..130...65L}) and a weight
according to the time-dependent MF are assigned. Then the integrated
spectra is obtained by summing up the contributions from the individual
stars. Here, we assume a well-populated MF, hence any stochastic effects
due to small number statistics, especially at the high-mass end of the
MF, are neglected. Hence, we model an {\sl average} star cluster.

 The treatment of stochastic effects would go well beyond the scope
of this paper. The impact of such stochastic effects was studied in
depth by Cervi{\~n}o and collaborators (see e.g.
\citealt{2002A&A...394..525C, 2004A&A...413..145C, 2006A&A...451..475C})
and \citet{2007A&A...462..107F},  who found these effects to be strongly
age- and wavelength-dependent. The strongest impact is found for red
passbands, which are dominated by few red supergiants (young clusters)
or very bright upper RGB and AGB stars (intermediate-age clusters). The
effects become smaller for older ages, for non-dissolving clusters.
However, the decreasing number of stars with age in our dissolving
cluster models likely counteracts this effects. We therefore caution
users about applying our models to a single cluster. The models
represent {\sl average} star clusters with the given parameters, hence
need to be applied to a whole star cluster system.

More generally, small number statistics is the likely origin of the
scatter seen in Figs. \ref{fig:slope1change} and \ref{fig:slope2change}.
However, as we use 19 BM03 models with a variety of parameters (i.e.
total masses and dissolution times) to model the dissolution we can
describe the {\sl average} cluster dissolution. The impact of the spread
seen in Figs. \ref{fig:slope1change} and \ref{fig:slope2change} will be
discussed in more detail in Sect. \ref{sec:spread}.

Due to computational restrictions we only calculated individual models
for MF slopes with 2 decimal places. If at any given age the MF slopes
were identical to within these 2 decimal places with the MF slopes of a
previously computed model we reused this older model. Due to this finite
step-size, some cluster colours exhibit small jumps of the order of
$\sim$0.001 mag (up to 0.004 mag for the most extreme cases).

The spectrophotometry is normalised to a luminous cluster mass as
described in Sect. \ref{sec:timemass}.

We calculate models for t$_{\rm 95\%}$:
\begin{itemize}
\item in the range of 100 -- 900 Myr: in 50 Myr steps
\item in the range of 1 -- 16 Gyr: in 500 Myr steps
\item for t$_{\rm 95\%}$ = 18, 20, 25, 30, 40, 60, 100, 150 and 200 Gyr
\end{itemize}
and for metallicities (limited by the metallicities provided by the
Padova isochrones):
\begin{itemize}
\item Z=0.0004 $\leftrightarrow$ [Fe/H] = -- 1.7
\item Z=0.004 $\leftrightarrow$ [Fe/H] = -- 0.7
\item Z=0.008 $\leftrightarrow$ [Fe/H] = -- 0.4
\item Z=0.02=Z$_\odot$ $\leftrightarrow$ [Fe/H] = 0.0
\item Z=0.05 $\leftrightarrow$ [Fe/H] = + 0.4
\end{itemize}

For t$_{\rm 95\%} \gta$ 200 Gyr, within a Hubble time the MF slopes
deviate from the universal \citet{2001MNRAS.322..231K} IMF by less than
0.005. 
For conditions similar to the Solar Neighbourhood, i.e. t$_4$ = 1.3 Gyr
as determined by \citealt{2005A&A...429..173L}, the range in total
disruption times corresponds to a cluster mass range 160 -- $3.4 \times
10^7$ M$_\odot$ (i.e. a 160 M$_\odot$ cluster needs 100 Myr to totally
disrupt, a 10$^4$ M$_\odot$ cluster needs 1.3 Gyr, and a $3.4 \times
10^7$ M$_\odot$ cluster needs 200 Gyr).  For the SMC, with t$_4$ =
$\sim$10 Gyr, the range in total disruption times would correspond to a
cluster mass range 6 -- $1.25 \times 10^6$ M$_\odot$.

The data are made publicly available at our webpages
http://www.phys.uu.nl/$\sim$anders/data/SSP\_varMF and
http://data.galev.org. We provide the user with integrated cluster
magnitudes in a variety of passbands plus cluster masses (total mass,
luminous mass and mass in stellar remnants) for each of the models.
Integrated spectra are available upon request.

\clearpage

\section{Results and implications}

In this section we present our new models for solar metallicity
(unless otherwise noted) and discuss its implications. Models for all
metallicities are available at our webpages
http://www.phys.uu.nl/$\sim$anders/data/SSP\_varMF and
http://data.galev.org. We would like to emphasize that the absolute
values of our models (and therefore also the results and implications
discussed in this section) depend on our choice of isochrones and other
input physics. In Sect. \ref{sec:comparison} we discuss some of these
uncertainties. The main results, the systematic differences induced by
the preferential mass loss, are hardly changed.

\label{sec:results}
\subsection{Photometry}
\label{sec:photo}

\begin{figure*}
\begin{center}
  \vspace{-0.5cm}
  \hspace{1.2cm}
  \begin{tabular}{cc}
	 \includegraphics[angle=270,width=0.4\linewidth]{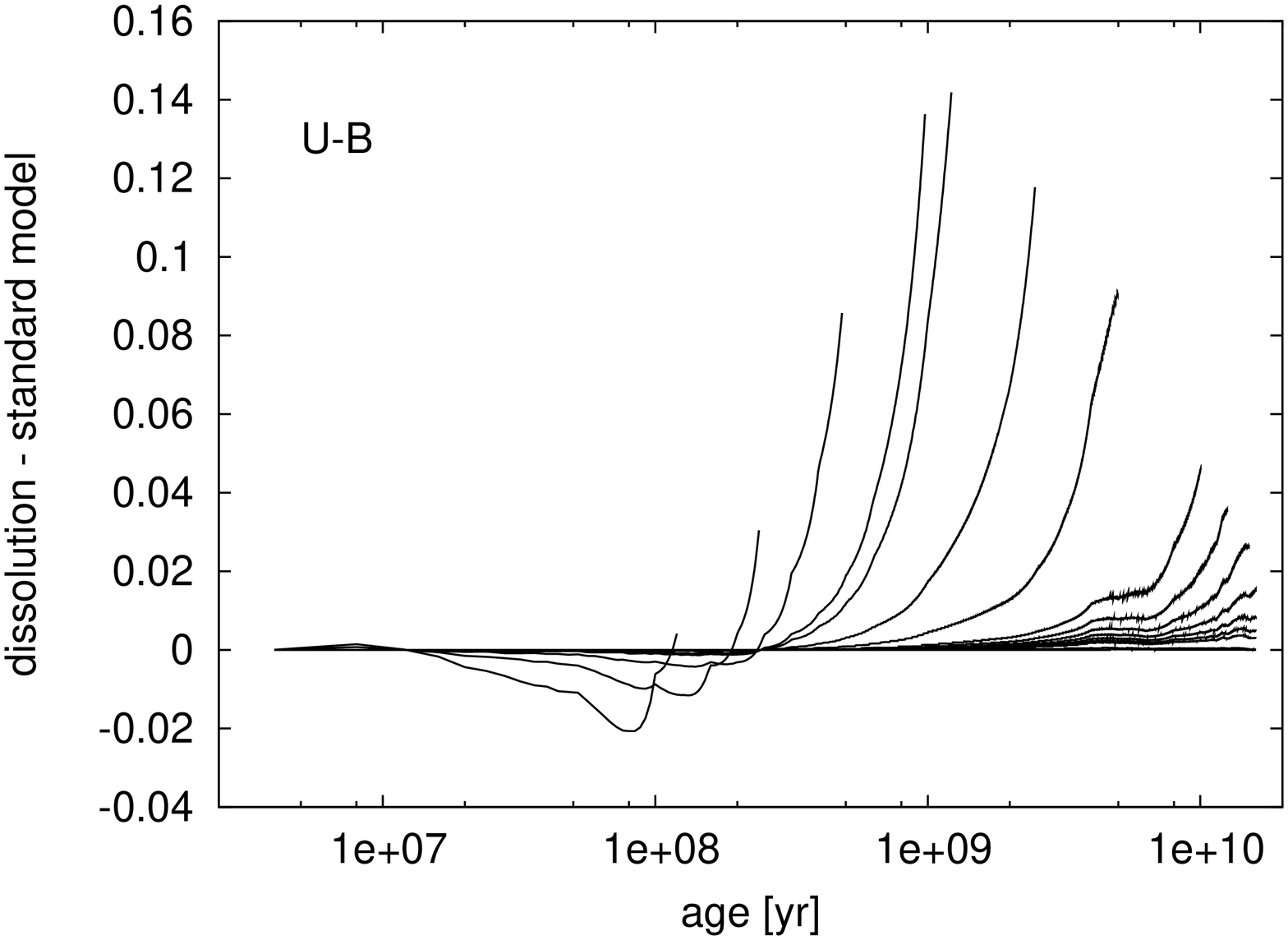} &
	 \includegraphics[angle=270,width=0.4\linewidth]{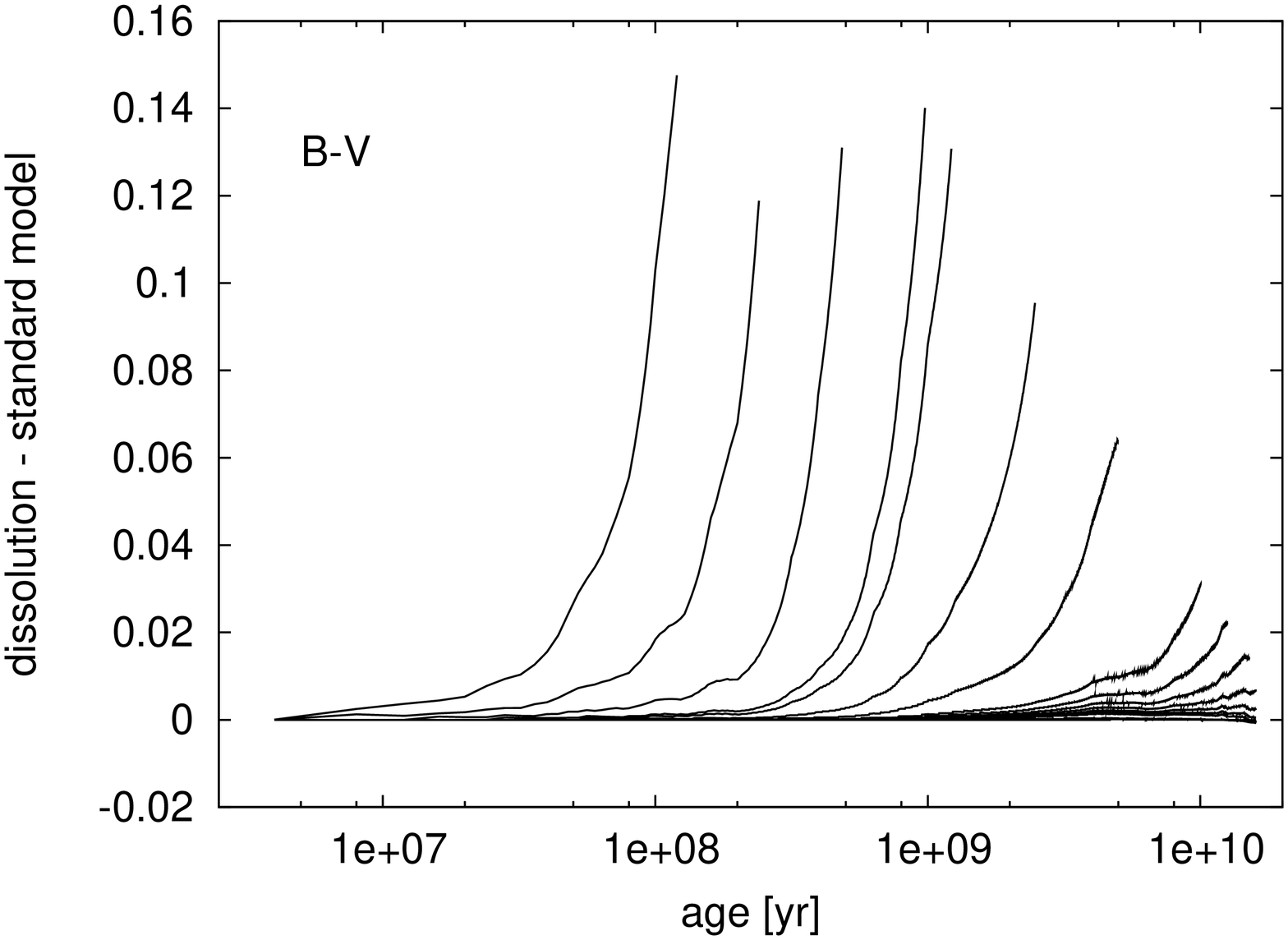} \\  
	 \includegraphics[angle=270,width=0.4\linewidth]{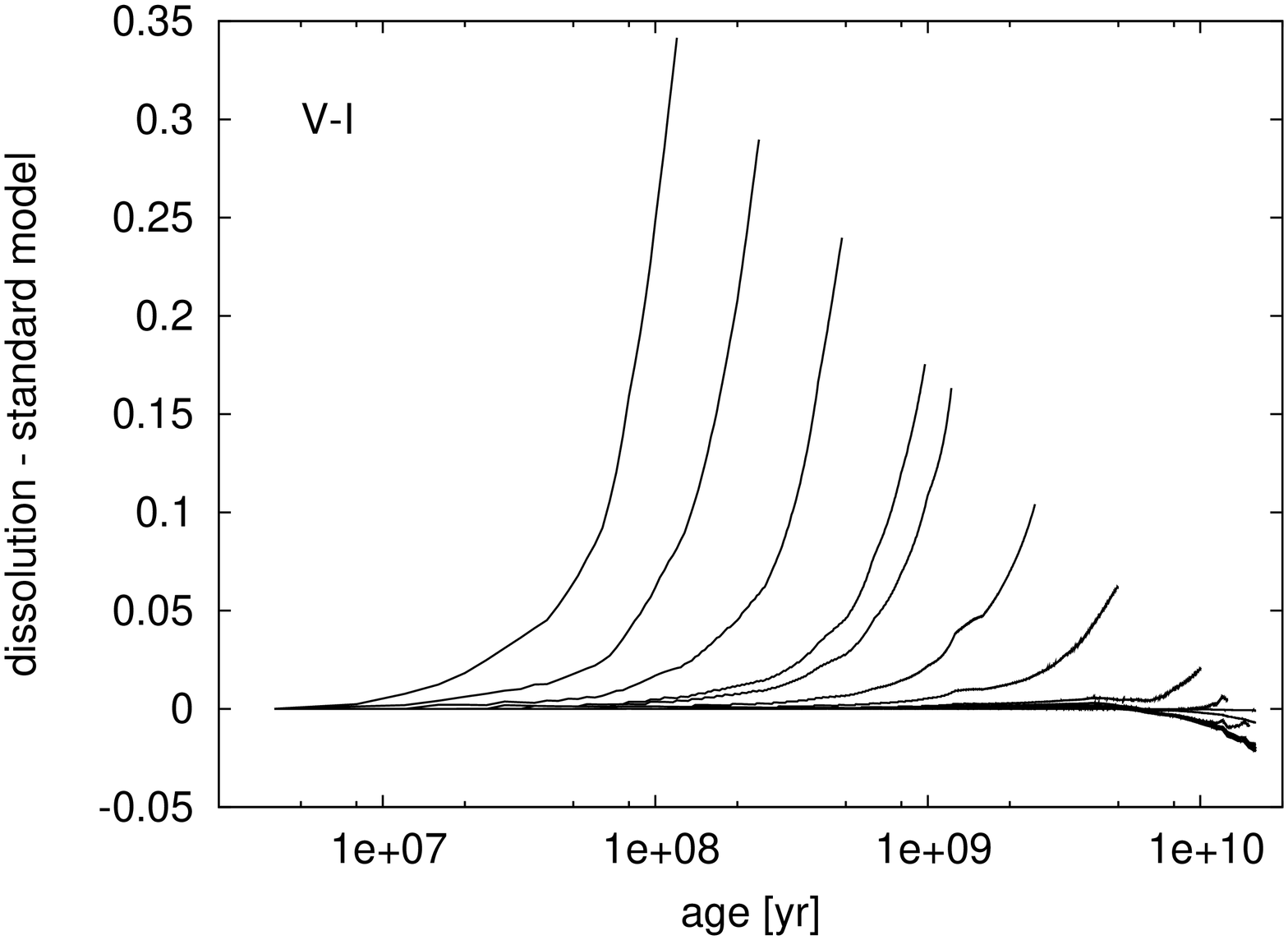} &
	 \includegraphics[angle=270,width=0.4\linewidth]{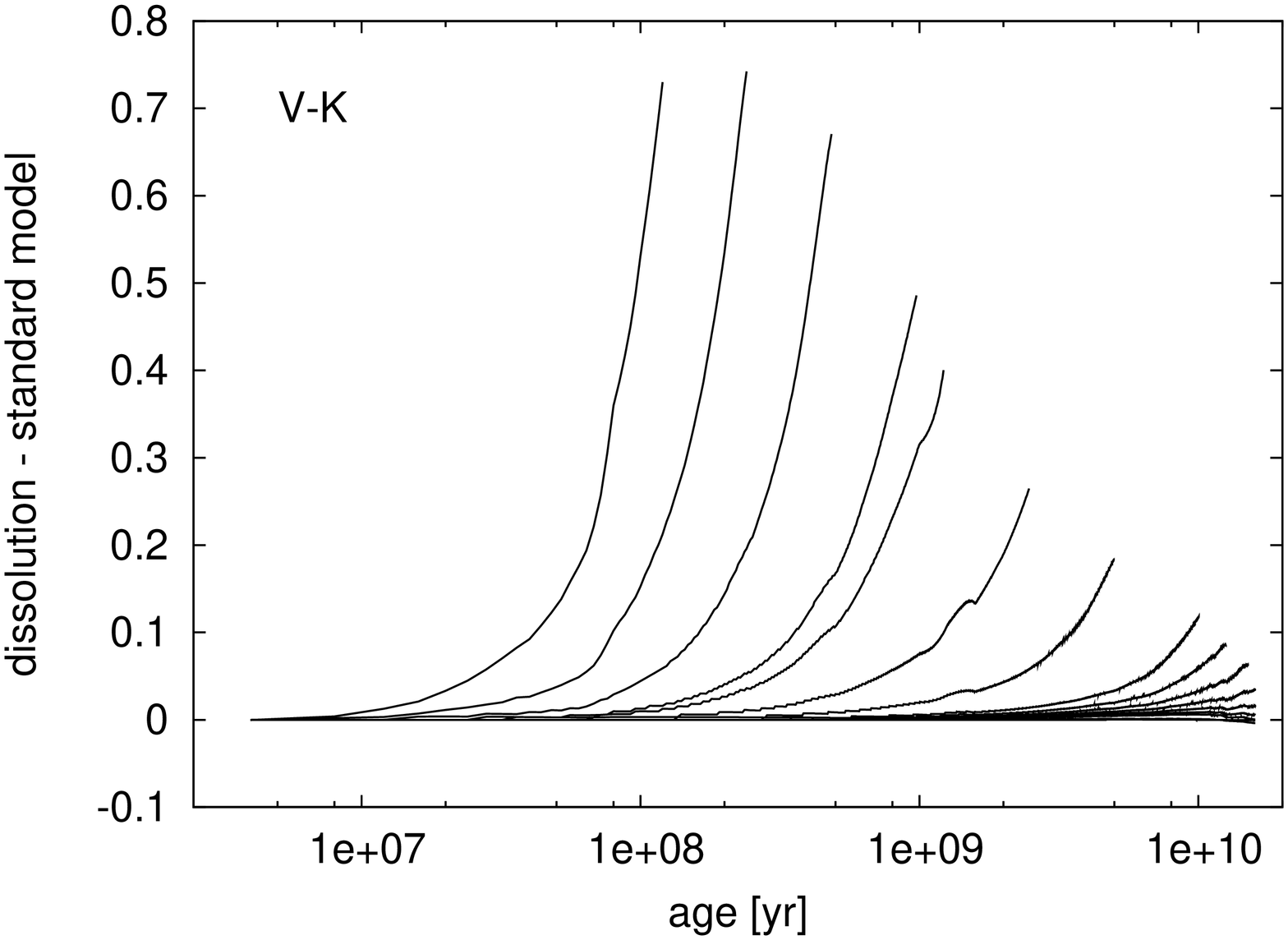} \\
	 \includegraphics[angle=270,width=0.4\linewidth]{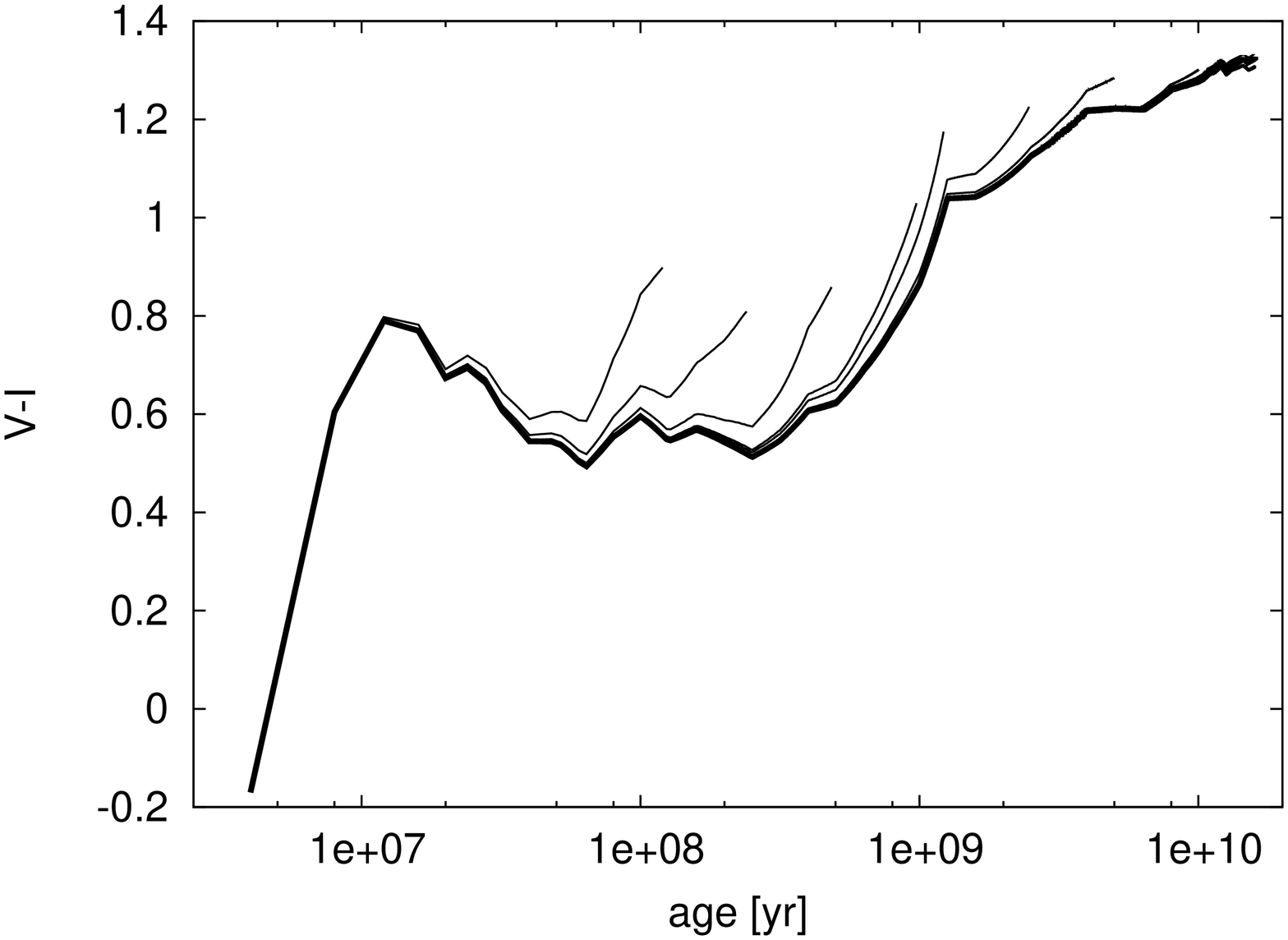} &
	 \includegraphics[angle=270,width=0.4\linewidth]{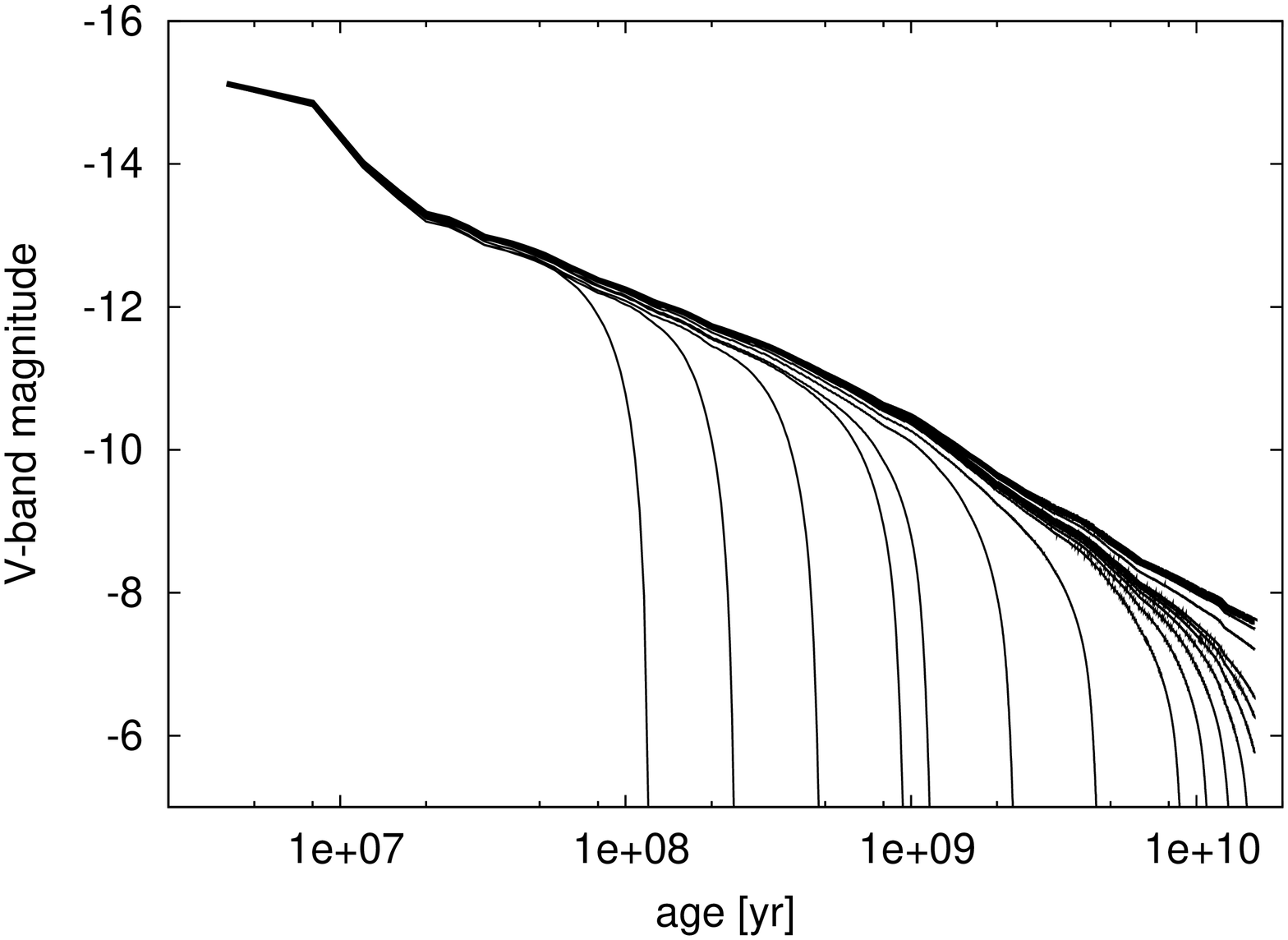} \\
  \end{tabular}
\end{center}

\caption{Solar metallicity models with the changing mass function
treatment, following equation (\ref{form:fit}) with t$_{\rm 95\%}$ in
the range 100 Myr -- 100 Gyr (for clarity only models at 100, 200, 400,
800 Myr and 1, 2, 4, 8, 10, 12, 14, 16, 18, 20, 40, 100 Gyr are shown),
going from left to right, respectively. Upper left panel: U-B colour,
top right: B-V, middle left: V-I, middle right: V-K. For these colours,
the  differences between the new models and the standard models are
displayed. Bottom left: the absolute values for V-I, bottom right:
V-band magnitude evolution of a 10$^6$ M$_\odot$ cluster. Thick lines
represent the standard models.}

\label{fig:colors}
\end{figure*}

The photometry for the new models is shown in Fig. \ref{fig:colors}. The
colours are shown as differences between the new models with changing MF
and the standard models with fixed (initial) MF. For illustrative
purposes, in the bottom left panel the absolute values for the V-I
colour are presented. The V-band magnitude evolution (bottom right
panel) is given in absolute values for a 10$^6$ M$_\odot$ cluster.

The V-band magnitude evolution shows the stellar evolution fading line
as bright limit for the new models, which they follow for young ages,
when the effect of mass loss is not yet pronounced. Already after
$\sim$10\% of their respective total disruption times, the new models
have evolved 0.1 mag away from the fiducial fading line, due to the loss
of stars. At $\sim$80\% ($\pm$ 10\%, depending on the model) of their
respective total disruption times, the new models are 1 mag fainter than
standard models predict, due to the loss of stars.

Except for the very earliest stages of cluster evolution (first few
Myrs), the flux in passbands redder than the V band is dominated by
stars initially more massive than the main contributors to the flux in
bluer passbands. This is caused by the flux in the red passbands being
dominated by red (super)giants, which are more luminous than the stars
at the low-mass end of the main sequence (MS), even after taking the
larger number of low-mass MS stars due to the IMF into account. For a
changing MF due to dynamical evolution, the contribution from low-mass
MS stars is even further reduced.

The dominant source of flux contribution in passbands including and
bluewards of the V band is a strong function of time: in early stages,
the flux is dominated by mid-MS stars (the evolution through the
Hertzsprung gap is too fast to significantly contribute). As the cluster
ages, the MS turn-off (MSTO) shifts successively redwards through the
filters, ever increasing its contribution to the band's flux. However,
the relative contribution of mid-MS stars and MSTO stars is also
strongly dependent on the MF, hence is dependent on the total disruption
time of our models.

As the selective mass loss preferentially removes the least massive
stars from the cluster (and therefore its integrated photometry), it
causes the cluster to generally become redder than the standard models
without cluster dissolution (i.e. with infinite total cluster disruption
time). The MF evolution and the resulting reddening speeds up while the
cluster approaches its final disruption, leading to the steep colour
evolutions towards the end of a cluster's lifetime, as seen in Fig.
\ref{fig:colors}.

Two exceptions have to be noted: 

\begin{itemize} 

\item the colour U-B (and similar colours) becomes bluer than the
standard models for total disruption times shorter than $\sim$ 1Gyr. At
these ages the B band is entirely dominated by mid-MS stars, while bluer
bands contain contributions from the higher MS stars and the MSTO stars.
As the mid-MS is stronger depopulated than the upper MS and MSTO due to
the dynamical cluster evolution, the colours become bluer. For longer
total disruption times the mid-MS is not sufficiently depopulated to
notice the effect until the B band gets contributions from the MSTO
stars. Redder passbands do not show this effect as they contain
contributions from bright red (super-)giants.

\item for colours like V-R and V-I and ages $\ga$ 6 Gyr the models get
slightly bluer than the standard models for total disruption times $\ga$
10 Gyr. Likely, this comes from the strong depopulation of the lower MS
(and the standard IMF containing a large number of stars at low masses),
which leaves an imprint even though a single lower-MS star is 3-4 mag
fainter than an RGB/AGB star of similar temperature. Redder passbands do
not show this effect as the magnitude difference between lower-MS stars
and RGB/AGB stars increases with increasing wavelength/decreasing
temperature, and the total contribution from MS stars decreases.

\end{itemize}

For long total disruption times, the maximum colour deviation of our
models for dissolving clusters from the standard models decreases with
increasing total disruption time.  This comes from several effects at
once:

\begin{itemize}
\item for such old ages, the MF covers only a narrow mass range in both
 cases
\item the integrated cluster flux is dominated by the upper-RGB/AGB
stars, as they are significantly brighter than the MSTO region (the
magnitude difference between upper-RGB/AGB and MSTO increases with
time), resulting in an even narrower ``effectively visible MF'' range
\item the temperature range these stars cover is significantly smaller
than at younger ages, resulting in a lower sensitivity of the colours to
the exact distributions of stars along the isochrone.
\end{itemize}

However, the changes in mass/absolute magnitude (see Fig.
\ref{fig:colors}, bottom right panel) and M/L ratio (see next
section) are significant in all cases.

The increasing maximum colour deviation of our dissolving clusters
models from the standard models for blue passbands and increasing total
disruption times $\lta$ 1 Gyr originates from the redward shifting of
the MSTO through the filters.

\subsection{Mass-to-light ratios}
\label{sec:ML}

\begin{figure}
\begin{center}
 \vspace{-0.5cm}
 \hspace{1.2cm}
 \includegraphics[angle=270,width=0.9\linewidth]{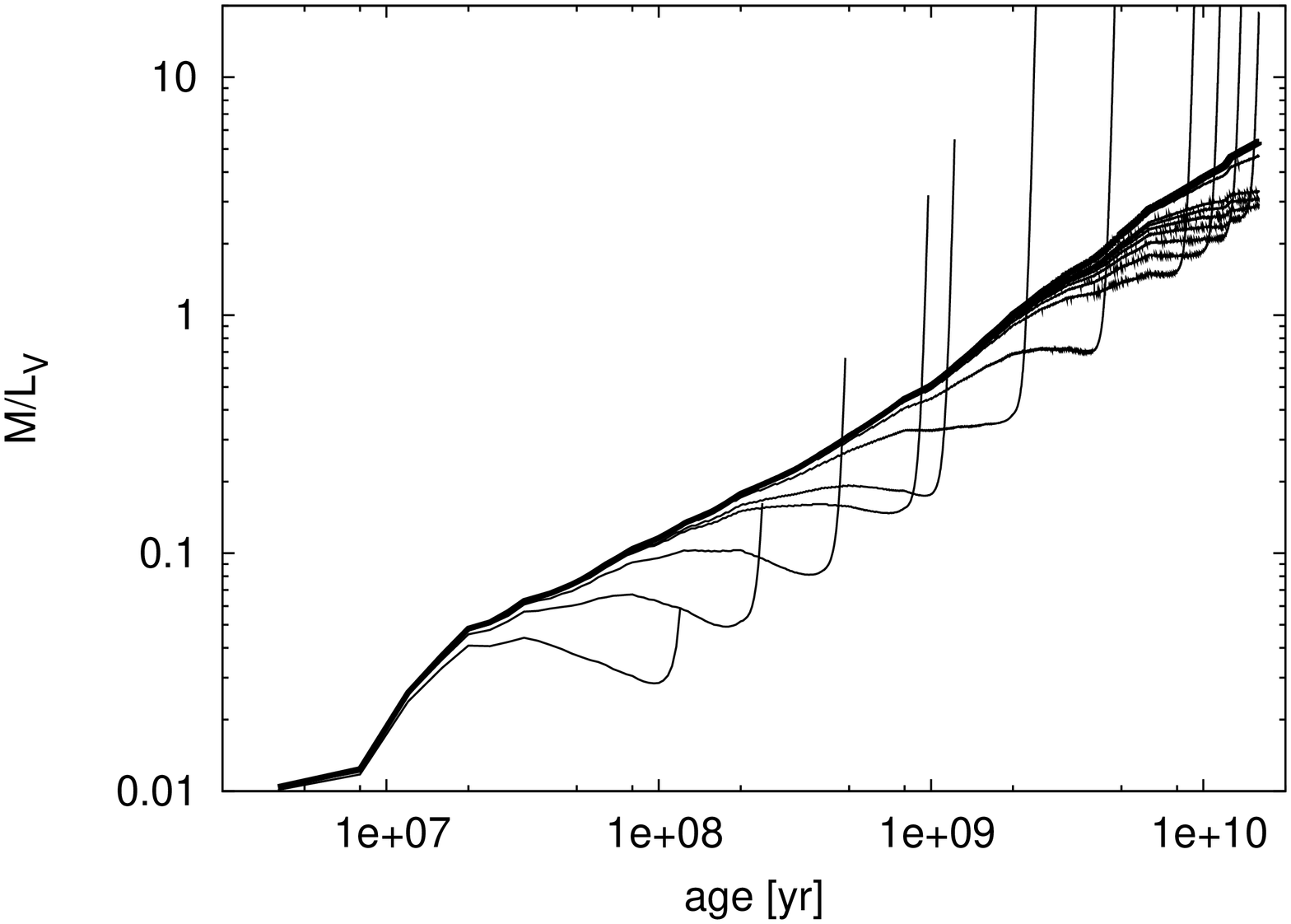}
 \includegraphics[angle=270,width=0.9\linewidth]{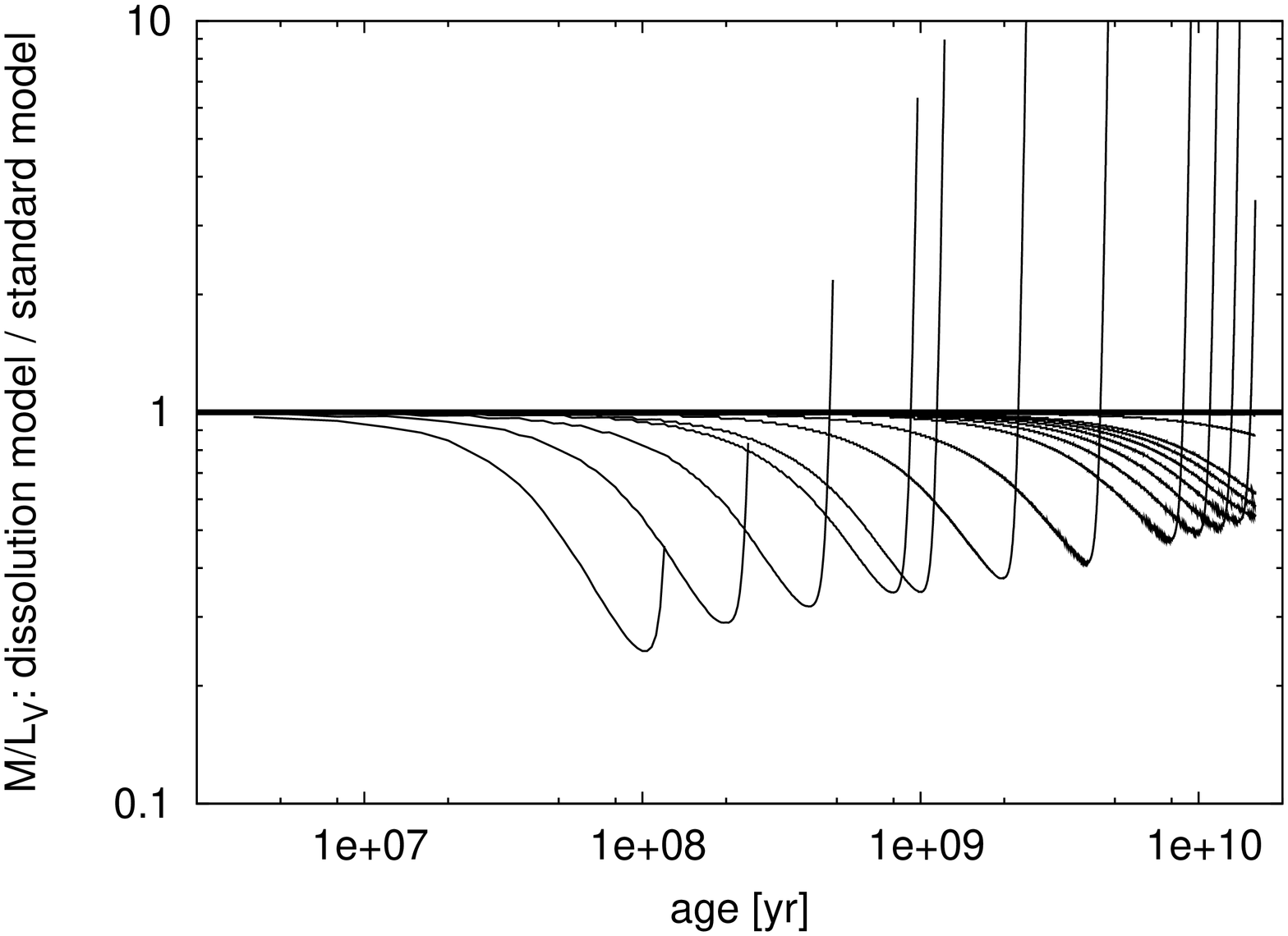}
 \includegraphics[angle=270,width=0.9\linewidth]{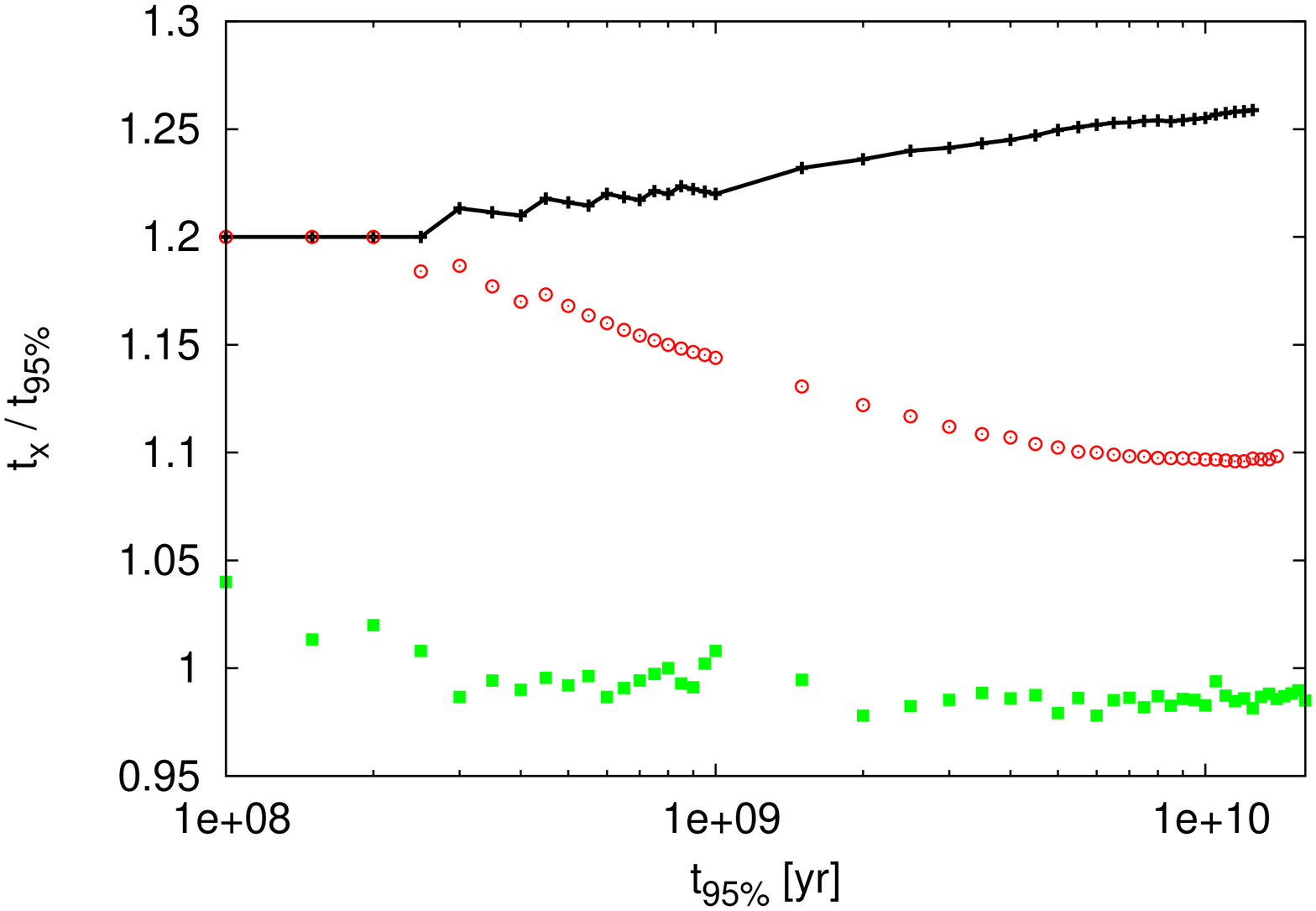}
\end{center}

\caption{Same models as Fig. \ref{fig:colors}, but M/L ratios. Upper
panel: time evolution of V-band M/L ratio for the new models. Middle
panel: ratio of new model's  V-band M/L ratio and V-band M/L ratio from
standard models. The non-dissolving/standard model is shown as thick
line. Bottom panel: characteristic times of the models as function of
total disruption time: black crosses and line = age at which the cluster
contains only 100 M$_\odot$ luminous matter (i.e. termination age of
model); red open circles = age at which the M/L ratio evolution crosses
the standard model; green filled squares = age at which the M/L ratio of
the dissolving cluster models is minimal w.r.t. the standard models
(i.e. the dip seen in the middle panel). Only models for which the
respective age is smaller than the maximum model age of 16 Gyr are
shown.}

\label{fig:MLratios}
\end{figure}

The low-mass stars preferentially removed in the course of cluster
dissolution have mass-to-light (M/L) ratios which are higher than the
M/L ratio of the average cluster star. On the other hand, as shown by
BM03, the fraction of (non-luminous) stellar remnants in dissolving
clusters is enhanced w.r.t. the standard models. These effects partially
cancel each other and lead to the time-dependent M/L ratios shown in
Fig. \ref{fig:MLratios}. It demonstrates that for each total disruption
time the M/L ratio of our dissolving cluster models is systematically
lower than a standard model would suggest for the majority of a
cluster's total disruption time. For the final up to $\sim$16\% of a
cluster's total disruption time, the M/L ratio can become enhanced
w.r.t. the standard models (see Fig. \ref{fig:MLratios}, bottom panel).
This is caused by the increasing fraction of stellar remnants inside the
cluster.

\begin{figure}
\begin{center}
 \vspace{-0.5cm}
 \hspace{1.2cm}
 \includegraphics[angle=270,width=1.0\linewidth]{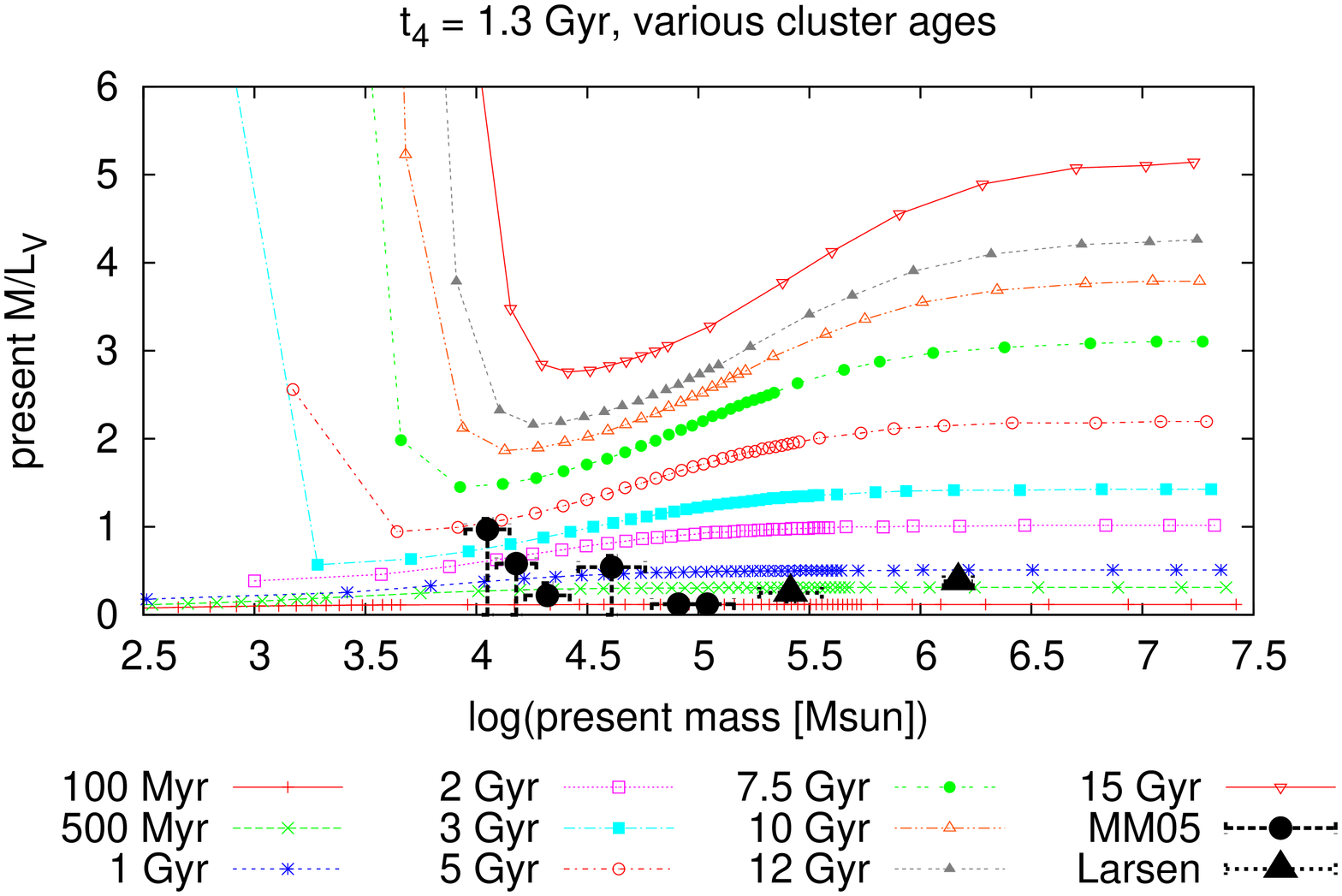}
 \includegraphics[angle=270,width=1.0\linewidth]{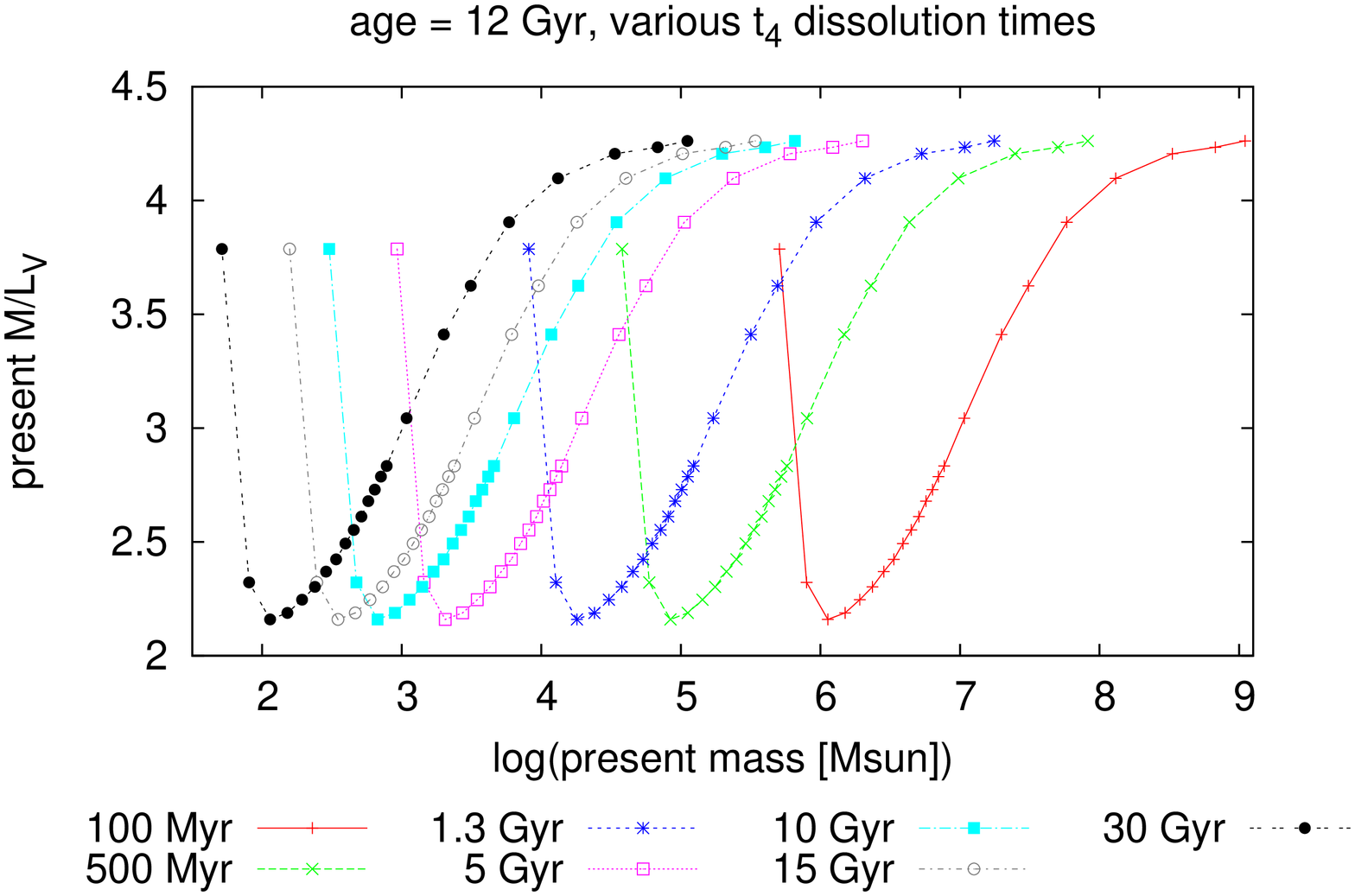}
 \includegraphics[angle=270,width=1.0\linewidth]{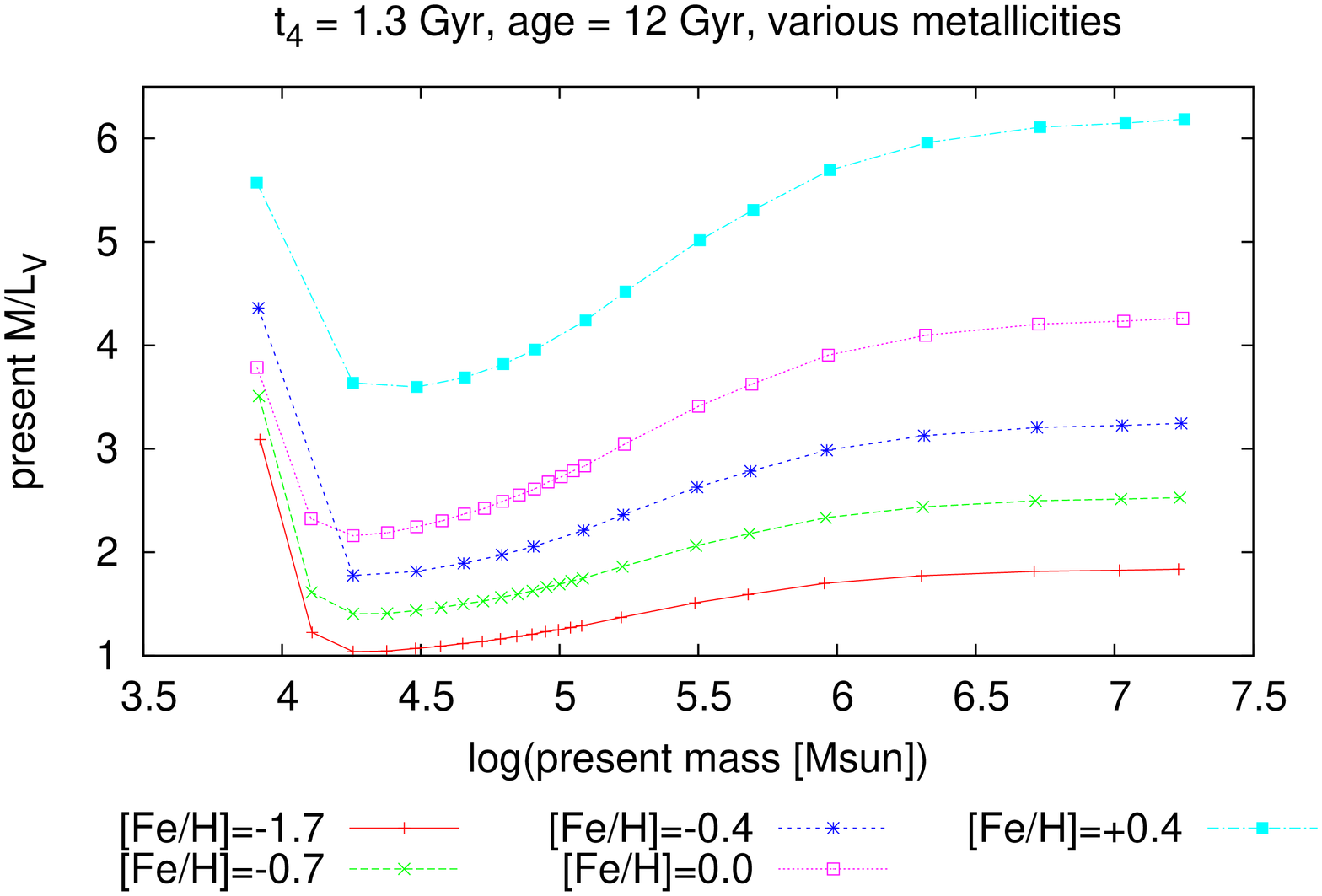}
\end{center}

\caption{Top panel: time evolution of V-band M/L ratio (for solar
metallicity) as a function of present-day cluster mass for a fixed local
gravitational field strength characteristic for the Solar Neighbourhood
(i.e. t$_4$=1.3 Gyr, see \citealt{2005A&A...441..117L}). Middle panel:
V-band M/L ratio (for solar metallicity) as a function of present-day
cluster mass for a range of local gravitational field strengths for
clusters observed at an age of 12 Gyr. Bottom panel: V-band M/L ratio as
a function of present-day cluster mass for a range in metallicity for
clusters observed at an age of 12 Gyr and experiencing a typical
disruption time t$_4$=1.3 Gyr.  In the top panel, observations from
\citet{2005ApJS..161..304M} (MM05) and from
\citealt{2004A&A...427..495L,2004AJ....128.2295L} are overplotted, for
young clusters with ages $<$ 1 Gyr.}

\label{fig:MLratios2}
\end{figure}

In Fig. \ref{fig:MLratios2} we present the dependence of the V-band M/L
ratio on the present cluster mass. The top panel shows how this
relation evolves with cluster age at a field strength (i.e. location in
a galaxy, characterised by t$_4$, the total disruption time of a 10$^4$
M$_\odot$ cluster, as described in Sect. \ref{sec:timemass})
representative for the  Solar Neighbourhood, as found by
\citet{2005A&A...441..117L}. As the clusters evolve, generally the M/L
ratio increases due to stellar evolution. In addition, the lowest-mass
clusters eventually disrupt (and drop out of this plot). The
highest-mass clusters lose mass, but still have M/L ratios close to the
canonical value from stellar evolution. Intermediate-mass clusters
suffer from the impact of the changing mass function which reduces their
M/L ratio significantly compared to the canonical value. A few cases of
{\sl enhanced} M/L ratios can be seen for clusters close to final
disruption (at the low-mass end of the curves). 

In Fig. \ref{fig:MLratios2} (top panel) we overplotted data of
young (ages $<$ 1 Gyr) LMC and SMC clusters by
\citet{2005ApJS..161..304M} (labelled ``MM05'', for details of this
dataset see following subsection) and Larsen and collaborators (labelled
``Larsen'') in 4 spiral and irregular galaxies, (see
\citealt{2004A&A...427..495L} and \citealt{2004AJ....128.2295L}). Out of
the 13 clusters in these samples, 8 have M/L ratios consistent with our
models for their respective ages (within their 1$\sigma$ uncertainty
ranges). The remaining 5 clusters all have too high M/L ratios for their
respective ages. 2 clusters are very young ($\sim$10 Myr), hence could
be out of equilibrium during their gas expulsion and readjustment phase,
and their velocity dispersions might not trace their dynamical masses
(see \citealt{2006MNRAS.373..752G}). The deviations of the remaining
clusters might be indications of errors in the models, or could be signs
that the age determination is uncertain or the velocity dispersions are
seriously affected by the orbital motions of binaries (or other
systematic observational effects, like macroturbulence in the stellar
atmospheres or instrumental resolution).

The middle panel shows the M/L ratio in the V band as a function of the
present-day mass of 12 Gyr old clusters, for a range of gravitational
field strengths (i.e. typical disruption times t$_4$). Within each line,
the cluster's total disruption time goes from 10 Gyr (low mass end;
clusters with shorter total disruption times have been disrupted by an
age of 12 Gyr) to 200 Gyr (upper end of available total disruption time
range). For example: a cluster located at a position in a galaxy
characterised by a field strength with t$_4$ = 1.3 Gyr (i.e. the blue
line, corresponding to the environment in the Solar Neighbourhood),
observed now (i.e. at an age of 12 Gyr) with a mass = 10$^6$ M$_\odot$
is expected to have a M/L$_{\rm V}$ of $\sim$4, while a cluster  with a
mass = 10$^4$ M$_\odot$ has a modelled M/L$_{\rm V}$ of $\sim$2.3. 

The bottom panel shows the impact of metallicity on the M/L ratios: with
increasing metallicity the M/L$_{\rm V}$ ratio based on stellar
evolution increases. Therefore all curves are stretched to reach higher
M/L$_{\rm V}$ ratios for higher metallicities, while the shape of the
curves is largely unaffected. At the high-mass end, all curves level off
to their respective values determined by stellar evolution alone. 

By comparing the top and bottom panels of Fig. \ref{fig:MLratios2},
both for our new models as for the non-dissolving standard models the
well-known age-metallicity degeneracy is apparent (see e.g.
\citealt{1994ApJS...95..107W}).

As the {\sc galev} code (like most other evolutionary synthesis
codes) is not capable to directly deal with stochastic effects
(especially not with the stochastic effects inherent to the selective
mass loss caused by dissolution), we employ the following estimation
scheme: 

\begin{itemize}
\item We assume, that the majority of stochasticity originates from
evolved stars (mainly RGB, AGB stars).
\item As a function of age, we determine the relative number of evolved
to unevolved stars (i.e. MS stars).
\item From this relative number we determine the total number of evolved
stars for clusters of different masses, and the stochastic scatter
(i.e. the square root of the total number of evolved stars).
\item We determine average properties of the evolved stars (mean
effective temperature, mean log(g) and mean luminosity).
\item We multiply the stochastic scatter with the spectrum of the mean
evolved star, and add resp. subtract this from our standard spectrum.
\end{itemize}

From this approach, we estimate the effect of IMF stochasticity on the
M/L ratios to be roughly: 15/5/1.5 \% uncertainty for clusters with
total mass 10,000/100,000/1,000,000 M$_\odot$, respectively. This test 
was only done for the standard, not depopulated IMF. The effects for
the depopulated MF will be smaller, as for the same total mass, the 
number of giant stars will be larger.

Our results show features similar to those presented by
\citet{2008A&A...490..151K} and applied by \citet{2008A&A...486L..21K}.
However, systematic differences are present, inherent to the underlying
assumptions, and discussed in Sect. \ref{sec:earlierwork}.

\subsubsection{Comparison with observations}

In the following section we want to compare our new models with old
globular clusters (and other old massive stellar systems) in the Milky
Way and other galaxies. This is a first step to validate our models.

\begin{figure}
\begin{center}
 \vspace{-0.5cm}
 \hspace{1.2cm}
 \includegraphics[angle=270,width=1.0\linewidth]{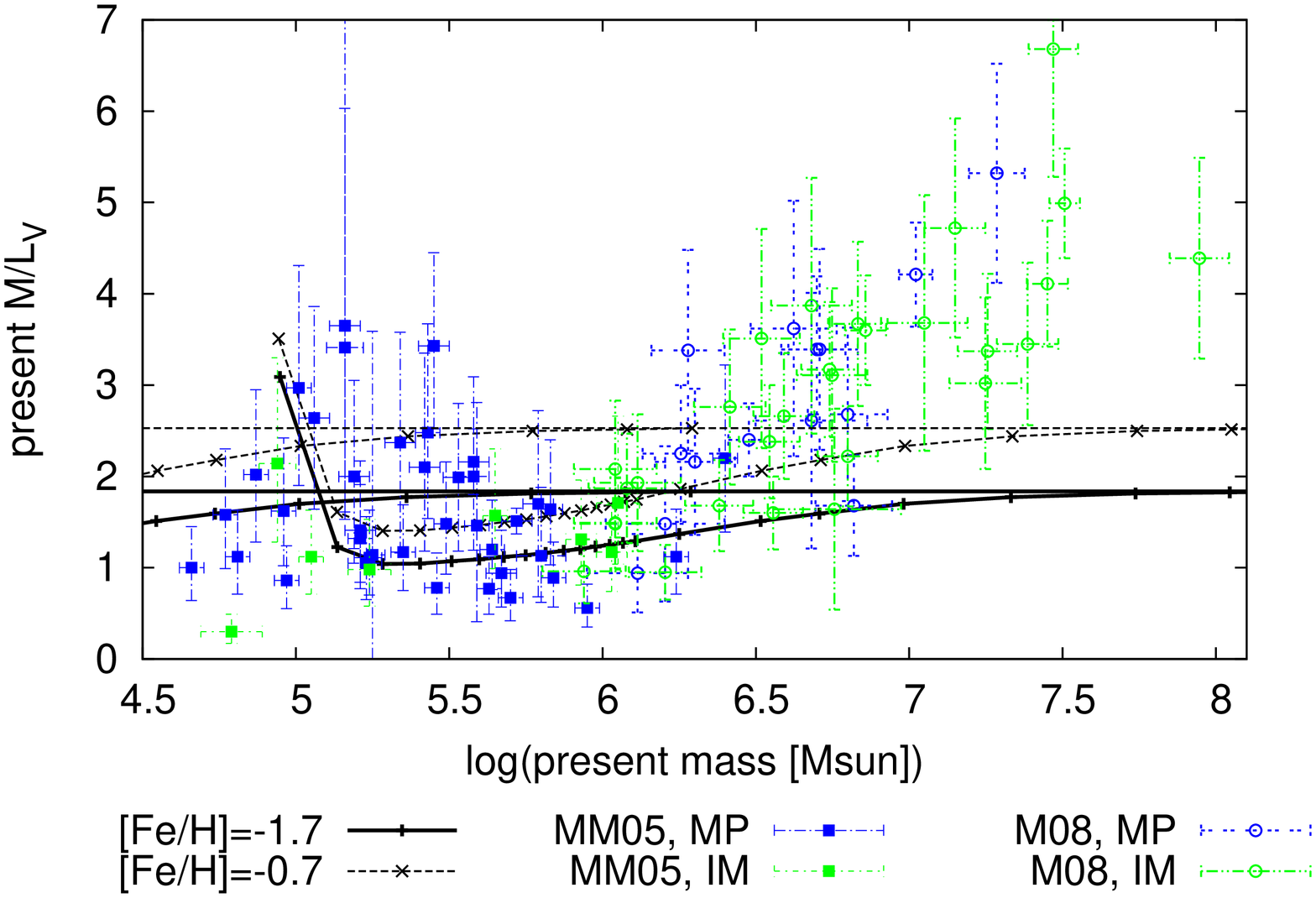}
 \includegraphics[angle=270,width=1.0\linewidth]{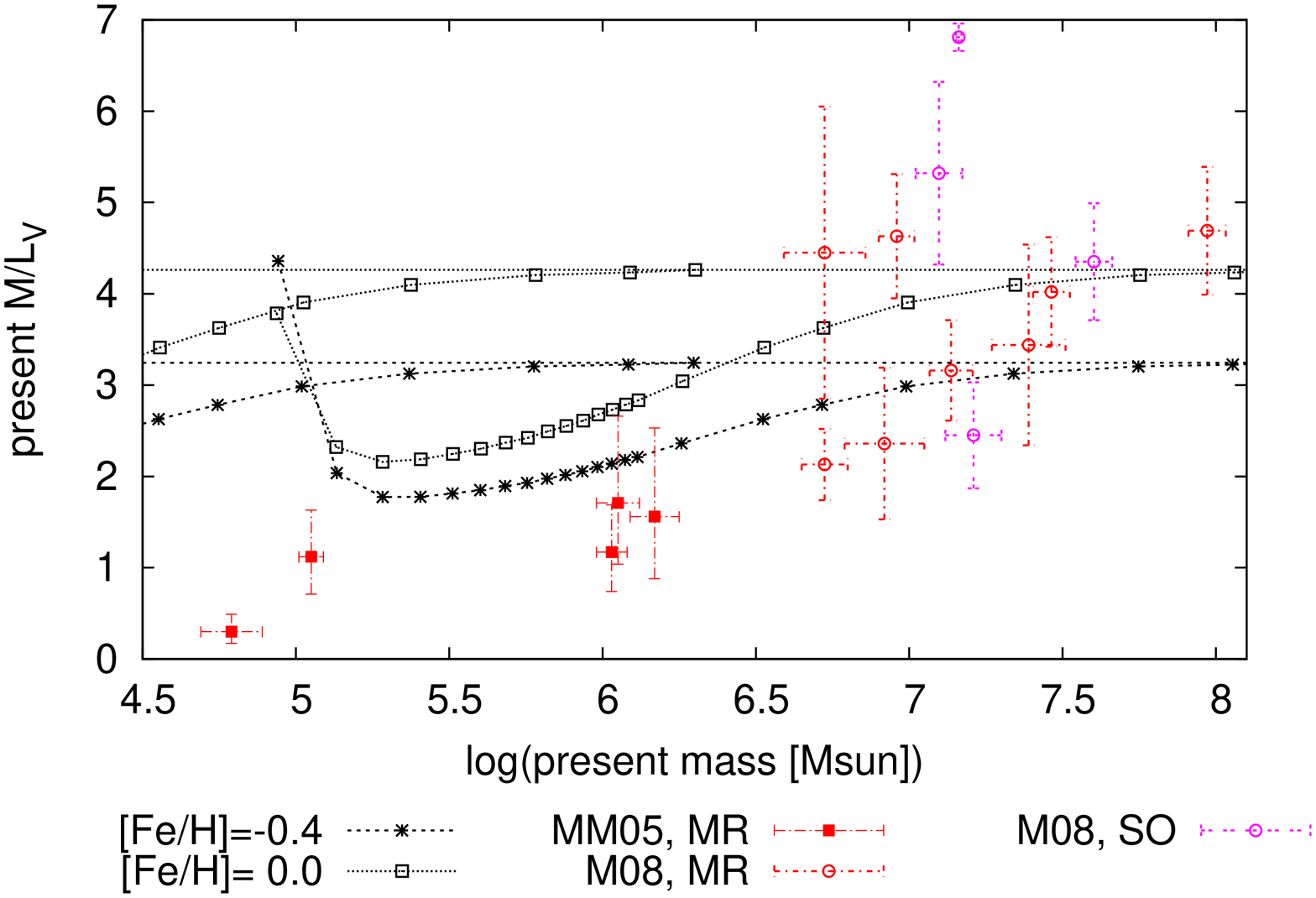}
\end{center}

\caption{Observations of old ($\sim$12 Gyr) objects from
\citet{2005ApJS..161..304M} (MM05) and \citealt{2008A&A...487..921M}
(M08). Top panel: objects with [Fe/H] $<$ -0.55. Bottom panel: objects
with [Fe/H] $\ge$ -0.55. Overplotted are our model V-band M/L ratio as a
function of present-day cluster mass for the appropriate metallicities,
at 12 Gyr and for typical disruption times t$_4$=300 Myr (lower/right
branches for each metallicity) and t$_4$=5 Gyr (upper/left branches).
Horizontal lines at the canonical values of M/L from stellar evolution
are also shown. They represent the limit of infinite disruption times.
These models are intended to illustrate the range covered by our
models, and the inconsistency of most obsverations with models of
infinite disruption time.} 

\label{fig:MLratios3}
\end{figure}

Fig. \ref{fig:MLratios3} compares our new models with observational data
from \citet{2005ApJS..161..304M} and \citet{2008A&A...487..921M}. The
colour coding of the data refers to their metallicity: 

\begin{itemize}

\item blue = [Fe/H] $<$ -1.2 = ``MP'' (metal-poor)
\item green = -1.2 $\le$ [Fe/H] $<$ -0.55 = ``IM''
(intermediate metallicity)
\item red = -0.55 $\le$ [Fe/H] $<$ -0.2 = ``MR'' (metal-rich)
\item magenta = -0.2 $\le$ [Fe/H] = ``SO'' (around solar). 

\end{itemize}

These ranges were chosen to be consistent with the metallicities of the
Padova isochrones/our models. 3 objects from the cited samples are not
shown in these plots due their high M/L ratios: LMC-NGC2257 and
MW-NGC6535 from \citet{2005ApJS..161..304M}, which have M/L ratios of
the order of 8-10 with error bars of the order of 4-5, and the Virgo
cluster object S999 (with M/L = 10.2, from the
\citealt{2008A&A...487..921M} sample), which might have a genuinely high
M/L ratio.

Overplotted are 2 bundles of models for metallicities in the range from
[Fe/H]= -- 1.7 to 0.0 (as is appropriate for the shown observational
data), for cluster ages of 12 Gyr, and local tidal field strengths with
t$_4$ = 5 Gyr (upper/left branches of models of a given metallicity,
representative for halo clusters) and t$_4$ = 300 Myr (lower/right
branches, representative for strong dissolution). The main point of
the comparison is to show the range of M/L$_V$ values reachable with our
models. As can be seen in Fig. \ref{fig:MLratios3}, most Galactic GCs
have M/L values compatible with our predictions and many, especially
low-mass ones are below those of the standard isochrones. The estimated
uncertainties of few per cent, as estimated above for clusters in this
mass range, are not sufficient to bring the observations into agreement
with the standard predictions from stellar evolution alone. We take
this as clear evidence for cluster evolution/dissolution and that our
evolving cluster models are a clear improvement over standard isochrone
fitting for GCs.

Data for the Milky Way and the LMC are taken from
\citet{2005ApJS..161..304M}, the most extensive homogenised compilation
of star cluster M/L ratios (and providing also other star cluster
properties) for these galaxies. The majority of the data for the Milky
Way originates from \citealt{1993ASPC...50..357P}. 
\citet{1993ASPC...50..357P} find a weak correlation of M/L ratio with
mass, consistent with our models, but with large scatter and
uncertainties (on average about 50-60\%) and M/L ratios outside the
accessible range of our models for some of their sample clusters
(\citealt{2005ApJS..161..304M} do not elaborate on this dependence).
They find no significant correlation of M/L ratio with the distance of
the cluster from the Galactic Center or the Galactic plane, contrary to
what might be expected from the BM03 simulations and our models
(although the mixture of clusters with different masses at different
Galactocentric radii, i.e. experiencing different tidal field strengths,
could erase any such signature). However, the present-day cluster
position within the Galaxy is likely less important for the total
disruption time (and therefore the M/L ratio evolution) than the
perigalactic distance and the number of past disk passages, which are
unknown for most clusters. In addition, stochastic effects of the MF
could induce additional scatter. The error bars are too large to find a
clear trend of M/L ratio with metallicity.

Of the 52 old clusters all but 8 are consistent within their 1$\sigma$
ranges with models for [Fe/H]= -- 1.7 or [Fe/H]= -- 0.7. All of these 8
clusters have significantly too low M/L ratios. NGC 2419\footnote{A
re-analysis of the velocity dispersion of NGC 2419 by Baumgardt et al.,
{\sl submitted}, indicates that the mass-to-light ratio is around 2,
which is in good agreement with a canonical mass-to-light ratio and no
dynamical cluster evolution.} and NGC 4590 both have
metallicities\footnote{Data taken  from the Harris catalogue
\citet{1996AJ....112.1487H}, available at\\
http://physwww.physics.mcmaster.ca/$\sim$harris/mwgc.dat} below [Fe/H]=
--1.7, the lowest metallicity for which we can provide models. Those
clusters could possibly be explained by models of even lower metallicity.
For the other clusters (NGC 5272, NGC 5286, NGC 5904, NGC 6366, NGC 6715,
NGC 7089), no immediate explanation (apart from underestimated
observational uncertainties or the impact of the unknown perigalactic
distance and past disk passages) is apparent. However, our new models
represent a significant improvement: while 6 clusters are not consistent
with our new models, 21 clusters are not consistent with the standard
constant-M/L models.

\begin{figure}
\begin{center}
 \vspace{-0.5cm}
 \hspace{1.2cm}
 \includegraphics[angle=270,width=0.9\linewidth]{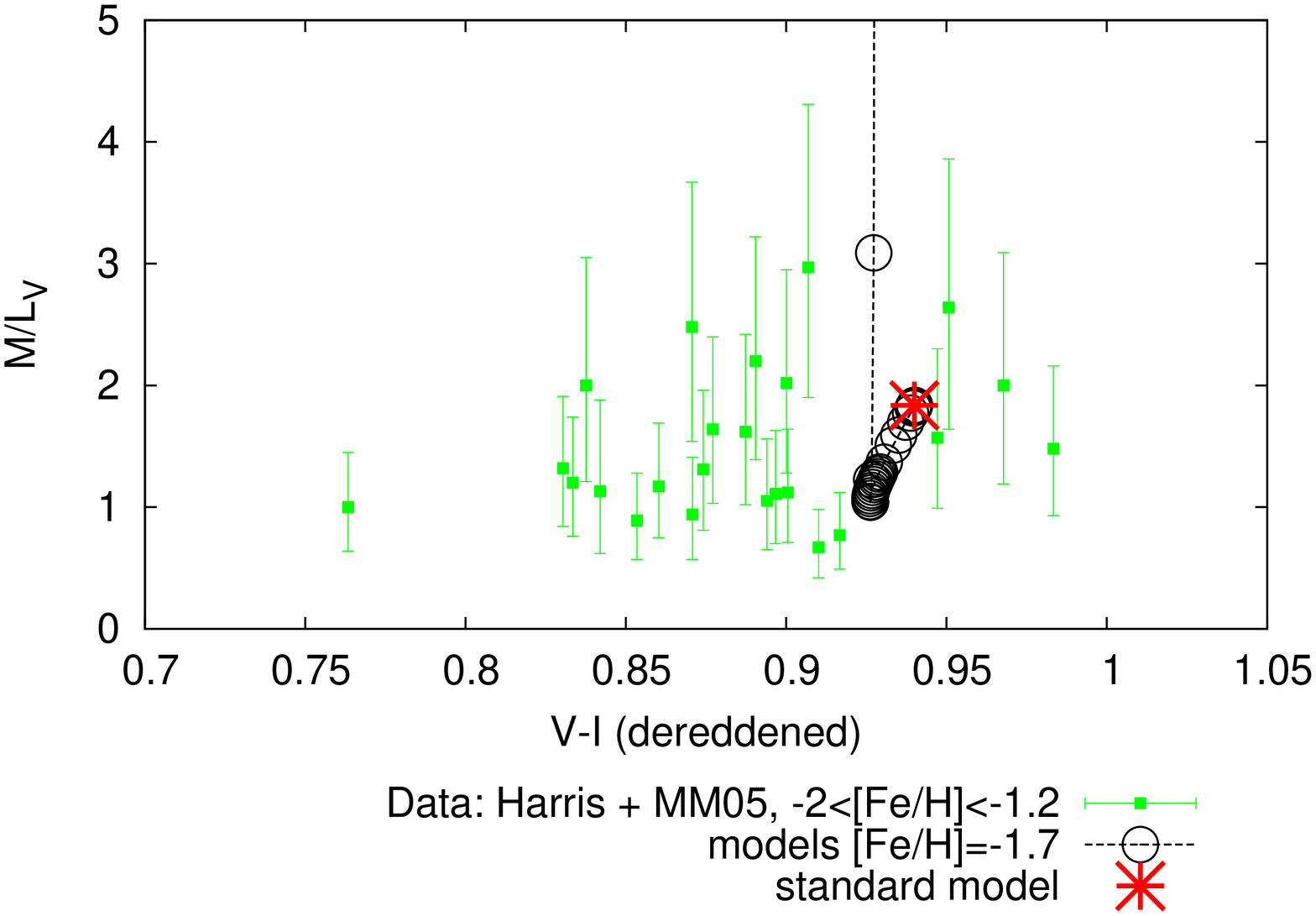}
\end{center}

\caption{ Observations of old ($\sim$12 Gyr) Milky Way globular clusters
from the Harris catalogue (colours) and \citet{2005ApJS..161..304M} (M/L
ratios). Shown is the relation between observed V-I colour (dereddened)
vs M/L$_V$ ratio for metal-poor clusters.} 

\label{fig:MLratios4}
\end{figure}

Another way to analyse the properties of Milky Way globular clusters is
Fig. \ref{fig:MLratios4}, where we compare their dereddened V-I colours
(taken from  the Harris catalogue) with their M/L$_V$ ratios (as given by
\citealt{2005ApJS..161..304M}). We overplot our models for a cluster age
of 12 Gyr. The models with the longest disruption time are equivalent to
the standard/non-dissolving model (marked with the red asterix). For
decreasing disruption time, the models' M/L ratios drop, before
drastically increasing again at the final stages of dissolution. We
restrict this analysis to metal-poor clusters (-2 $<$ [Fe/H] $<$ -1.2),
as the number of higher-metallicity clusters with the required data is
too low to draw strong conclusions. We find good agreement between the
observational data and our models for these clusters concerning their
M/L$_V$ ratios: the observational data are clearly spread out over a
wider range than the standard model could account for, while our new
models cover this range much better. However, the observed cluster
colours span a wide range in V-I (though no colour uncertainties are
available), which we cannot fully account for with our models. Our models
might be about 0.05mag too red for the observations. This could originate
from our choice of isochrones: see Sect. \ref{sec:isochrones} where we
find that other isochrones give results that are bluer than our set of
isochrones. However, the other isochrones are than {\sl bluer} than the
observations, by again $\sim$0.05mag. Other possible causes include:
uncertainties in the reddening estimates, filter curve mismatch, etc.

Data for massive star clusters in NGC 5128 (= Cen A) as well as for
massive objects (commonly referred to as ``Ultra-Compact Dwarf
galaxies'' = UCDs) in the Virgo and Fornax galaxy cluster are taken from
\citet{2008A&A...487..921M}. These data include earlier observations by
\citet{2007A&A...469..147R} for the star clusters in Cen A, and
observations by a variety of authors for the UCDs (see
\citealt{2008A&A...487..921M} for details). While 27 of these
clusters are not consistent with our new models, 47 clusters are not
consistent with the standard constant-M/L models. Also for this sample,
the new models are a significant improvement. 

We therefore conclude that the samples are better described by our new
models with preferential loss of low-mass stars, and we see ongoing
cluster dissolution. In future cluster modelling, this effect has
clearly to be taken into account.

Nonetheless, the sample of massive Cen A clusters and UCDs shows a very
clear and strong trend of increasing M/L ratio with object mass,
especially for metal-poor/intermediate metallicity objects (metal-rich
objects are reasonably well covered by our models, except for the Virgo
cluster UCD S490). This trend can not be reproduced by our models: for
masses larger than $\approx$10$^7$ M$_\odot$ only 3 out of 12 objects
are consistent with our models within their respective 1$\sigma$
uncertainties (one further object is marginally consistent). Such
massive systems are not expected to be mass segregated due to their
large relaxation time, let alone close to disruption (which in our
models is the only possibility to reach M/L ratios higher than the
predictions from standard stellar evolution). While the models do have
inherent sources of uncertainties (e.g. the assumed initial-final mass
relation for remnants, uncertainties in the underlying stellar
isochrones, which will be studied in more detail in Sect.
\ref{sec:comparison}), they are unlikely to raise the model M/L ratios
sufficiently to accommodate a significant fraction of the currently
unexplained observations (especially without removing the agreement for
objects with lower M/L ratios). Two possible explanations for the high
M/L ratios would remain: either a stellar mass function significantly
deviating from the universal \citet{2001MNRAS.322..231K} IMF (see also
\citealt{2008MNRAS.386..864D,2008ApJ...677..276M}), or dark matter (see
\citealt{2008MNRAS.tmp.1255B} for how dark matter can explain the high
M/L ratios of UCDs). 

\subsection{Impact on age determination}
\label{sec:ages}

Evolutionary synthesis models are regularly used to derive the physical
parameters of star clusters (and galaxies) from observed
spectrophotometry. Derived quantities are age, mass and metallicity of
the star cluster as well as the extinction in front of the star cluster
(see e.g., among many others,
\citealt{2002A&A...387..412B,2003AJ....126.1276K,2004MNRAS.347...17A,
2004MNRAS.352..263D,2005ApJ...634L..41K,2006MNRAS.366..295D,
2007ApJ...667L.145S}). Our {\sc galev} models provide a model grid of
SEDs in age/metallicity/extinction. The ``AnalySED tool'' (which we
developed and tested in \citealt{2004MNRAS.347..196A}) compares these
model SEDs with the observed SED of a star cluster using a $\chi^2$
algorithm, to derive the best-fitting parameter combination and their
respective uncertainty ranges from integrated multi-band cluster
photometry.

\begin{figure*}
\begin{center}
  \vspace{-0.5cm}
  \hspace{1.2cm}
  \begin{tabular}{cc}
	 \includegraphics[angle=270,width=0.4\linewidth]{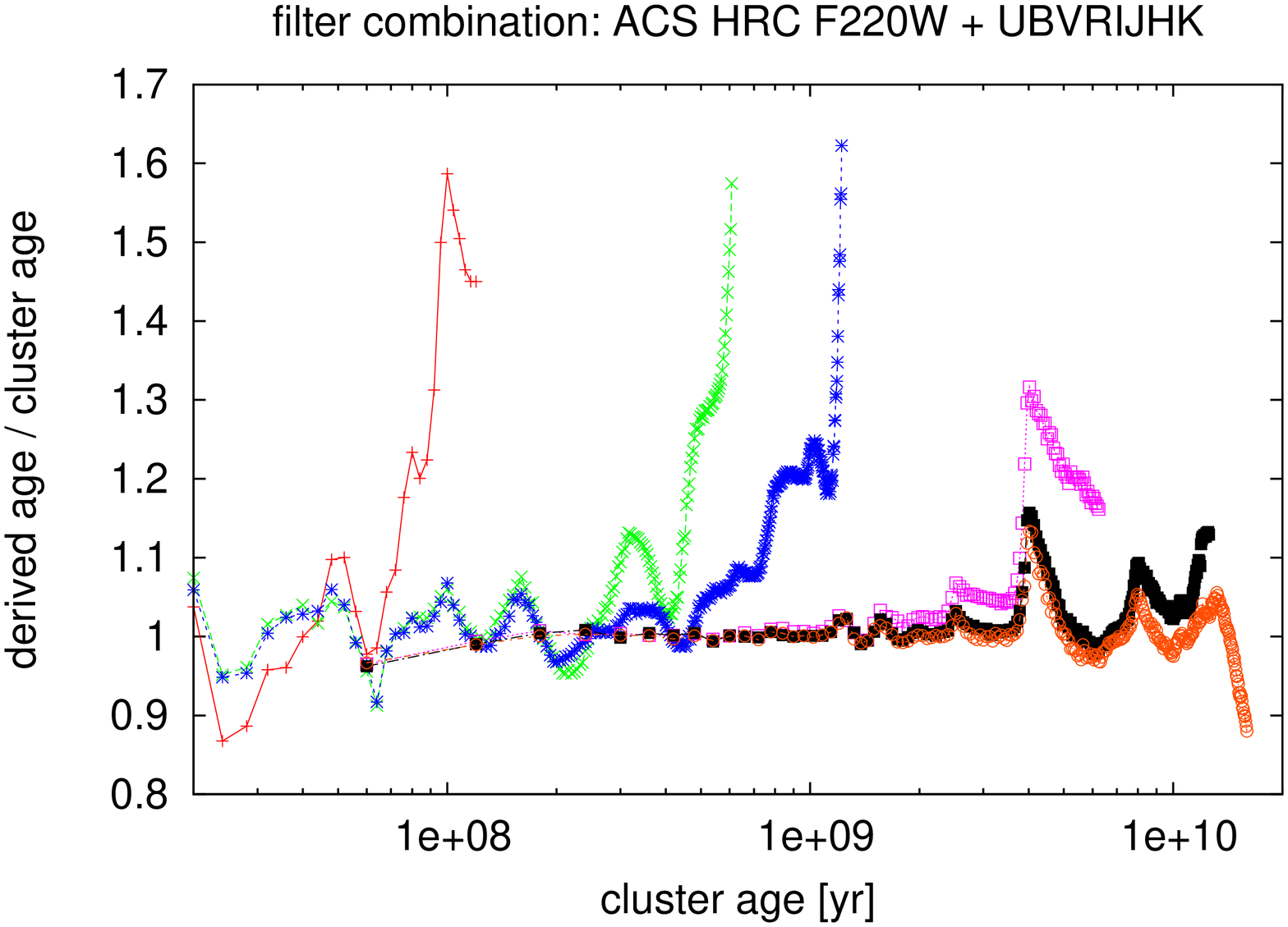} &
	 \includegraphics[angle=270,width=0.4\linewidth]{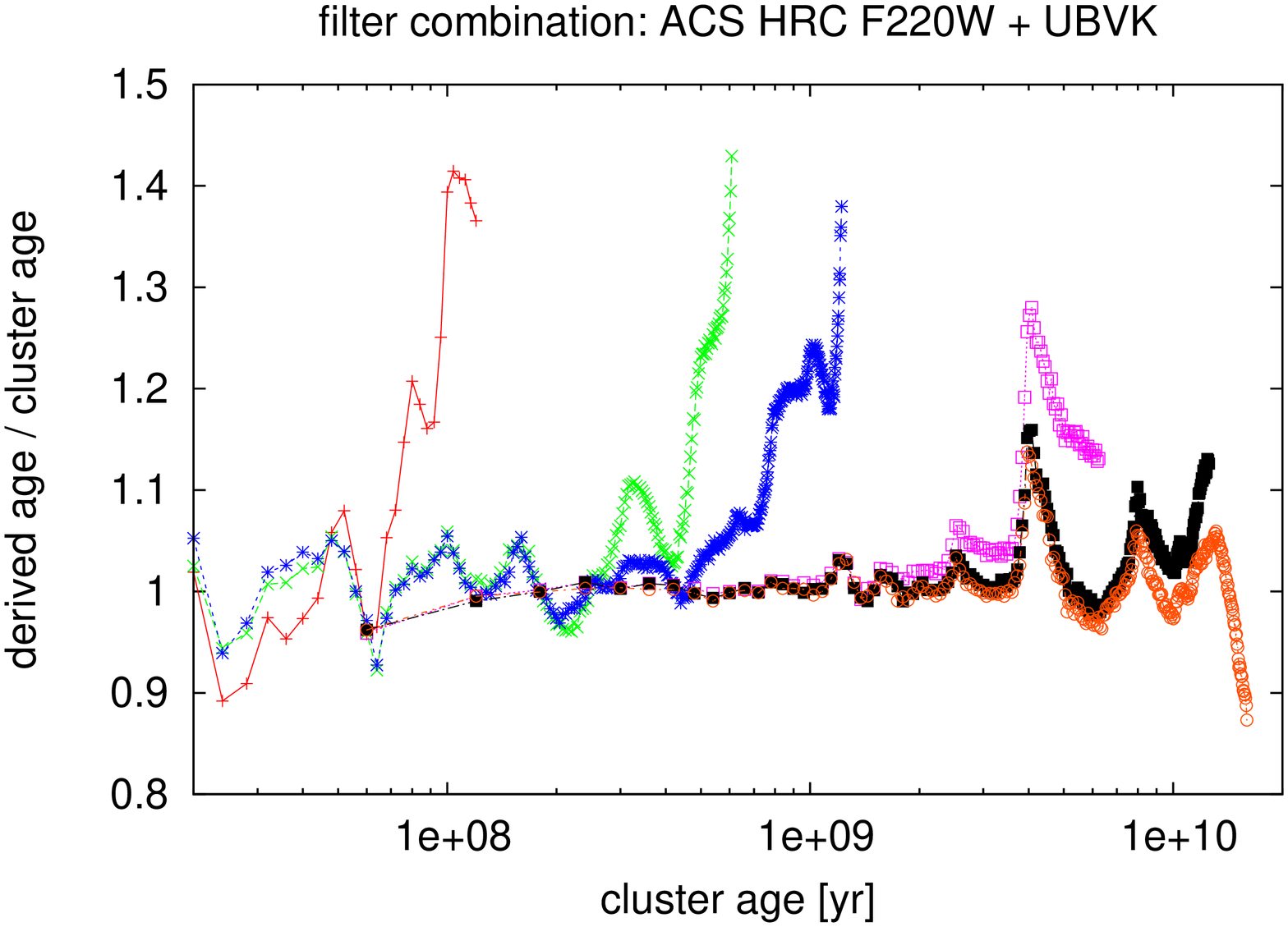} \\  
	 \includegraphics[angle=270,width=0.4\linewidth]{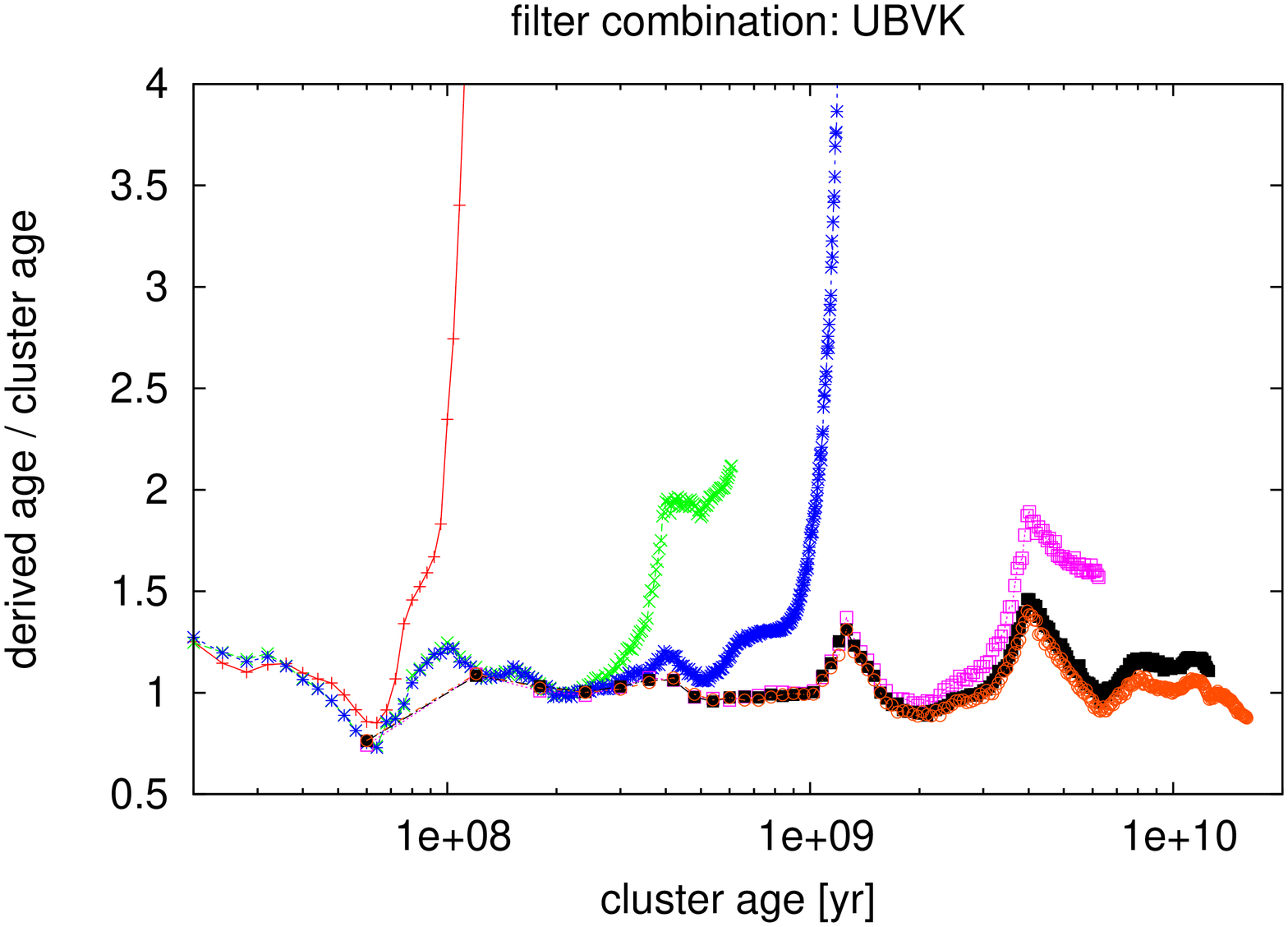} &
	 \includegraphics[angle=270,width=0.4\linewidth]{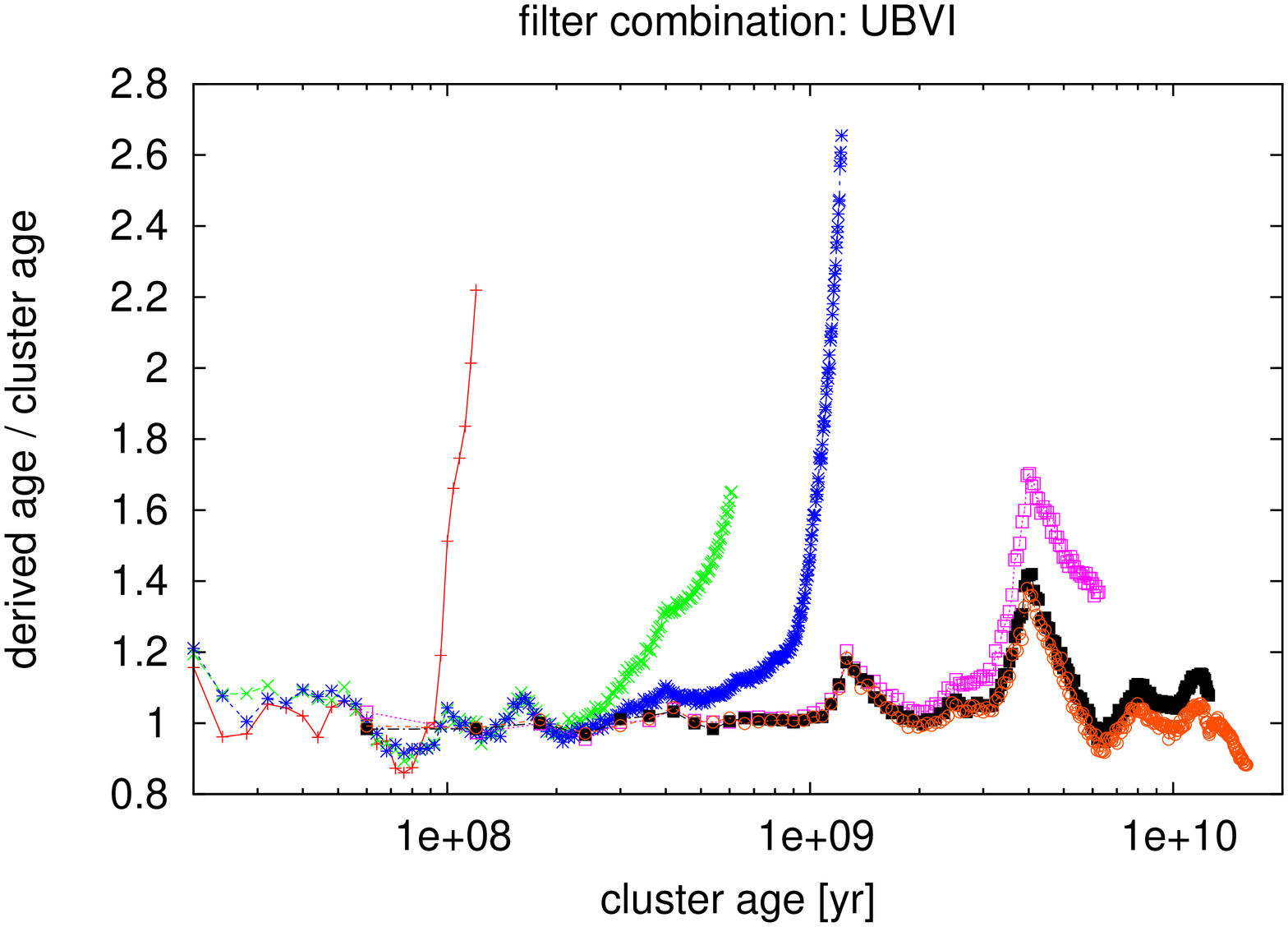} \\
  \end{tabular}
\end{center}

\caption{Derived ages of dissolving star clusters with t$_{\rm 95\%}$ in
the range 100 Myr -- 60 Gyr (from left to right: 100, 500 Myr, 1, 5, 10,
20 Gyr), using standard models, for a number of filter combinations.
Upper left panel: ACS HRC F220W + standard UBVRIJHK, top right: ACS HRC
F220W + standard UBVK,  bottom left: standard UBVK, bottom right:
standard UBVI. Shown is the ratio between the age derived using standard
models and the age the dissolving cluster has.} 

\label{fig:ages} 
\end{figure*}

Here we employ the ``AnalySED tool'' to quantify the differences between
the true ages of dissolving clusters (with a time-dependent MF) and the
ages derived using the standard evolutionary synthesis models (with a MF
fixed to the IMF slopes). We take the cluster photometry from the
dissolving cluster models (for a number of filter combinations), apply
Gaussian noise (with $\sigma$ = 0.1mag) to the photometry in the
individual passbands, and analyse them using the standard,
non-dissolving cluster models. The analysis is done for fixed solar
metallicity and zero extinction (leaving these parameters free to vary
would lead to even stronger deviations from the standard models and
larger uncertainties, as shown in \citealt{2004MNRAS.347..196A}). For
each filter combination, total disruption time and age we generate 1000
test clusters, derive their physical parameters and determine the mean
of the derived ages. The results in terms of the ratio between the
derived mean age and the true cluster age are shown in Fig.
\ref{fig:ages}.

In general, for all models the ages get overestimated for a significant
fraction of the cluster lifetime (for some ages and models the ages can
also get severely underestimated). This agrees well with the discussion
concerning the cluster colours in Sect. \ref{sec:photo}: generally, when
the cluster colours become redder than the standard models, the ages
become overestimated. A direct comparison is not appropriate, though, as
``AnalySED'' uses the whole available Spectral Energy Distribution (SED,
i.e. the dataset containing all magnitudes in a given set of filters for
a given cluster) to determine the model with the best-matching
parameters, hence differences in different filters can either cancel or
amplify each other.

Datasets including the mid-UV (here represented by the ACS HRC F220W
filter) show only modest deviations from the standard models (Fig.
\ref{fig:ages}, top panels). However, 20\% deviations are regularly
found. Datasets lacking the mid-UV, and especially those including
near-IR data, are more sensitive to the changes in the mass functions
(Fig. \ref{fig:ages}, bottom panels). For those datasets, deviations of
50\% or even a factor $\sim$2-3 are found.

\clearpage

\section{Validation of the models}
\label{sec:comparison}

In this section we will investigate several uncertainties inherent to
our models, as well as comparing our new models with models previously
released by our group.

All the following values are {\sl maximum} differences from the
standard models of a given parameter in a given time interval, unless
otherwise noted. In many cases the maximum deviations occur in the
final stages of cluster dissolution, and for the longest total
disruption times $<$ maximum model age. If we would have chosen a
maximum model age of 13 Gyr instead of 16 Gyr, the maximum deviations
would generally be slightly smaller.

For the different issues discussed in this section, we also publish
a few test cases on our webpage, illustrating the impact of different
initial-final mass relations, isochrones and parametrisations of  the
mass function evolution on colours and masses/ mass-to-light ratios. For
the parametrisations of the mass function evolution we select a few
disruption times for presentation on the webpage, while for the
initial-final mass relations and isochrones we present only data without
disruption (i.e. pure stellar evolution) to  avoid confusion.

\subsection{Parametrisation of mass function slope evolution}
\label{sec:spread}

As discussed earlier, the mass function slope evolution is derived from
a subset of N-body simulations by BM03. The subset was selected to
cover the parameter space well, while limiting the impact of low-number
statistics (see Sect. \ref{sec:MF}). 

The fit to the data of the time evolution of the mass function slopes
has formally a very high accuracy due to the large number of data
points. However, as shown in Figs. \ref{fig:slope1change} and
\ref{fig:slope2change}, the N-body models show an intrinsic spread
around the fitted function. We quantified this spread to have a median
value $\le$ $\pm$15\% for ages $\ge$ 1/3 t$_{\rm 95\%}$. For younger
ages, this {\sl relative} spread is larger, however the median of the
{\sl absolute} spread is small, $\le$ $\pm$0.02-0.03 change in the slope.

We test the impact of this spread by calculating models for which the
time-dependent part of the mass function slope is reduced/increased by
15\%. We find, as expected, the change in the ``high-mass slope'' (i.e.
for masses $\ge$ 0.3 M$_\odot$) to be of primary importance, while the
time-dependent contribution from stars with masses $\le$ 0.3 M$_\odot$
changes the photometry only mildly.

As the mass evolutions of the cluster (total, luminous and remnant mass)
were derived independent of the mass function evolution, the masses are
not affected.

The impact of this uncertainty on the colors is small: the changes
induced for the models with the shortest disruption time t$_{\rm 95\%}$
(i.e. 100 Myr) and the colors with the longest wavelength coverage (i.e.
V-K) reach $\approx$ 0.07mag at final disruption. These changes
decrease rapidly with increasing disruption time and decreasing
wavelength coverage. 

For ages ${\rm t \le t_{95\%}}$ the magnitudes change by $\sim$ 0.15 --
0.2 mag (with the changes slightly larger for the shortest t$_{\rm
95\%}$ and red passbands). As the mass is unaltered, this directly
translates into a change in the M/L ratio by 15 -- 20\%.

For ages ${\rm t > t_{95\%}}$ both magnitudes and M/L ratios diverge
from the models using the best fitting relation for the time evolution
of the mass function slopes. Models with a weaker time evolution are
increasingly brighter and have consequently lower M/L ratios.

\begin{figure}
\begin{center}
  \vspace{-0.5cm}
  \hspace{1.2cm}
	 \includegraphics[angle=270,width=1.0\linewidth]{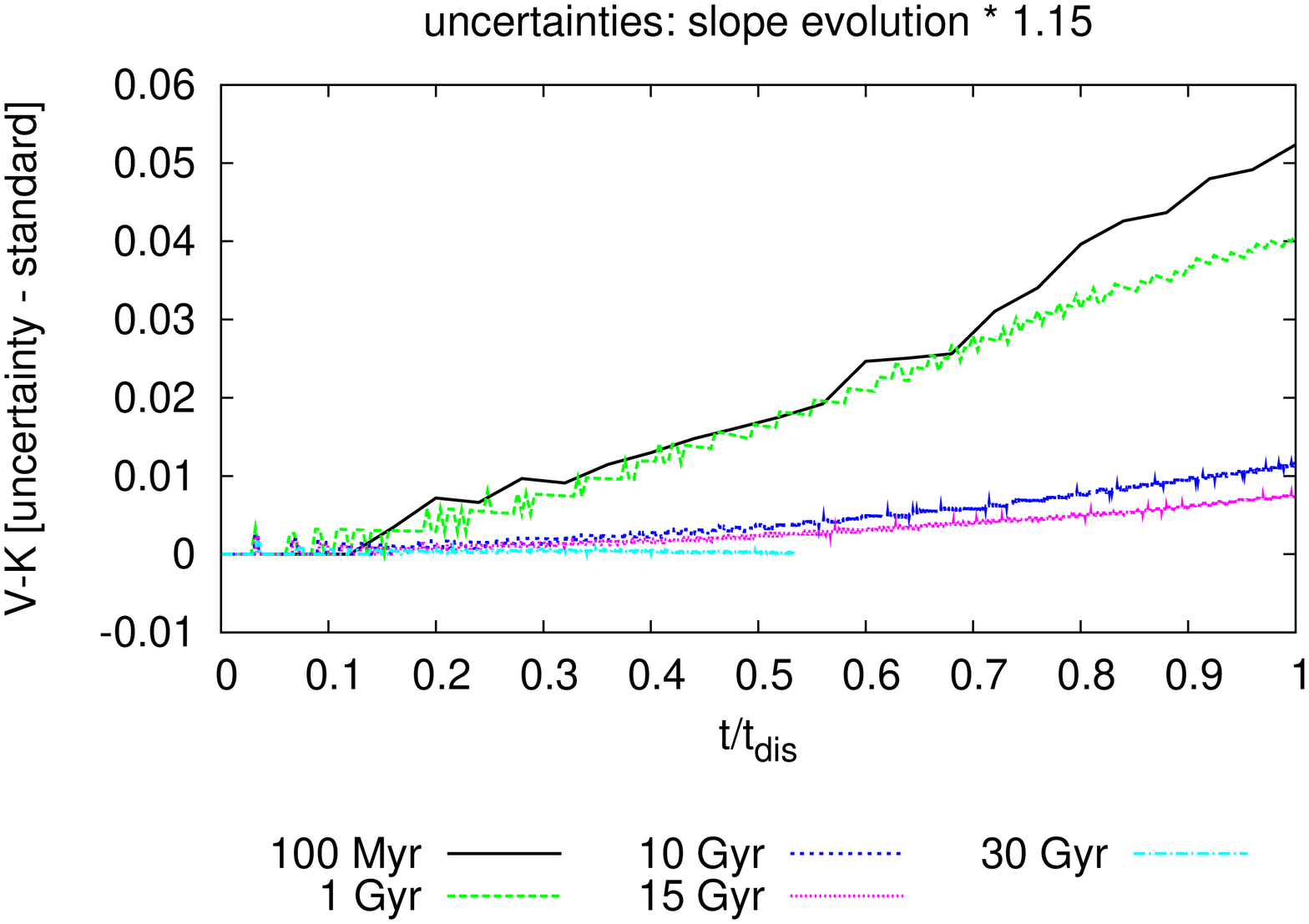} 
	 \includegraphics[angle=270,width=1.0\linewidth]{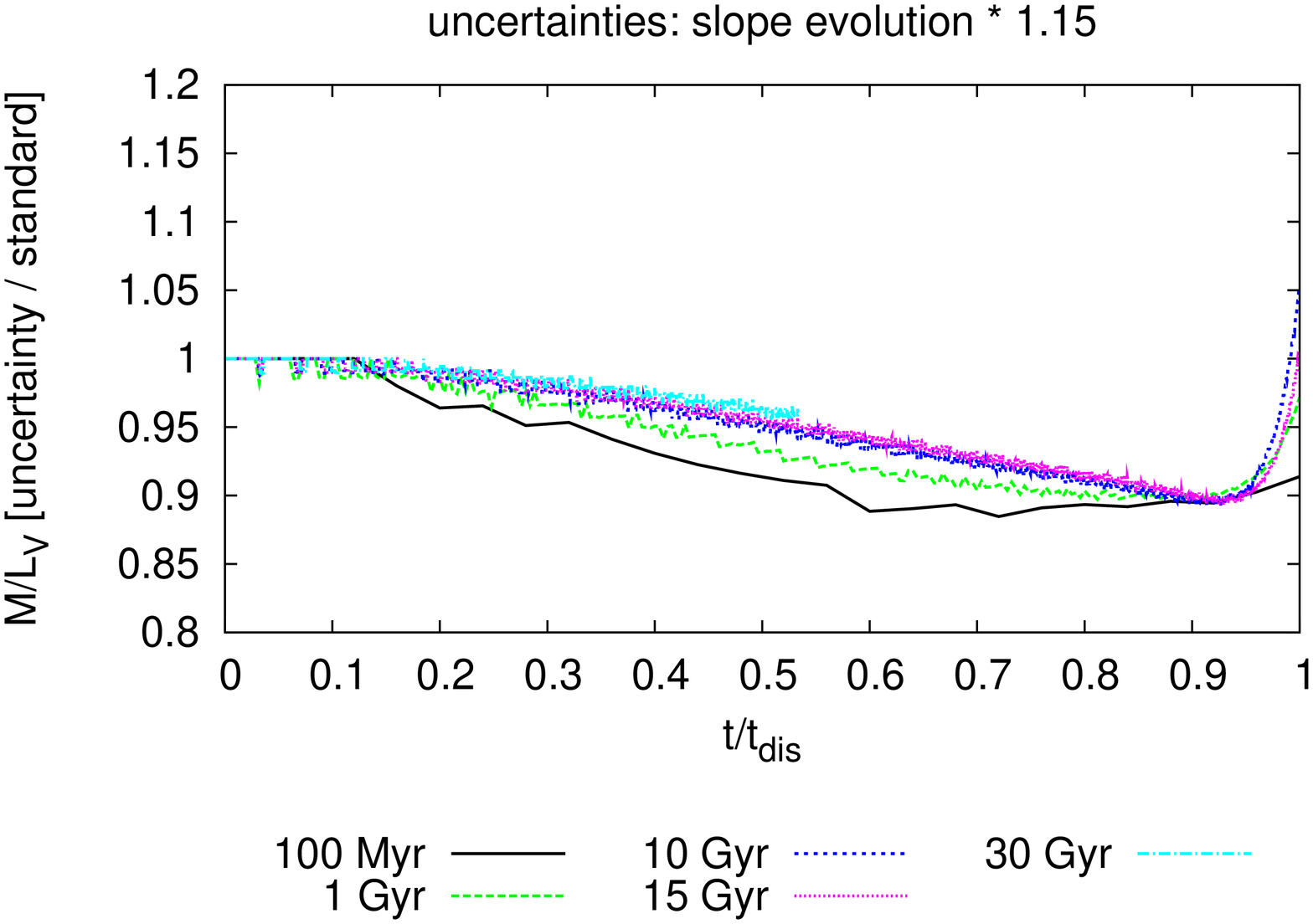} 
\end{center}

\caption{Integrated V-K colour (left panel) and M/L$_V$ ratio (right
panel) for a Kroupa IMF, solar metallicity, our standard isochrones, a
range of cluster disruption times, with the effect of MF evolution 
enhanced by 15\%  w.r.t.the standard models. Shown are the quantities
relative to the respective quantities of our standard models.
Diminishing the effect of MF evolution by 15\% yields quantitatively
similar results, however, the changes are in the opposite direction.} 

\label{fig:slope_uncert} 
\end{figure}

In Fig. \ref{fig:slope_uncert} we illustrate the effects of
enhancing the MF evolution by 15\%. Diminishing the effects of MF
evolution by 15\% gives quantitatively similar results with the changes
w.r.t. the standard models going in the opposite direction. The V-K 
colour evolutions are the most extreme cases: Effects become smaller
for shorter wavelength coverage and longer disruption times.

In summary, the uncertainty induced by the spread of N-body model data
around the fitted time evolution of the mass function slopes has an
impact on the model predictions. However, for ages ${\rm t \le t_{95\%}}$
the induced uncertainties are much smaller than the error one makes by
not taking into account the effect of preferential mass loss and cluster
dissolution. In addition, we aim at describing the {\sl average}
cluster.

\subsection{Initial-final mass relations}
\label{sec:ifmr}

Our model results, especially the M/L ratios discussed in Sect.
\ref{sec:ML}, depend on the treatment of stellar remnants. The remnants
mass is calculated from the progenitor star's initial mass and the
adopted initial-final mass relation (IFMR), which accounts for mass loss
during the life of the progenitor star and due to the ``death'' of the
star and the formation of the remnant.

The IFMR for white dwarfs used in this work is based on the work by
\citet{1983A&A...121...77W} (hereafter ``Weidemann83''). As the IFMR is
still uncertain, we tested our choice by adopting different IFMRs for
white dwarfs, namely by \citet{2000A&A...363..647W} (hereafter
``Weidemann00''), by \citet{2008ApJ...676..594K} (hereafter
``Kalirai08'') and the prescription by \citet{2000MNRAS.315..543H}
(hereafter ``HPT00''). For the latter one we also adopt their IFMR for
neutron stars, while for all other IFMRs we adopt
\citealt{1988PhR...163...13N}. 

Changes discussed below are w.r.t. our standard IFMR Weidemann83.

We find the IFMR to be of minor influence on the results for ages ${\rm
t \le t_{95\%}}$: the total mass changes by maximum 2 -- 3\%, while the
luminous mass changes by 4\% and 6\% (for Weidemann00/Kalirai08 and
HPT00, respectively). This translates into magnitude changes of
$\sim$0.05mag and $\sim$0.07 mag (for Weidemann00/Kalirai08 and HPT00,
respectively). The associated effect on the M/L ratios is $\sim$4.5\%
and $\sim$6.5\% (for Weidemann00/Kalirai08 and HPT00, respectively).

For larger ages ${\rm t \ge t_{95\%}}$ the results eventually diverge.
However, only in the last 5\% of a cluster's lifetime the total mass
differs by more than 10\%, regardless of the choice of IFMR. 

\begin{figure}
\begin{center}
  \vspace{-0.5cm}
  \hspace{1.2cm}
	 \includegraphics[angle=270,width=1.0\linewidth]{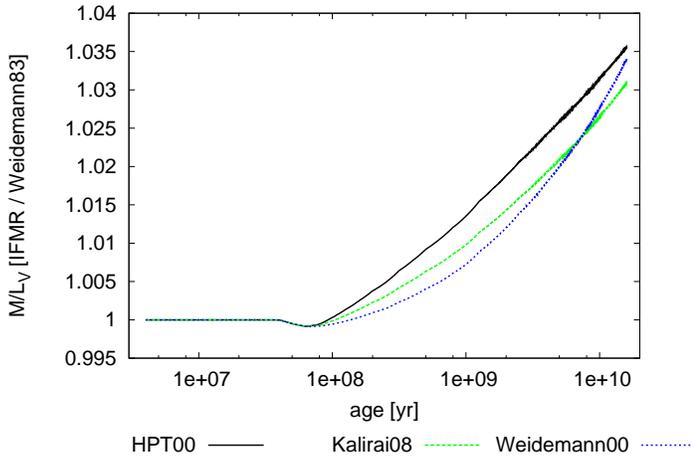}
\end{center}

\caption{M/L$_V$ ratio for a Kroupa IMF, solar metallicity, no cluster
dissolution and the 3 alternative IFMRs discussed in the text. Shown are
the quantities relative to the respective quantities of our standard
models.} 

\label{fig:IFMR} 
\end{figure}

In Fig. \ref{fig:IFMR} we show the impact of the chosen IFMR on the
M/L$_V$ ratio for infinite disruption time.

\subsection{Isochrones}
\label{sec:isochrones}

Our choice of isochrones (i.e. isochrones from the Pavoda group, first
presented in \citealt{1994A&AS..106..275B} with later updates of the
Padova group concerning the TP-AGB phase = ``updated Padova94'') is driven by
the following points:

\begin{enumerate}

\item For consistency with {\sc galev} models of galaxies we require
isochrones which cover the full mass range up to high masses (ideally up
to $\sim$ 120 M$_\odot$) to properly model ongoing star formation in
galaxies.

\item Likewise models covering a wide range in metallicities is desired
to consistently model old/metal-poor globular clusters and
young/metal-rich star clusters formed in nearby starbursts, as well as
to model galaxies consistently from the onset of star formation to their
present stage.

\item The models should cover all relevant evolutionary stages of these
stars, especially the very luminous phases (for our study especially the
TP-AGB phase is of prime importance, but also early stages like
supergiants are important).

\end{enumerate}

We regret that recent high-quality isochrone calculations (e.g.
\citealt{2000A&AS..141..371G, 2001ApJS..136..417Y, 2004A&A...421.1121C,
2004ApJ...612..168P, 2006ApJ...642..797P, 2008A&A...484..815B}) are all
focussing on ``low-mass'' stars (maximum up to $\sim$ 10 M$_\odot$,
though \citealt{2008A&A...484..815B} announce models up to 20 M$_\odot$
for the near future) and/or do not fulfil one or more criteria mentioned
above. However, stars more massive than $\approx$ 10 M$_\odot$
contribute significantly to the chemical enrichment and the
light of young star clusters and most galaxies.

The only models fulfilling all mentioned criteria are models by the
Padova group (\citealt{1994A&AS..106..275B} plus TP-AGB updates) and by
the Geneva group (\citealt{1992A&AS...96..269S, 1993A&AS..101..415C,
1993A&AS...98..523S}).

As the main focus in this paper is on systems older than $\sim$ 100 Myr
we prefer the updated Padova94 isochrones over the Geneva isochrones.

The alternative solution, to combine isochrones from different
groups/epochs, was rejected as consistency cannot be ensured.

We tested solar-metallicity isochrones by \citet{2004A&A...421.1121C,
2004ApJ...612..168P, 2008A&A...482..883M} (also known as ``Pisa/GIPSY'',
``BASTI'' and ``new Padova'', respectively) with respect to the updated
Padova94 isochrones we used in this study, and derived star cluster
models for test purposes. While the Pisa isochrones are offset from all
other isochrones (they are generally significantly hotter, but are based
on more limiting input physics), the other isochrones are in overall
good agreement with the updated Padova94 isochrones. Small differences
include:

\begin{itemize}

\item For increasing age, the BASTI main-sequence turn-off temperature
goes from slightly cooler than the updated Padova94 to slightly hotter
(by a few per cent). This results in an increasing deviation of
U-/B-band magnitudes compared to the updated Padova94 isochrones by up
to 0.5mag. The new Padova isochrones show much smaller deviations $\la$
0.15mag in these passbands. Contrary, for both BASTI and new Padova,
colours like U-B or B-V deviate for most of the time by $\la$ 0.1mag,
and for the majority of time by $\la$ 0.05mag from the updated Padova94
models.

\item Overall, the RGBs and AGBs in the BASTI and new Padova isochrones 
are hotter than in the updated Padova94 isochrones. Especially stars
with the highest luminosities on the RGB/AGB are treated differently.
For ages younger than $\sim$1 Gyr, the test models deviate
significantly, both from our standard model as well as from each other.
For ages $\ga$2 Gyr, the BASTI and new Padova isochrones give comparable
optical/NIR colours V-I and V-K, but are offset from the updated
Padova94 models by $\sim$ -0.1mag (V-I) and $\sim$ -0.6mag (V-K), in the
sense that the updated Padova94 models are redder.

\item The BASTI ``non-canonical models'' (i.e. with core convective
overshooting during the H-burning phase) are closer to the updated
Padova94 isochrones than their ``canonical models'' (i.e. without
overshooting).

\item The mass lost due to stellar evolution differs by 2\% (new Padova)
-- 7\% (BASTI) when compared to the updated Padova94 isochrones.

\item The {\sl relative} effects induced by the preferential mass loss
(i.e. the difference between models with and without the effects of
cluster dissolution) are qualitatively robust against the choice of
isochrones. Small quantitative differences are present. However, they
tend to be even stronger for the new isochrones than for the updated
Padova94 isochrones.

\end{itemize}

\begin{figure}
\begin{center}
  \vspace{-0.5cm}
  \hspace{1.2cm}
	 \includegraphics[angle=270,width=1.0\linewidth]{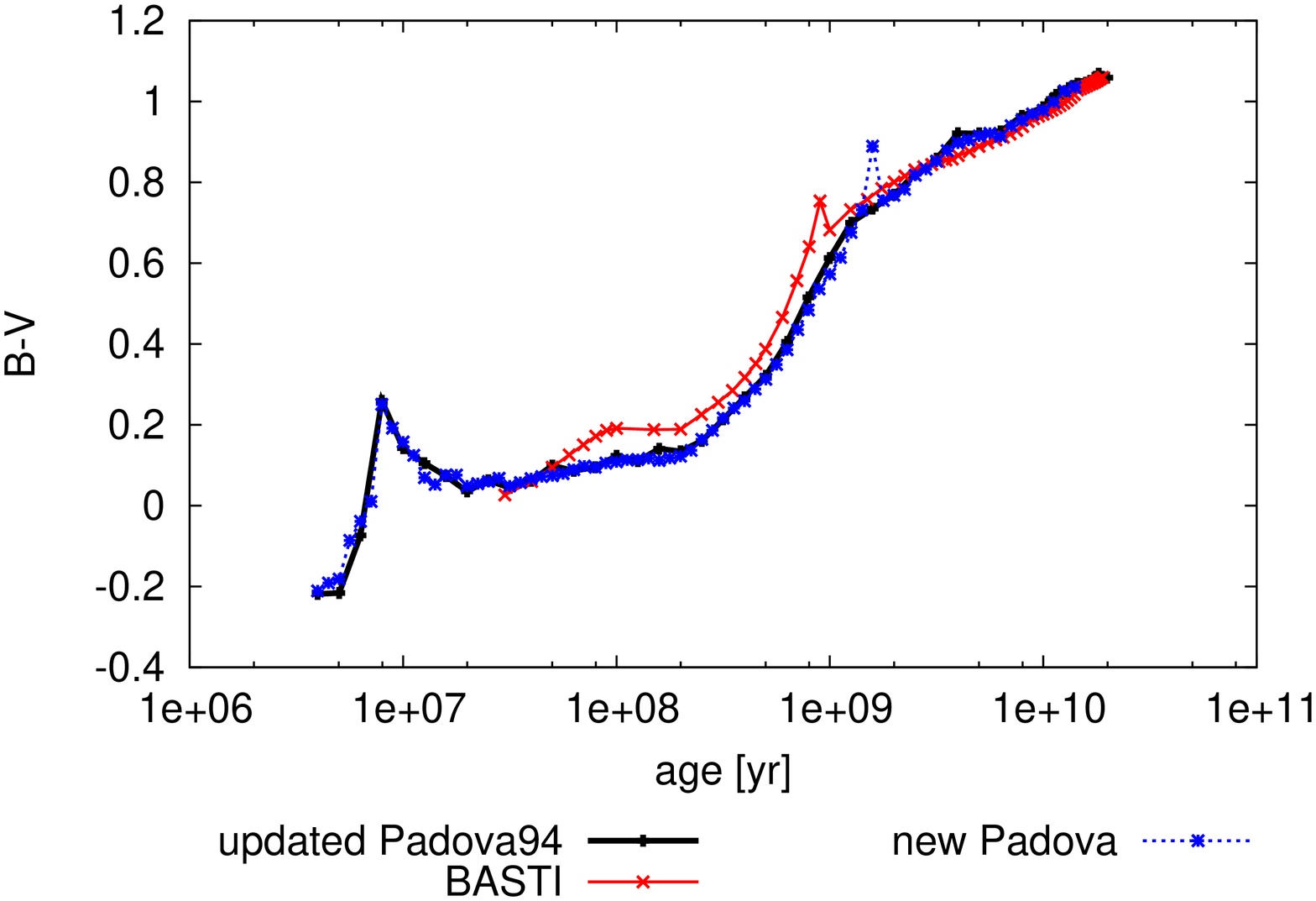}
	 \includegraphics[angle=270,width=1.0\linewidth]{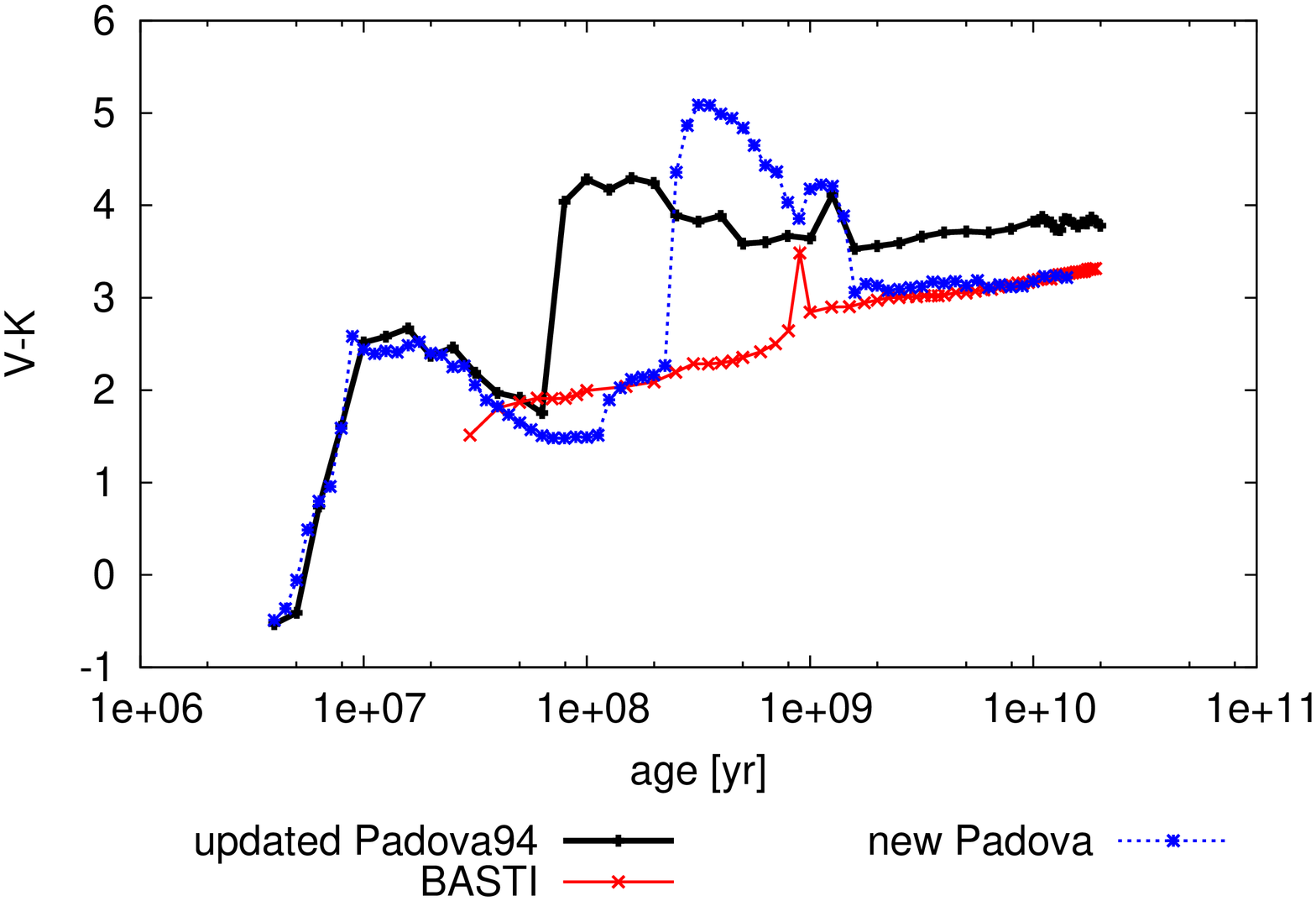}
\end{center}

\caption{Integrated B-V (left panel) and V-K colour (right panel) for a
Kroupa IMF, solar metallicity, no cluster dissolution and 3 different
sets of isochrones: the ``upgraded Padova94'' (our standard models,
black lines), the ``BASTI'' isochrones (red lines,
\citealt{2004ApJ...612..168P}) and the ``new Padova'' isochrones (blue
dotted lines, \citealt{2008A&A...482..883M}).} 

\label{fig:BASTI} 
\end{figure}

For two example colours (B-V and V-K), the time evolution for a standard
SSP (i.e. without cluster dissolution) at solar metallicity is shown in
Fig. \ref{fig:BASTI}. The blue colour B-V, dominated by hot stars mainly
on the main sequence, shows good agreement between the investigated
isochrones. The V-K colour for ages younger than $\approx$1.5 Gyr is in
strong disagreement between all three isochrones, up to 2.8mag
difference around an age of 300 Myr, showing strong differences in the
treatment of the AGB phase. For older ages, the new Padova agrees well
with the BASTI isochrones, but both are offset from the upgraded
Padova94 by  $\approx$0.6mag.

To account for the uncertainty in the choice of isochrones, we will
build grids of dissolving cluster models, based on the BASTI, the new
Padova and the \citet{2008A&A...484..815B} (once the extension to
higher masses is published) isochrones, respectively, and release it 
on our webpage. These models will also employ the 
\citet{2008ApJ...676..594K} initial-final mass relation.

\subsection{Comparison with earlier work}
\label{sec:earlierwork}

The new models presented here supersede our earlier work
(\citealt{2006A&A...452..131L}). In \citet{2006A&A...452..131L} we
approximated the changes in the mass function by a time-dependent lower
mass limit (i.e. assuming that only the lowest-mass stars are removed
from the cluster, while higher-mass stars might only be removed by
stellar evolution) and scaled our models to match the total mass in
stars with M $<$ 2 M$_\odot$ with the BM03 simulations. Recently, this
approach has been improved by \citet{2008A&A...490..151K} by
incorporating the effects of stellar remnants for clusters of different
initial masses and different total disruption times for a range of
metallicities. In that paper, the consequences of various physical
effects on the photometry and M/L ratios have been investigated, e.g.
initial mass segregation, the role of white dwarfs and neutron stars,
the role of metallicity etc..

Due to the normalisation procedure, the {\sl total} masses of the
earlier models differs negligibly from the new models. However, the
number of bright stars in the new models decreases slower than in the
old models by \citet{2006A&A...452..131L} (the models by
\citealt{2008A&A...490..151K} represent already an improvement over the
earlier work, and are more consistent with the work presented here).
Hence the new models are brighter than the old models, especially for
short disruption times. Consequently, the new mass-to-light ratio is
lower, by 20 -- 40\%.

The new models are redder than the older ones, because the changing
slope of the mass function slowly depopulates the (blue) main-sequence
turn-off region already early on. Contrary, in the older models stars in
the main-sequence turn-off region are removed more abruptly when the
lower mass limit reaches the turn-off mass. 

Before the lower mass limit reaches the turn-off mass in the old models,
colors get slightly bluer for a short time, because almost all main
sequence stars redder than the turn-off have been removed by then. This
feature is not present in the new models, due to the more gradual mass
loss. In addition, the old models show a strong reddening in their final
stages, as the star cluster contains exclusively red giants/AGB stars
(plus stellar remnants). Also this feature is not strongly present in
the new models, as the mass function even close to total disruption
covers a wider range.

\section{Conclusions}
\label{sec:conclusions}

We presented a novel suite of evolutionary synthesis models, which
accounts for the dynamical evolution of star clusters in a tidal field
in a realistic manner. The dynamically induced changes in the stellar MF
within the cluster and the overall mass loss of stars from the cluster
into the surrounding field population is consistently taken into
account. The models are made publicly available on our webpages 
http://www.phys.uu.nl/$\sim$anders/data/SSP\_varMF/ and
http://data.galev.org for general use.

Based on the simulations presented in BM03, we improved the
parametrisation of the time evolution of the MF slope changes. We then
combined this new description of the MF slopes with our {\sc galev}
evolutionary synthesis models. The resulting models, calculated for a
range in metallicities and total cluster disruption times, were shown to
deviate significantly from the canonical evolutionary synthesis models
which neglect the effects of dynamical cluster evolution. Depending on
the total cluster disruption time and the colour index under
investigation, differences up to 0.7mag (and in a large number of cases
exceeding 0.1mag) were found. These deviation were shown to lead to
significant misinterpretations of observation. E.g. cluster age
determinations can be wrong by 20 -- 50\%, in extreme cases by up to a
factor $\sim$2 -- 3. These deviations were found to depend strongly on
the filter combination used to derive the ages: combinations including
near-IR filters tend to be more sensitive to the changing MF, while for
large wavelength coverage and/or large numbers of filters the deviations
are still significant but generally smaller.

Also the M/L ratios are strongly affected, and therefore photometric
cluster masses derived from observations. For the largest part of a
cluster's lifetime the M/L ratios are significantly below the canonical
values (by up to a factor $\sim$3 -- 7). In late stages of cluster
dissolution the M/L ratios exceed the standard values, as the cluster
mass gets increasingly dominated by stellar remnants. This period can
last for up to $\sim$16\% of the cluster's total disruption time. In
both cases, the M/L ratios are strongly time-dependent. For fixed
cluster age and/or fixed local disruption time, the dependence of M/L
ratios on the presently observed cluster mass was investigated. They are
broadly consistent with observations, although the observations show
large scatter and uncertainties.

Our results confirm the trends in the evolution of colour and
mass-to-light ratios of dissolving clusters, obtained  by
\citet{2008A&A...490..151K} and \citet{2008A&A...486L..21K}, who used a
simplified description of the changes in the mass function due to the
preferential loss of low-mass stars in star clusters.

While the absolute values of our results depend on our choice of
input physics, the general behaviour is robust against these choices. We
will update our models whenever better input physics becomes available.

\section{Acknowledgements}  

PA acknowledges funding by NWO (grant 614.000.529) and the European
Union (Marie Curie EIF grant MEIF-CT-2006-041108). PA and HL would like
to thank the ISSI in Bern/Switzerland for their hospitality and support.
PA would like to acknowledge fruitful discussions with Ines Brott and
Rob Izzard, as well as with Diederik Kruijssen. Many thanks to Marina
Rejkuba for a critical reading of the paper and asking the right
questions. In addition, thanks to Marina Rejkuba and Steffen Mieske for
kindly providing part of the observational data. PA is in Uta Fritzes
debt for many years of teaching, advice and fruitful collaboration.
This research was supported in part by the National Science Foundation 
under Grant No. PHY05-51164.

\bibliographystyle{aa}
\bibliography{Bibliography}

\begin{thebibliography}{79}
\expandafter\ifx\csname natexlab\endcsname\relax\def\natexlab#1{#1}\fi

\bibitem[{{Anders} {et~al.}(2004{\natexlab{a}}){Anders}, {Bissantz},
  {Fritze-v.~Alvensleben}, \& {de Grijs}}]{2004MNRAS.347..196A}
{Anders}, P., {Bissantz}, N., {Fritze-v.~Alvensleben}, U., \& {de Grijs}, R.
  2004{\natexlab{a}}, \mnras, 347, 196

\bibitem[{{Anders} {et~al.}(2004{\natexlab{b}}){Anders}, {de Grijs},
  {Fritze-v.~Alvensleben}, \& {Bissantz}}]{2004MNRAS.347...17A}
{Anders}, P., {de Grijs}, R., {Fritze-v.~Alvensleben}, U., \& {Bissantz}, N.
  2004{\natexlab{b}}, \mnras, 347, 17

\bibitem[{{Anders} \& {Fritze-v.~Alvensleben}(2003)}]{2003A&A...401.1063A}
{Anders}, P. \& {Fritze-v.~Alvensleben}, U. 2003, \aap, 401, 1063

\bibitem[{{Bastian} \& {Goodwin}(2006)}]{2006MNRAS.369L...9B}
{Bastian}, N. \& {Goodwin}, S.~P. 2006, \mnras, 369, L9

\bibitem[{{Baumgardt} {et~al.}(2008){Baumgardt}, {De Marchi}, \&
  {Kroupa}}]{2008ApJ...685..247B}
{Baumgardt}, H., {De Marchi}, G., \& {Kroupa}, P. 2008, \apj, 685, 247

\bibitem[{{Baumgardt} \& {Makino}(2003)}]{2003MNRAS.340..227B}
{Baumgardt}, H. \& {Makino}, J. 2003, \mnras, 340, 227

\bibitem[{{Baumgardt} \& {Mieske}(2008)}]{2008MNRAS.tmp.1255B}
{Baumgardt}, H. \& {Mieske}, S. 2008, \mnras, 1255

\bibitem[{{Bertelli} {et~al.}(1994){Bertelli}, {Bressan}, {Chiosi}, {Fagotto},
  \& {Nasi}}]{1994A&AS..106..275B}
{Bertelli}, G., {Bressan}, A., {Chiosi}, C., {Fagotto}, F., \& {Nasi}, E. 1994,
  \aaps, 106, 275

\bibitem[{{Bertelli} {et~al.}(2008){Bertelli}, {Girardi}, {Marigo}, \&
  {Nasi}}]{2008A&A...484..815B}
{Bertelli}, G., {Girardi}, L., {Marigo}, P., \& {Nasi}, E. 2008, \aap, 484, 815

\bibitem[{{Bicker} {et~al.}(2002){Bicker}, {Fritze-v.~Alvensleben}, \&
  {Fricke}}]{2002A&A...387..412B}
{Bicker}, J., {Fritze-v.~Alvensleben}, U., \& {Fricke}, K.~J. 2002, \aap, 387,
  412

\bibitem[{{Bicker} {et~al.}(2004){Bicker}, {Fritze-v.~Alvensleben},
  {M{\"o}ller}, \& {Fricke}}]{2004A&A...413...37B}
{Bicker}, J., {Fritze-v.~Alvensleben}, U., {M{\"o}ller}, C.~S., \& {Fricke},
  K.~J. 2004, \aap, 413, 37

\bibitem[{{Boutloukos} \& {Lamers}(2003)}]{2003MNRAS.338..717B}
{Boutloukos}, S.~G. \& {Lamers}, H.~J.~G.~L.~M. 2003, \mnras, 338, 717

\bibitem[{{Bruzual} \& {Charlot}(2003)}]{2003MNRAS.344.1000B}
{Bruzual}, G. \& {Charlot}, S. 2003, \mnras, 344, 1000

\bibitem[{{Cariulo} {et~al.}(2004){Cariulo}, {Degl'Innocenti}, \&
  {Castellani}}]{2004A&A...421.1121C}
{Cariulo}, P., {Degl'Innocenti}, S., \& {Castellani}, V. 2004, \aap, 421, 1121

\bibitem[{{Cervi{\~n}o} \& {Luridiana}(2004)}]{2004A&A...413..145C}
{Cervi{\~n}o}, M. \& {Luridiana}, V. 2004, \aap, 413, 145

\bibitem[{{Cervi{\~n}o} \& {Luridiana}(2006)}]{2006A&A...451..475C}
{Cervi{\~n}o}, M. \& {Luridiana}, V. 2006, \aap, 451, 475

\bibitem[{{Cervi{\~n}o} \& {Moll{\'a}}(2002)}]{2002A&A...394..525C}
{Cervi{\~n}o}, M. \& {Moll{\'a}}, M. 2002, \aap, 394, 525

\bibitem[{{Charbonnel} {et~al.}(1993){Charbonnel}, {Meynet}, {Maeder},
  {Schaller}, \& {Schaerer}}]{1993A&AS..101..415C}
{Charbonnel}, C., {Meynet}, G., {Maeder}, A., {Schaller}, G., \& {Schaerer}, D.
  1993, \aaps, 101, 415

\bibitem[{{Chen} {et~al.}(2007){Chen}, {de Grijs}, \&
  {Zhao}}]{2007AJ....134.1368C}
{Chen}, L., {de Grijs}, R., \& {Zhao}, J.~L. 2007, \aj, 134, 1368

\bibitem[{{Dabringhausen} {et~al.}(2008){Dabringhausen}, {Hilker}, \&
  {Kroupa}}]{2008MNRAS.386..864D}
{Dabringhausen}, J., {Hilker}, M., \& {Kroupa}, P. 2008, \mnras, 386, 864

\bibitem[{{de Grijs} \& {Anders}(2006)}]{2006MNRAS.366..295D}
{de Grijs}, R. \& {Anders}, P. 2006, \mnras, 366, 295

\bibitem[{{de Grijs} {et~al.}(2004){de Grijs}, {Smith}, {Bunker}, {Sharp},
  {Gallagher}, {Anders}, {Lan{\c c}on}, {O'Connell}, \&
  {Parry}}]{2004MNRAS.352..263D}
{de Grijs}, R., {Smith}, L.~J., {Bunker}, A., {et~al.} 2004, \mnras, 352, 263

\bibitem[{{Fagiolini} {et~al.}(2007){Fagiolini}, {Raimondo}, \&
  {Degl'Innocenti}}]{2007A&A...462..107F}
{Fagiolini}, M., {Raimondo}, G., \& {Degl'Innocenti}, S. 2007, \aap, 462, 107

\bibitem[{{Fioc} \& {Rocca-Volmerange}(1997)}]{1997A&A...326..950F}
{Fioc}, M. \& {Rocca-Volmerange}, B. 1997, \aap, 326, 950

\bibitem[{{Gieles} {et~al.}(2007){Gieles}, {Athanassoula}, \& {Portegies
  Zwart}}]{2007MNRAS.376..809G}
{Gieles}, M., {Athanassoula}, E., \& {Portegies Zwart}, S.~F. 2007, \mnras,
  376, 809

\bibitem[{{Gieles} {et~al.}(2005){Gieles}, {Bastian}, {Lamers}, \&
  {Mout}}]{2005A&A...441..949G}
{Gieles}, M., {Bastian}, N., {Lamers}, H.~J.~G.~L.~M., \& {Mout}, J.~N. 2005,
  \aap, 441, 949

\bibitem[{{Gieles} \& {Baumgardt}(2008)}]{2008MNRAS.389L..28G}
{Gieles}, M. \& {Baumgardt}, H. 2008, \mnras, 389, L28

\bibitem[{{Gieles} {et~al.}(2006){Gieles}, {Portegies Zwart}, {Baumgardt},
  {Athanassoula}, {Lamers}, {Sipior}, \& {Leenaarts}}]{2006MNRAS.371..793G}
{Gieles}, M., {Portegies Zwart}, S.~F., {Baumgardt}, H., {et~al.} 2006, \mnras,
  371, 793

\bibitem[{{Giersz} \& {Heggie}(1997)}]{1997MNRAS.286..709G}
{Giersz}, M. \& {Heggie}, D.~C. 1997, \mnras, 286, 709

\bibitem[{{Gill} {et~al.}(2008){Gill}, {Trenti}, {Miller}, {van der Marel},
  {Hamilton}, \& {Stiavelli}}]{2008ApJ...686..303G}
{Gill}, M., {Trenti}, M., {Miller}, M.~C., {et~al.} 2008, \apj, 686, 303

\bibitem[{{Girardi} {et~al.}(2000){Girardi}, {Bressan}, {Bertelli}, \&
  {Chiosi}}]{2000A&AS..141..371G}
{Girardi}, L., {Bressan}, A., {Bertelli}, G., \& {Chiosi}, C. 2000, \aaps, 141,
  371

\bibitem[{{Goodwin} \& {Bastian}(2006)}]{2006MNRAS.373..752G}
{Goodwin}, S.~P. \& {Bastian}, N. 2006, \mnras, 373, 752

\bibitem[{{Gouliermis} {et~al.}(2004){Gouliermis}, {Keller}, {Kontizas},
  {Kontizas}, \& {Bellas-Velidis}}]{2004A&A...416..137G}
{Gouliermis}, D., {Keller}, S.~C., {Kontizas}, M., {Kontizas}, E., \&
  {Bellas-Velidis}, I. 2004, \aap, 416, 137

\bibitem[{{Harris}(1996)}]{1996AJ....112.1487H}
{Harris}, W.~E. 1996, \aj, 112, 1487

\bibitem[{{Henon}(1969)}]{1969A&A.....2..151H}
{Henon}, M. 1969, \aap, 2, 151

\bibitem[{{Hurley} {et~al.}(2000){Hurley}, {Pols}, \&
  {Tout}}]{2000MNRAS.315..543H}
{Hurley}, J.~R., {Pols}, O.~R., \& {Tout}, C.~A. 2000, \mnras, 315, 543

\bibitem[{{Hurley} {et~al.}(2004){Hurley}, {Tout}, {Aarseth}, \&
  {Pols}}]{2004MNRAS.355.1207H}
{Hurley}, J.~R., {Tout}, C.~A., {Aarseth}, S.~J., \& {Pols}, O.~R. 2004,
  \mnras, 355, 1207

\bibitem[{{Kalirai} {et~al.}(2008){Kalirai}, {Hansen}, {Kelson}, {Reitzel},
  {Rich}, \& {Richer}}]{2008ApJ...676..594K}
{Kalirai}, J.~S., {Hansen}, B.~M.~S., {Kelson}, D.~D., {et~al.} 2008, \apj,
  676, 594

\bibitem[{{Kassin} {et~al.}(2003){Kassin}, {Frogel}, {Pogge}, {Tiede}, \&
  {Sellgren}}]{2003AJ....126.1276K}
{Kassin}, S.~A., {Frogel}, J.~A., {Pogge}, R.~W., {Tiede}, G.~P., \&
  {Sellgren}, K. 2003, \aj, 126, 1276

\bibitem[{{Kroupa}(2001)}]{2001MNRAS.322..231K}
{Kroupa}, P. 2001, \mnras, 322, 231

\bibitem[{{Kruijssen}(2008)}]{2008A&A...486L..21K}
{Kruijssen}, J.~M.~D. 2008, \aap, 486, L21

\bibitem[{{Kruijssen} \& {Lamers}(2008)}]{2008A&A...490..151K}
{Kruijssen}, J.~M.~D. \& {Lamers}, H.~J.~G.~L.~M. 2008, \aap, 490, 151

\bibitem[{{Kundu} {et~al.}(2005){Kundu}, {Zepf}, {Hempel}, {Morton}, {Ashman},
  {Maccarone}, {Kissler-Patig}, {Puzia}, \& {Vesperini}}]{2005ApJ...634L..41K}
{Kundu}, A., {Zepf}, S.~E., {Hempel}, M., {et~al.} 2005, \apjl, 634, L41

\bibitem[{{K{\"u}pper} {et~al.}(2008){K{\"u}pper}, {Kroupa}, \&
  {Baumgardt}}]{2008MNRAS.389..889K}
{K{\"u}pper}, A.~H.~W., {Kroupa}, P., \& {Baumgardt}, H. 2008, \mnras, 389, 889

\bibitem[{{Lada} \& {Lada}(2003)}]{2003ARA&A..41...57L}
{Lada}, C.~J. \& {Lada}, E.~A. 2003, \araa, 41, 57

\bibitem[{{Lamers} {et~al.}(2006){Lamers}, {Anders}, \& {de
  Grijs}}]{2006A&A...452..131L}
{Lamers}, H.~J.~G.~L.~M., {Anders}, P., \& {de Grijs}, R. 2006, \aap, 452, 131

\bibitem[{{Lamers} {et~al.}(2005{\natexlab{a}}){Lamers}, {Gieles}, {Bastian},
  {Baumgardt}, {Kharchenko}, \& {Portegies Zwart}}]{2005A&A...441..117L}
{Lamers}, H.~J.~G.~L.~M., {Gieles}, M., {Bastian}, N., {et~al.}
  2005{\natexlab{a}}, \aap, 441, 117

\bibitem[{{Lamers} {et~al.}(2005{\natexlab{b}}){Lamers}, {Gieles}, \&
  {Portegies Zwart}}]{2005A&A...429..173L}
{Lamers}, H.~J.~G.~L.~M., {Gieles}, M., \& {Portegies Zwart}, S.~F.
  2005{\natexlab{b}}, \aap, 429, 173

\bibitem[{{Larsen} {et~al.}(2004){Larsen}, {Brodie}, \&
  {Hunter}}]{2004AJ....128.2295L}
{Larsen}, S.~S., {Brodie}, J.~P., \& {Hunter}, D.~A. 2004, \aj, 128, 2295

\bibitem[{{Larsen} \& {Richtler}(2004)}]{2004A&A...427..495L}
{Larsen}, S.~S. \& {Richtler}, T. 2004, \aap, 427, 495

\bibitem[{{Leitherer} {et~al.}(1999){Leitherer}, {Schaerer}, {Goldader},
  {Delgado}, {Robert}, {Kune}, {de Mello}, {Devost}, \&
  {Heckman}}]{1999ApJS..123....3L}
{Leitherer}, C., {Schaerer}, D., {Goldader}, J.~D., {et~al.} 1999, \apjs, 123,
  3

\bibitem[{{Lejeune} {et~al.}(1997){Lejeune}, {Cuisinier}, \&
  {Buser}}]{1997A&AS..125..229L}
{Lejeune}, T., {Cuisinier}, F., \& {Buser}, R. 1997, \aaps, 125, 229

\bibitem[{{Lejeune} {et~al.}(1998){Lejeune}, {Cuisinier}, \&
  {Buser}}]{1998A&AS..130...65L}
{Lejeune}, T., {Cuisinier}, F., \& {Buser}, R. 1998, \aaps, 130, 65

\bibitem[{{Maccarone} \& {Servillat}(2008)}]{2008MNRAS.389..379M}
{Maccarone}, T.~J. \& {Servillat}, M. 2008, \mnras, 389, 379

\bibitem[{{Maraston}(2005)}]{2005MNRAS.362..799M}
{Maraston}, C. 2005, \mnras, 362, 799

\bibitem[{{Marigo} {et~al.}(2008){Marigo}, {Girardi}, {Bressan}, {Groenewegen},
  {Silva}, \& {Granato}}]{2008A&A...482..883M}
{Marigo}, P., {Girardi}, L., {Bressan}, A., {et~al.} 2008, \aap, 482, 883

\bibitem[{{Marks} {et~al.}(2008){Marks}, {Kroupa}, \&
  {Baumgardt}}]{2008MNRAS.386.2047M}
{Marks}, M., {Kroupa}, P., \& {Baumgardt}, H. 2008, \mnras, 386, 2047

\bibitem[{{McLaughlin} \& {van der Marel}(2005)}]{2005ApJS..161..304M}
{McLaughlin}, D.~E. \& {van der Marel}, R.~P. 2005, \apjs, 161, 304

\bibitem[{{Mieske} {et~al.}(2008){Mieske}, {Hilker}, {Jord{\'a}n}, {Infante},
  {Kissler-Patig}, {Rejkuba}, {Richtler}, {C{\^o}t{\'e}}, {Baumgardt}, {West},
  {Ferrarese}, \& {Peng}}]{2008A&A...487..921M}
{Mieske}, S., {Hilker}, M., {Jord{\'a}n}, A., {et~al.} 2008, \aap, 487, 921

\bibitem[{{Mieske} \& {Kroupa}(2008)}]{2008ApJ...677..276M}
{Mieske}, S. \& {Kroupa}, P. 2008, \apj, 677, 276

\bibitem[{{Nomoto} \& {Hashimoto}(1988)}]{1988PhR...163...13N}
{Nomoto}, K. \& {Hashimoto}, M. 1988, \physrep, 163, 13

\bibitem[{{Odenkirchen} {et~al.}(2003){Odenkirchen}, {Grebel}, {Dehnen}, {Rix},
  {Yanny}, {Newberg}, {Rockosi}, {Mart{\'{\i}}nez-Delgado}, {Brinkmann}, \&
  {Pier}}]{2003AJ....126.2385O}
{Odenkirchen}, M., {Grebel}, E.~K., {Dehnen}, W., {et~al.} 2003, \aj, 126, 2385

\bibitem[{{Pietrinferni} {et~al.}(2004){Pietrinferni}, {Cassisi}, {Salaris}, \&
  {Castelli}}]{2004ApJ...612..168P}
{Pietrinferni}, A., {Cassisi}, S., {Salaris}, M., \& {Castelli}, F. 2004, \apj,
  612, 168

\bibitem[{{Pietrinferni} {et~al.}(2006){Pietrinferni}, {Cassisi}, {Salaris}, \&
  {Castelli}}]{2006ApJ...642..797P}
{Pietrinferni}, A., {Cassisi}, S., {Salaris}, M., \& {Castelli}, F. 2006, \apj,
  642, 797

\bibitem[{{Pryor} \& {Meylan}(1993)}]{1993ASPC...50..357P}
{Pryor}, C. \& {Meylan}, G. 1993, in Astronomical Society of the Pacific
  Conference Series, Vol.~50, Structure and Dynamics of Globular Clusters, ed.
  S.~G. {Djorgovski} \& G.~{Meylan}, 357--+

\bibitem[{{Rejkuba} {et~al.}(2007){Rejkuba}, {Dubath}, {Minniti}, \&
  {Meylan}}]{2007A&A...469..147R}
{Rejkuba}, M., {Dubath}, P., {Minniti}, D., \& {Meylan}, G. 2007, \aap, 469,
  147

\bibitem[{{Salpeter}(1955)}]{1955ApJ...121..161S}
{Salpeter}, E.~E. 1955, \apj, 121, 161

\bibitem[{{Schaerer} {et~al.}(1993){Schaerer}, {Meynet}, {Maeder}, \&
  {Schaller}}]{1993A&AS...98..523S}
{Schaerer}, D., {Meynet}, G., {Maeder}, A., \& {Schaller}, G. 1993, \aaps, 98,
  523

\bibitem[{{Schaller} {et~al.}(1992){Schaller}, {Schaerer}, {Meynet}, \&
  {Maeder}}]{1992A&AS...96..269S}
{Schaller}, G., {Schaerer}, D., {Meynet}, G., \& {Maeder}, A. 1992, \aaps, 96,
  269

\bibitem[{{Schulz} {et~al.}(2002){Schulz}, {Fritze-v.~Alvensleben},
  {M{\"o}ller}, \& {Fricke}}]{2002A&A...392....1S}
{Schulz}, J., {Fritze-v.~Alvensleben}, U., {M{\"o}ller}, C.~S., \& {Fricke},
  K.~J. 2002, \aap, 392, 1

\bibitem[{{Smith} {et~al.}(2007){Smith}, {Bastian}, {Konstantopoulos},
  {Gallagher}, {Gieles}, {de Grijs}, {Larsen}, {O'Connell}, \&
  {Westmoquette}}]{2007ApJ...667L.145S}
{Smith}, L.~J., {Bastian}, N., {Konstantopoulos}, I.~S., {et~al.} 2007, \apjl,
  667, L145

\bibitem[{{Spitzer} \& {Shull}(1975)}]{1975ApJ...201..773S}
{Spitzer}, Jr., L. \& {Shull}, J.~M. 1975, \apj, 201, 773

\bibitem[{{Tinsley}(1968)}]{1968ApJ...151..547T}
{Tinsley}, B.~M. 1968, \apj, 151, 547

\bibitem[{{Tinsley}(1980)}]{1980FCPh....5..287T}
{Tinsley}, B.~M. 1980, Fundamentals of Cosmic Physics, 5, 287

\bibitem[{{Tinsley} \& {Gunn}(1976)}]{1976ApJ...203...52T}
{Tinsley}, B.~M. \& {Gunn}, J.~E. 1976, \apj, 203, 52

\bibitem[{{Weidemann}(2000)}]{2000A&A...363..647W}
{Weidemann}, V. 2000, \aap, 363, 647

\bibitem[{{Weidemann} \& {Koester}(1983)}]{1983A&A...121...77W}
{Weidemann}, V. \& {Koester}, D. 1983, \aap, 121, 77

\bibitem[{{Worthey}(1994)}]{1994ApJS...95..107W}
{Worthey}, G. 1994, \apjs, 95, 107

\bibitem[{{Yi} {et~al.}(2001){Yi}, {Demarque}, {Kim}, {Lee}, {Ree}, {Lejeune},
  \& {Barnes}}]{2001ApJS..136..417Y}
{Yi}, S., {Demarque}, P., {Kim}, Y.-C., {et~al.} 2001, \apjs, 136, 417

\end{thebibliography}

\end{document}